\documentclass[symmetry,review,accept,moreauthors,pdftex]{mdpi}

\firstpage{1} 
\pubvolume{1}
\issuenum{1}
\articlenumber{0}
\pubyear{2021}
\copyrightyear{2020}
\datereceived{} 
\dateaccepted{} 
\datepublished{} 
\hreflink{https://doi.org/} 
\pdfoutput=1

\graphicspath{{../}}

\Title{Collider Searches for Dark Matter through the Higgs Lens}

\TitleCitation{Collider Searches for Dark Matter through the Higgs Lens}

\Author{Spyros Argyropoulos $^{1}$\orcidA{}, Oleg Brandt $^{2,*}$\orcidB{} and Ulrich Haisch $^{3}$\orcidC{}}

\AuthorNames{Spyros Argyropoulos, Oleg Brandt and Ulrich Haisch}

\AuthorCitation{Argyropoulos, S.; Brandt, O.; Haisch, U.}

\address{%
$^{1}$ \quad Physikalisches Institut, Albert-Ludwigs Universit\"at Freiburg; spyros.argyropoulos@cern.ch\\
$^{2}$ \quad Cavendish Laboratory, University of Cambridge; obrandt@hep.phy.cam.ac.uk\\
$^{3}$ \quad Max Planck Institut f\"ur Physik in M\"unchen; haisch@mpp.mpg.de}

\corres{Corresponding author}

\abstract{Despite the fact that dark matter constitutes one of the cornerstones of the standard cosmological paradigm, its existence has so far only been inferred from astronomical observations and its microscopic nature remains elusive. Theoretical arguments suggest that dark matter might be connected to the symmetry-breaking mechanism of the electroweak interactions or of other symmetries extending the Standard Model of particle physics. The resulting Higgs bosons, including the $125 \, {\rm GeV}$ spin-0 particle discovered recently at the Large Hadron Collider therefore represent a unique tool to search for dark matter candidates at collider experiments. This article reviews some of the relevant theoretical models as well as the results from the searches for dark matter in signatures that involve a Higgs-like particle at the Large Hadron Collider.}


\def\bm#1{\mbox{\boldmath$#1$\unboldmath}} 

\begin{document}

\tableofcontents

\addtocontents{toc}{\protect\setcounter{tocdepth}{2}}

\section{Introduction}
\label{sec:intro}

The concept of dark matter~(DM) was originally introduced to reconcile the observations of the high velocity dispersion of galactic clusters~\cite{Zwicky:1937zza} and the flat rotational curves of spiral galaxies~\cite{Rubin:1970zza} with the predictions of Newton's gravity. Since then, precise cosmological and astrophysical observations~\cite{Zyla:2020zbs} have strengthened the evidence for the existence of DM, establishing that around 25\% of the energy budget of the observable Universe consists of a matter component that in the standard DM picture is electrically neutral, weakly interacting and non-relativistic. 

Since the Standard Model (SM) of particle physics does not provide any candidate particle with the above characteristics, the existence of DM constitutes evidence for physics beyond the SM~(BSM). The detection and identification of the nature of DM at the microscopic level constitutes therefore one of the major challenges for particle physics, with complementary searches pursued by direct detection (DD), indirect detection~(ID) and collider experiments. The former two types of searches rely on the observation of recoils from the elastic scattering of DM particles on the nuclei in the detector material or on the observation of annihilation products of DM pairs such as monochromatic photons. For recent reviews of DD and ID search strategies see for instance~\cite{Schumann:2019eaa,Billard:2021uyg}. Collider searches for DM on the other hand rely on the production of DM particles in high-energy particle collisions and can be separated in two broad classes according to the experimental signature that they produce: (i)~searches for missing transverse momentum~($E_T^{\mathrm{miss}}$) plus $X$ signatures, also known as mono-$X$, where the $E_T^{\mathrm{miss}}$ resulting from the DM particles leaves the detectors unnoticed and the visible, i.e.~detectable, final state $X$ is used for triggering, and (ii)~searches containing only visible particles such as pairs of leptons or jets that aim to detect the particles mediating the interactions between the DM and the SM  particles through observation of a new resonance or a modification of the kinematics of the final-state particles. Both types of searches have been pursued at the Large~Hadron~Collider~(LHC) since its very beginning and previously at the Tevatron. See recent~reviews~\cite{Kahlhoefer:2017dnp,Penning:2017tmb,Boveia:2018yeb} for general overviews on DM collider phenomenology.

The discovery of a SM-like Higgs boson by the ATLAS and CMS collaborations~\cite{Aad:2012tfa,Chatrchyan:2012ufa} has opened up a new avenue in the searches for DM, allowing to probe the possible connection between the Higgs boson and the  dark sector, i.e.~a BSM sector that contains the DM particle and is almost decoupled from the SM.  In fact, there are both experimental and theoretical arguments that suggest that the Higgs boson partakes in the mediation between the  dark and the visible sector. Experimentally, the Higgs sector is compared to the gauge or fermionic sector of the SM far less explored and constrained, while theoretically the SM Higgs doublet plays a special role because it is the only SM field that allows to write down a renormalisable coupling to the dark sector, if DM is uncharged under the SM gauge group. The Higgs sector therefore offers an interesting portal to the hidden sector~\cite{Silveira:1985rk,Veltman:1989vw,Schabinger:2005ei,Patt:2006fw,OConnell:2006rsp,Kim:2006af,Barger:2007im,Kanemura:2010sh,Djouadi:2011aa,Mambrini:2011ik,Djouadi:2012zc,Alanne:2017oqj,Arcadi:2017kky,Balkin:2018tma,Ruhdorfer:2019utl,Arcadi:2019lka,Das:2020ozo,Arcadi:2021mag,Lebedev:2021xey} and dark sectors in particular. The goal of this review is to discuss experimental and theoretical aspects of BSM models that feature a DM-Higgs connection.

Given that the microscopic nature of DM is essentially unknown, there is a lot of freedom in the description of the interactions between the dark  and the visible sector, leading to a vast array of phenomenological models. The models discussed in this review can be broadly classified into four categories according to the structure of the dark and visible sector and the type of particle(s) mediating between the two sectors: 
\begin{itemize}

\item[(i)] Models in which the SM-like Higgs boson itself mediates between the dark sector and the~SM. This model class represents the simplest realisation of the Higgs portal idea with  a minimal particle content, containing besides the SM states only a single DM field that can be of spin-$0$, spin-$1/2$ or spin-$1$. The DM-SM interactions can be   formulated in terms of composite operators of dimension four and higher in an effective field theory~(EFT) framework, which allows to describe the DD, the ID and the collider phenomenology in a model-independent fashion. In~Section~\ref{sec:EFTHiggsPortal} we discuss in detail the simplest realisation of these Higgs portal EFTs, namely the case of a real singlet scalar field, considering both a marginal dimension-four and a derivative dimension-six coupling. Special emphasise  is thereby put on highlighting the complementary of the different non-collider and collider search strategies  in constraining the parameter space of the two models. We also briefly discuss the case of the Higgs portal model with fermionic DM that is a benchmark model used by both ATLAS and CMS to interpret their invisible Higgs decay searches. 

\item[(ii)] Models with an extended Higgs sector  in which a spin-0 particle that mixes with one of the visible non-SM Higgs bosons mediates between the dark and the visible sector. Our discussion of this  class of models is presented in Section~\ref{sec:extendedhiggsportals} and focuses on the 2HDM+$a$~\cite{Ipek:2014gua,No:2015xqa,Goncalves:2016iyg,Bauer:2017ota} and the 2HDM+$s$~\cite{Bell:2016ekl,Bell:2017rgi} models. These two models  are the simplest gauge-invariant and renormalisable models in this class that feature a fermionic DM candidate, and therefore represent   the natural extension of the simplified pseudoscalar and scalar DM models as defined in~\cite{Abdallah:2015ter,Abercrombie:2015wmb}. The particle content of these models involves besides the DM candidate four additional BSM spin-$0$ states. The~extra spin-0 states have an important impact on the collider phenomenology of the 2HDM+$a$ and  2HDM+$s$ models, as they allow for resonant mono-Higgs, mono-$Z$ and $tW+E_T^{\rm miss}$ production, thereby leading to a far richer mono-$X$ phenomenology at the LHC when compared to the spin-0 simplified DM models. We review the state-of-the-art of the phenomenology and experimental constraints on the  2HDM+$a$ and  2HDM+$s$ models, pointing out in particular similarities and difference between the two models. 

\item[(iii)] Models with extended Higgs and gauge sectors in which both a spin-0 and a spin-1 portal connect the dark and the visible sector. The resulting theories fall into the class of  dark Higgs or dark $Z^\prime$ models, which
  typically have a rich collider and DM phenomenology. In Section~\ref{sec:extendedhiggsgaugeportal} we discuss two representative models in more detail: the 2HDM+$Z^\prime$ model~\cite{Berlin:2014cfa} and the dark Higgs two mediator DM~(2MDM) model~\cite{Duerr:2016tmh}, which were used to guide and interpret existing mono-Higgs searches by ATLAS and CMS. While the scalar sectors of this models are quite different, both the 2HDM+$Z^\prime$ and  the 2MDM model contain a spin-1 mediator that couples the DM particles to some of the SM states. This feature allows to the test the models by searching for resonant production of non-$E_T^{\rm miss}$ final states such as dijets, $t \bar t$ and $Zh$. Using the latest available LHC data, we derive the constraints on the 2HDM+$Z^\prime$ and the 2MDM model that arise from the relevant searches for resonant SM final states and the mono-jet signature. In both cases we show that for the benchmark scenarios  considered by ATLAS and CMS, non-$E_T^{\rm miss}$ searches exclude additional parameter space not probed by the existing mono-Higgs interpretations.
   
\item[(iv)] Models that feature exotic decays of the $125 \, {\rm GeV}$ Higgs boson into hidden sector particles with macroscopic proper decay lengths of $c \tau>10^{-2} \, {\rm m}$. Due to the long-lived nature of the hidden particles, their decays into SM particles are displaced from the primary interaction vertex, which represents a striking experimental signature. In~Section~\ref{sec:LLP} we discuss three representative models that feature such long-lived particles~(LLPs). These models derive either from the idea of neutral naturalness~\cite{Chacko:2005pe,Burdman:2006tz,Cai:2008au,Craig:2014aea,Curtin:2015fna} or involve a hypercharge portal~\cite{Curtin:2013fra,Curtin:2014cca} or a hypercharge and a fermion portal~\cite{Arkani-Hamed:2008kxc,Baumgart:2009tn,Cheung:2009su,Falkowski:2010cm,Falkowski:2010gv}. While the spin of the particle that connects the hidden and the visible sector differs, all three models share a common feature: they can lead to both prompt and displaced signatures depending on the strength of the interaction that links the two sectors. ATLAS, CMS and in some cases also LHCb have already searched for exotic Higgs decay signals, and we compare the available findings in a set of summary plots. Whenever possible, we include in these plots also other experimental limits to emphasise the unique role that searches for LLPs can play in constraining~the parameter space of the~considered theories.

\end{itemize}

\begin{table}[t!]
\def\arraystretch{1.25}
\centering
\begin{tabular}{|c|c|c|} 
 \hline
Signature & Model & References \\
 \hline \hline
\multirow{5}{*}{$h \to {\rm inv}$} & EFT Higgs portals & Sections~\ref{subsec:EFTHiggsPortal1} and \ref{subsec:further}, Figures~\ref{fig:summaryportal} and \ref{fig:ATLASsummary} \\ 
 & 2HDM+$a$ & Section~\ref{sec:2HDMa}  \\ 
 & Neutral naturalness & Section~\ref{sec:NN}, Figure~\ref{fig:HV} \\ 
 & Dark photons & Section~\ref{sec:darkphoton}, Figures~\ref{fig:darkphoton1} and~\ref{fig:darkphoton2} \\  
 & Vector plus fermion portal & Section~\ref{sec:FRVZ}, Figure~\ref{fig:FRVZsummary} \\  \hline 
\multirow{1}{*}{${\rm VBF} + E_T^{\rm miss}$} &EFT Higgs portals & Section~\ref{subsec:EFTHiggsPortal1}, Figure~\ref{fig:summaryportal} \\ \hline
\multirow{2}{*}{$t\bar t+ E_T^{\rm miss}$} &EFT Higgs portals & Section~\ref{subsec:EFTHiggsPortal1}, Figure~\ref{fig:summaryportal} \\ 
& 2MDM & Section~\ref{sec:darkHiggs} \\\hline
\multirow{2}{*}{$tW+ E_T^{\rm miss}$} &EFT Higgs portals& Section~\ref{subsec:EFTHiggsPortal1}, Figure~\ref{fig:summaryportal} \\ 
& 2HDM+$a$ & Section~\ref{sec:2HDMa}, Figures~\ref{fig:2hdma_2D_exclusion} and \ref{fig:2hdma_tanb_exclusion}  \\ \hline
\multirow{3}{*}{$h+ E_T^{\rm miss}$} & 2HDM+$a$ & Section~\ref{sec:2HDMa}, Figures~\ref{fig:2hdma_2D_exclusion}, \ref{fig:2hdma_tanb_exclusion}, \ref{fig:2hdma_sinp_exclusion}  and \ref{fig:2hdma_mdm}\\ 
& 2HDM+$s$ & Section~\ref{sec:2HDMs}, Figure~\ref{fig:2hdmsexclusion} \\ 
& 2HDM+$Z^\prime$ & Section~\ref{sec:2HDMZp}, Figure~\ref{fig:2HDMZpMM} \\ \hline
\multirow{2}{*}{$Z+ E_T^{\rm miss}$} & 2HDM+$a$ & Section~\ref{sec:2HDMa}, Figures~\ref{fig:2hdma_2D_exclusion}, \ref{fig:2hdma_tanb_exclusion}, \ref{fig:2hdma_sinp_exclusion}  and \ref{fig:2hdma_mdm} \\ 
& 2HDM+$s$ & Section~\ref{sec:2HDMs}, Figure~\ref{fig:2hdmsexclusion} \\  \hline
\multirow{1}{*}{$s+ E_T^{\rm miss}$} &  2MDM &  Section~\ref{sec:darkHiggs}, Figure~\ref{fig:monoS}  \\ \hline 
\multirow{1}{*}{$j + E_T^{\rm miss}$} & 2MDM & Section~\ref{sec:darkHiggs}, Figure~\ref{fig:monoS} \\ \hline \hline
\multirow{1}{*}{$H^{\pm} \to  t b$} & 2HDM+$a$ & Section~\ref{sec:2HDMa}, Figures~\ref{fig:2hdma_2D_exclusion} and \ref{fig:2hdma_tanb_exclusion} \\   \hline
\multirow{1}{*}{$h \to 4 f$} & 2HDM+$a$ & Section~\ref{sec:2HDMa}  \\ \hline
\multirow{2}{*}{Dijets} & 2HDM+$Z^\prime$ & Section~\ref{sec:2HDMZp}, Figures~\ref{fig:2HDMZpMM} and~\ref{fig:2HDMZpMg} \\ 
& 2MDM & Section~\ref{sec:darkHiggs}, Figure~\ref{fig:monoS} \\ \hline
\multirow{2}{*}{$Z^\prime \to t \bar t$}  & 2HDM+$Z^\prime$ & Section~\ref{sec:2HDMZp}, Figures~\ref{fig:2HDMZpMM} and~\ref{fig:2HDMZpMg} \\ 
& 2MDM & Section~\ref{sec:darkHiggs}, Figure~\ref{fig:monoS} \\ \hline
\multirow{1}{*}{$Z^\prime \to Zh$}  & 2HDM+$Z^\prime$ & Section~\ref{sec:2HDMZp}, Figures~\ref{fig:2HDMZpMM} and~\ref{fig:2HDMZpMg} \\ \hline \hline
\multirow{3}{*}{$h \to 4 f$} & Neutral naturalness & Section~\ref{sec:NN}, Figure~\ref{fig:HV} \\
& Dark photons & Section~\ref{sec:darkphoton}, Figures~\ref{fig:darkphoton1} and~\ref{fig:darkphoton2}  \\ 
& Vector plus fermion portal & Section~\ref{sec:FRVZ}, Figure~\ref{fig:FRVZsummary} \\ \hline
\multirow{2}{*}{$h \to {\rm inv}, {\rm undet}$} & Neutral naturalness & Section~\ref{sec:NN}, Figure~\ref{fig:HV} \\
& Dark photons & Section~\ref{sec:darkphoton}, Figure~\ref{fig:darkphoton1}  \\  \hline
\multirow{2}{*}{Dileptons} & Dark photons & Section~\ref{sec:darkphoton}, Figure~\ref{fig:darkphoton2}  \\  
& Vector plus fermion portal & Section~\ref{sec:FRVZ}, Figure~\ref{fig:FRVZsummary} \\ \hline
\end{tabular}
\vspace{4mm}
\caption{\label{tab:summaryforthebusyexpermentalist}  The list of LHC signatures most relevant to this review, the models that they constrain, and the section(s) and/or figures(s) where the corresponding discussion can be found in the manuscript. The~LHC signatures are grouped by double lines into three classes: (i) processes with a significant amount of $E_T^{\rm miss}$, (ii) prompt signals involving SM final states and (iii) signatures relevant in the context of LLP searches. In each class the signatures are ordered as they appear in the text.}
\end{table}

Each of the four sections mentioned above is structured in a similar fashion. We~first present the most relevant theoretical aspects of the considered models and then discuss the experimental constraints that apply in each case, focusing in many but not all cases on the collider bounds. The relevant constraints are combined into state-of-the-art summary plots in various benchmark scenarios of the examined DM models. Whenever several results that address a particular signature for a given model are available, we focus on the most recent measurements that typically provide the highest sensitivity. Hence, most ATLAS and CMS searches presented here are based on the full LHC~Run~2 data set of around $140 \, {\rm fb}^{-1}$ collected at $13 \, {\rm TeV}$. For the readers mostly interested in collider phenomenology, Table~\ref{tab:summaryforthebusyexpermentalist} provides a list of the LHC signatures that are discussed in this review, indicating which model is constrained by a given search and the place(s) where the corresponding discussion can be found. Our whole review is tied together in~Section~\ref{sec:conclusions} where we present an outlook. We commence without further ado.

\section{Higgs portal models} 
\label{sec:EFTHiggsPortal}

One of the special features of the SM Higgs doublet $H$ is that $H^\dagger H$ is the only Lorentz and gauge invariant operator with mass dimension of two. The operator~$H^\dagger H$  therefore furnishes a portal to the dark or hidden sector~\cite{Silveira:1985rk,Veltman:1989vw,Schabinger:2005ei,Patt:2006fw,OConnell:2006rsp,Kim:2006af,Barger:2007im,Kanemura:2010sh,Djouadi:2011aa,Mambrini:2011ik,Djouadi:2012zc,Alanne:2017oqj,Arcadi:2017kky,Balkin:2018tma,Ruhdorfer:2019utl,Arcadi:2019lka,Das:2020ozo,Arcadi:2021mag,Lebedev:2021xey}. In~particular, at the level of dimension-four operators one can write down couplings of~$H^\dagger H$   to dark spin-0 and spin-1 fields, while the leading interactions with dark spin-$1/2$ fields are of dimension five. If the resulting EFT is equipped with a suitable symmetry the dark field becomes stable giving rise to a scalar, vector and fermionic DM candidate, respectively. Such a symmetry can for instance be a $ \mathbb{Z}_2$ exchange symmetry. 

\subsection{Theory}
 
In this section, we will consider the simplest possibility of these Higgs portal models, namely the case of a real scalar~$\phi$ that is a singlet under the SM gauge group, but odd under a $ \mathbb{Z}_2$ symmetry, i.e.~$\phi \to -\phi$. This guarantees the stability of $\phi$ making it a suitable DM candidate. The~interactions between the dark sector and the SM that we consider are 
 \begin{equation} \label{eq:LphiH}
 {\cal L}_{\phi H} = c_m \phi^2 (H^\dagger H) + \frac{c_d}{\Lambda^2}  \left ( \partial_\mu \phi^2 \right )  \left ( \partial^\mu  (H^\dagger H) \right ) \,, 
 \end{equation}
where the first (second) term is the so-called marginal (derivative) Higgs portal, the parameter $c_m$~($c_d$)~denotes the corresponding coupling or Wilson coefficient and $\Lambda$~is a mass scale that suppresses the derivative Higgs portal that corresponds to a dimension-six operator. Such a derivative coupling with the Higgs field arises in models where DM is a pseudo Nambu-Goldstone boson~(pNGB)~\cite{Frigerio:2012uc,Chala:2012af,Marzocca:2014msa,Fonseca:2015gva,Brivio:2015kia,Chala:2016ykx,Barducci:2016fue,Wu:2017iji,Balkin:2017aep,Balkin:2017yns,Gross:2017dan,Balkin:2018tma,Ishiwata:2018sdi,Davoli:2019tpx,Ruhdorfer:2019utl,Ramos:2019qqa,Xing:2020uaf,Coito:2021fgo,Haisch:2021ugv}. In~such a case $\Lambda$ is associated with the scale of global symmetry breaking that gives rise to the appearance of the pNGB(s). Besides the two types of interactions introduced in~(\ref{eq:LphiH}), explicit spin-0 ultraviolate (UV) completions of Higgs portal models can contain additional operators (see~\cite{Alanne:2017oqj,Ruhdorfer:2019utl} for a full classification of operators up to dimension six). In order to highlight the complementarity of collider and non-collider bounds on Higgs portal models in a simple fashion, we focus in what follows on the subclass of models in which the leading effects are well captured by the EFT Lagrangian ${\cal L}_{\phi H}$. After drawing our general conclusions we will however also briefly discuss the possible impact of other operators not included in~(\ref{eq:LphiH}). Recent detailed phenomenological studies of Higgs portal models with vector and fermionic DM can be found for instance in~\cite{Arcadi:2019lka,Arcadi:2021mag}. See also the ATLAS and CMS publications~\cite{CMS:2016dhk,CMS:2019bke,CMS:2018yfx,ATLAS:2018bnv,ATLAS:2019cid,ATLAS:2020kdi}.

 \subsection{Collider constraints}
 \label{subsec:EFTHiggsPortal1}
 
\begin{figure}[t!]
\centering
\includegraphics[width=.6\linewidth]{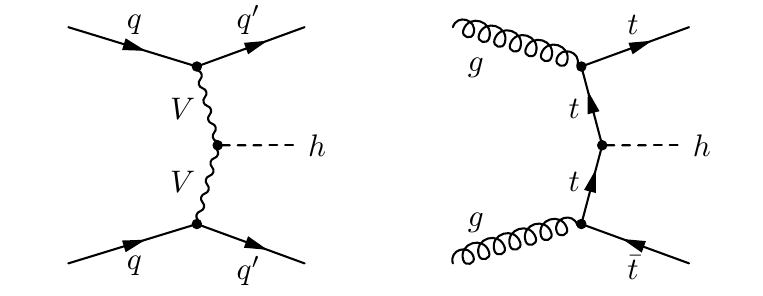}

\vspace{4mm}

\hspace{3mm} \includegraphics[width=.65\linewidth]{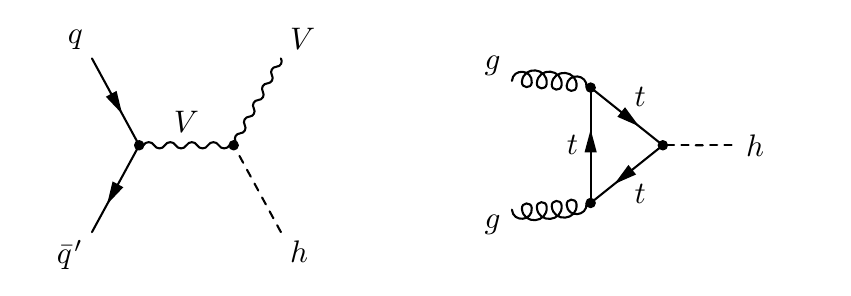}

\vspace{2mm}
\caption{\label{fig:higgsinv} Example Feynman diagrams of Higgs production in vector-boson fusion~(VBF), in association with a pair of top quarks, in association with a vector boson~($Vh$) and in gluon-gluon fusion~(ggF) in the upper left, upper right, lower left and lower right, respectively. 
}
\end{figure}
 
A common feature of Higgs portal models is that they predict Higgs to invisible decays if the DM candidate is kinematically accessible,~i.e.~$m_\phi < m_h/2$ in the case of the real scalar~$\phi$ with mass $m_\phi$  and $m_h \simeq 125 \, {\rm GeV}$  the mass of the SM-like Higgs boson. The~most important Higgs production channels  for searches of Higgs to invisible decays are displayed in~Figure~\ref{fig:higgsinv}. For the effective interactions~(\ref{eq:LphiH}) the relevant partial Higgs decay width reads 
\begin{equation} \label{eq:hDMDMwidth}
\Gamma \left ( h \to \phi \phi  \right ) = \frac{v^2}{8 \pi  m_h} \left ( 1 -  \frac{4 m_\phi^2}{m_h^2} \right )^{1/2} \,  \left (c_m + \frac{m_h^2 c_d }{\Lambda^2}  \right )^2 \,.
\end{equation}
where $v \simeq 246 \, {\rm GeV}$ denotes the vacuum expectation value~(VEV) of the $125 \, {\rm GeV}$ Higgs boson. The formula~(\ref{eq:hDMDMwidth}) can be used to translate experimental limits on the Higgs to invisible branching ratio ${\rm BR} \left (h \to {\rm inv} \right)$ into constraints on the strength of the marginal and derivative Higgs portals. In~fact, in the limit $m_\phi \ll m_h$ the best existing 95\% confidence level~(CL) exclusion LHC bound~\cite{ATLAS:2020kdi} of
\begin{equation} \label{eq:BRhinv}
{\rm BR} \left (h \to {\rm inv} \right) < 0.11 \,,
\end{equation}
leads to 
\begin{equation}\label{eq:hinvbound}
|c_m| < 5.1 \cdot 10^{-3} \,, \qquad 
\frac{\Lambda}{\sqrt{|c_d|}} > 1.7 \, {\rm TeV} \,,
\end{equation}
when the SM value $\Gamma_h^{\rm SM} \simeq 4.07 \, {\rm MeV}$~\cite{Zyla:2020zbs} of the total Higgs decay width is used. Notice that the bound~(\ref{eq:BRhinv}) results from a statistical combination of searches for invisible Higgs decays, where the Higgs is produced according to the SM via VBF (see~the upper left Feynman diagram in~Figure~\ref{fig:higgsinv}) or in association with a pair of top quarks (see~the upper right Feynman diagram in~Figure~\ref{fig:higgsinv}) in final states with zero or two leptons. At the high-luminosity upgrade of the LHC~(HL-LHC) it may be possible to set a limit on the Higgs to invisible branching ratio of ${\rm BR} \left (h \to {\rm inv} \right) < 2.5 \cdot 10^{-2}$~\cite{Cepeda:2019klc}. This implies that the bounds~(\ref{eq:hinvbound}) may be improved to $2.3 \cdot 10^{-3}$ and $2.6 \, {\rm TeV}$ by the end of the LHC era.

\begin{figure}[t!]
\centering
\includegraphics[width=.85\linewidth]{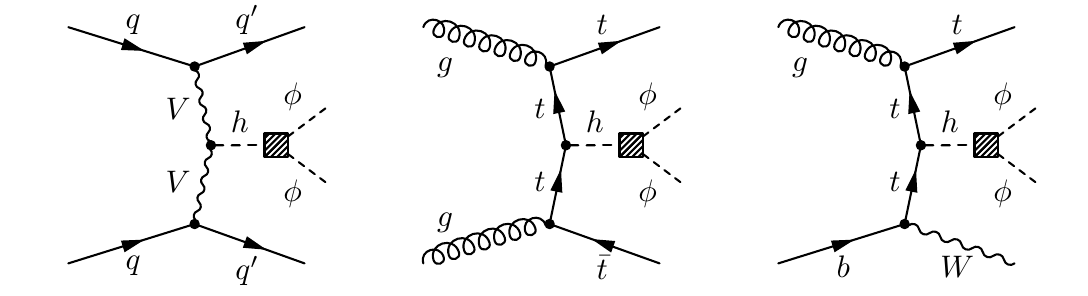}
\vspace{4mm}
\caption{\label{fig:higgsportal} Examples of Feynman diagrams that give rise to $E_T^{\rm miss}$ signatures in VBF Higgs~(left), $t \bar t$~(middle) and $tW$~(right) production. The shaded squares indicate an insertion of an effective  operator~(\ref{eq:LphiH}) while the black dotes correspond to  a SM interaction. See text for further explanations.}
\end{figure}

If the DM candidate is too heavy to be  pair produced as a real particle in the decay of the $125 \, {\rm GeV}$ Higgs boson, $E_T^{\rm miss}$ signatures still arise from off-shell Higgs production. Possible channels to search for signals of this kind are VBF Higgs production in the $jj + E_T^{\rm miss}$ channel as well as $t \bar t + E_T^{\rm miss}$ and $t W + E_T^{\rm miss}$ production. Relevant Feynman diagrams are shown in~Figure~\ref{fig:higgsportal}. The LHC reach of these channels in the context of~(\ref{eq:LphiH})  has been studied recently in~\cite{Ruhdorfer:2019utl,Haisch:2021ugv}, and we will summarise the main findings of these articles~below. 

 \subsection{DM phenomenology}

DM states that couple to the $125 \, {\rm GeV}$ Higgs typically lead to spin-independent~(SI) DM-nucleon cross sections~($\sigma_{\rm SI}$), which are severely constrained by the existing DM~DD experiments like XENON1T. While the marginal Higgs portal leads to an unsuppressed SI~DM-nucleon cross section, the DM-nucleon interactions that are mediated by the derivative Higgs portal are suppressed by $q^2/\Lambda^2 \lesssim (100 \, {\rm MeV})^2/\Lambda^2$ where $q^2$ characterises the momentum transfer in DM scattering with heavy nuclei. Explicitly one finds
\begin{equation} \label{eq:sigmaSI}
\sigma_{\rm SI}^{\phi N} = \frac{c_m^2 m_N^4 f_N^2}{\pi m_h^4 \left ( m_\phi + m_N \right )^2 } \,, 
\end{equation}
where $m_N \simeq 939 \, {\rm MeV}$ is the average of the nucleon mass and $f_N \simeq 0.31$~\cite{Alarcon:2011zs,Alarcon:2012nr,Junnarkar:2013ac,Hoferichter:2015dsa} parameterises the  strength of the Higgs-nucleon interactions. For $m_\phi =100 \, {\rm GeV}$ the latest XENON1T  90\%~CL upper limit on the SI~DM-nucleon cross section reads $\sigma_{\rm SI} < 9.12 \cdot 10^{-47} \, {\rm cm^2}$~\cite{XENON:2018voc}. By means of~(\ref{eq:sigmaSI})  this bound can be translated into a limit on the marginal Higgs portal coupling:
\begin{equation} \label{eq:DDmar}
|c_m| < 5.0 \cdot 10^{-3} \,.
\end{equation}
In contrast, the derivative Higgs portal coupling $c_d$ remains unconstrained by DD experiments due to the aforementioned momentum suppression. In fact, it turns out that up to dimension-six the derivative Higgs portal is the only spin-0 DM-Higgs operator that does naturally satisfy the constraints imposed by $\sigma_{\rm SI}$ once radiative corrections are considered~\cite{Haisch:2021ugv}.

In order to understand the physics of DM ID and thermal-freeze out in  Higgs portal models described by~(\ref{eq:LphiH}), let us write the velocity-averaged cross section for the annihilation of DM into a SM final state $X$ as
\begin{equation} \label{eq:annxsec}
\left \langle  \sigma \left (\phi \phi \to X \right ) v \right  \rangle \left ( T \right ) =  a_X + T  b_X \,,
\end{equation}
where $T$ denotes the DM temperature. Notice that in today's Universe $T_0 \simeq 0$, while at freeze-out $T_f \simeq m_\phi/25$. The $p$-wave coefficient $b_X$ can therefore usually be neglected in the calculation of the~ID constraints. However, it can be relevant in the case of the DM relic density~($\Omega_{\rm DM} h^2$), in particular, if the $s$-wave coefficient $a_X$ is parametrically suppressed. 

For $m_b < m_\phi \lesssim m_W$ with $m_b \simeq 4.2 \, {\rm GeV}$ ($m_W\simeq 80.4 \, {\rm GeV}$)   the bottom-quark ($W$-boson) mass, DM annihilation into bottom-antibottom quark pairs is   the dominant contribution to~$\Omega_{\rm DM} h^2$. The~corresponding  $s$-wave coefficient reads 
\begin{equation} \label{eq:abbbbb}
a_{b \bar b}  = \frac{3 \hspace{0.25mm} m_b^2}{\pi} \left |   \frac{1}{4  m_\phi^2 - m_h^2 + i  m_h\Gamma_h} \, \left ( c_m + \frac{4   m_\phi^2   c_d}{\Lambda^2}   \right )  \right |^2  \,,
\end{equation}
with $\Gamma_h$ denoting the total  decay width  of the $125 \, {\rm GeV}$ Higgs boson including contributions from $h \to \phi \phi$~$\big($see~(\ref{eq:hDMDMwidth})$\big)$. In the case $m_\phi \gtrsim m_W$ the $\phi \phi \to W^+ W^-, ZZ, hh, t \bar t$ channels  dominate DM annihilation. These~processes all receive unsuppressed $s$-wave contributions. For DM masses sufficiently far above $v$  the relevant coefficients take the following form 
\begin{equation} \label{eq:aVVahhatt} 
a_{X}  = \frac{N_X  m_\phi^2}{\pi} \left ( \frac{c_m}{4 m_\phi^2}  + \frac{c_d}{\Lambda^2}  \right )^2 \,, \qquad 
a_{t \bar t}  = \frac{3  m_t^2}{\pi} \left ( \frac{c_m}{4 m_\phi^2}  + \frac{c_d}{\Lambda^2}   \right )^2  \,, 
\end{equation}
with $X = W^+ W^-, ZZ, hh$ and $N_{W^+W^-} = 2$, $N_{ZZ} = N_{hh} = 1$. Notice that in the case of $m_\phi \gg v$, DM annihilation to $W$ and $Z$ bosons reduces to three times the contribution from annihilation to the $125 \, {\rm GeV}$ Higgs boson. This is an expected feature in the $SU(2)_L \times U(1)_Y$ symmetric~limit. In addition to the DM annihilation channels discussed above, DM annihilation into monochromatic photons can  also be relevant   in the context of~(\ref{eq:hDMDMwidth}). The corresponding formulas can be found for instance in~\cite{Haisch:2021ugv}. 

In terms of~(\ref{eq:abbbbb}) and~(\ref{eq:aVVahhatt}),  today's  DM relic density is approximately given by
\begin{equation} \label{eq:omegapocket}
\frac{\Omega_{\rm DM} h^2}{0.12}=  \frac{3 \cdot 10^{-26} \, {\rm cm}^3/{\rm s}}{ \langle \sigma v \rangle_f} \,, \qquad 
\langle \sigma v \rangle_f = \sum_{X} \left \langle  \sigma \left (\phi \phi \to X \right ) v \right  \rangle \big ( T_f \big ) \,,
\end{equation}
where the sum over $X$ involves all annihilation channels that are kinematically open at a given value of $m_{\phi}$. While the above formulas represent useful expressions to estimate~$\Omega_{\rm DM} h^2$, we will  use~{\tt micrOMEGAs}~\cite{Belanger:2018ccd} in our numerical analysis to obtain the constraints on the parameter space of~(\ref{eq:LphiH}) that follow from the   PLANCK measurement $\Omega_{\rm DM} h^2 = 0.120 \pm 0.001$~\cite{Planck:2018vyg}. The ID exclusions shown below in Figure~\ref{fig:summaryportal} are also determined with the help of  {\tt micrOMEGAs}. 

\subsection{Summary plots}
\label{sec:2.4}

\begin{figure}[!t]
\begin{center}
\includegraphics[width=.65\linewidth]{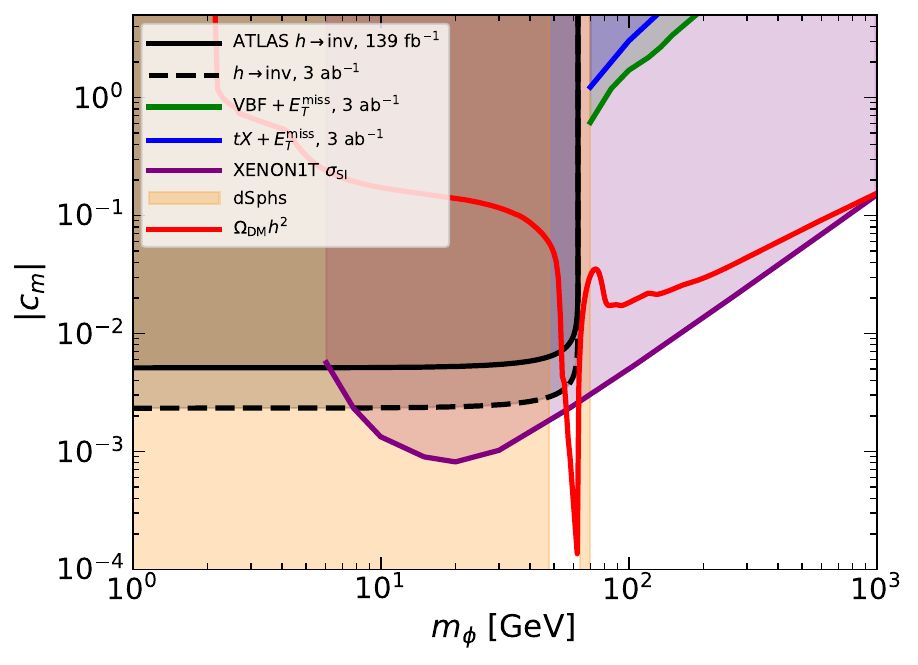} 

\vspace{2mm} 

\includegraphics[width=.65\linewidth]{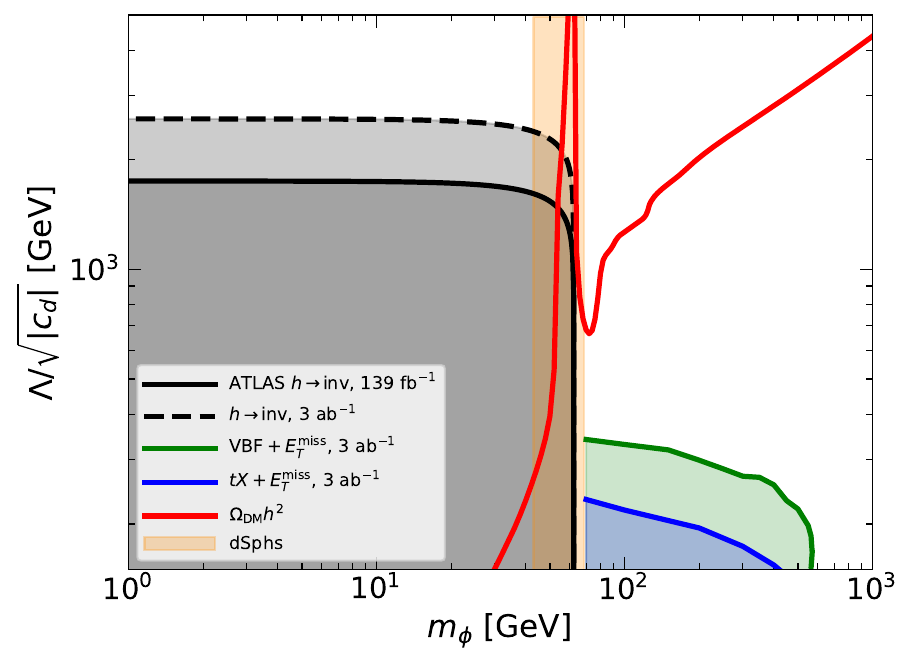} 
\vspace{2mm} 
\caption{\label{fig:summaryportal} Constraints for the marginal (derivative) Higgs portal model in the upper~(lower) panel. The~solid black contours correspond to~(\ref{eq:BRhinv}) while the  interpretations of the HL-LHC 95\%~CL limit ${\rm BR} \left (h \to {\rm inv} \right) < 2.5 \cdot 10^{-2}$~\cite{Cepeda:2019klc} are indicated by dashed black lines. The purple region in the upper panel is disfavoured by the 90\%~CL bound on the SI~DM-nucleon cross section $\sigma_{\rm SI}$  set by XENON1T~\cite{XENON:2018voc}. The~vertical orange shaded bands display the range of DM masses that is excluded at 95\%~CL by Fermi-LAT and~DES~\cite{Fermi-LAT:2016uux}. The red curves correspond to the value~$\Omega_{\rm DM} h^2 = 0.12$~\cite{Planck:2018vyg}. In~the parameter space below (above) the red curve the Universe is overclosed in the case of the marginal (derivative) Higgs portal model. The green regions indicate the 95\%~CL exclusion limit obtained in~\cite{Ruhdorfer:2019utl} from a study of off-shell Higgs production in the~${\rm VBF} + E_T^{\rm miss}$ channel, while the blue regions represent the corresponding exclusion limits derived  in~\cite{Haisch:2021ugv} 
from a study of $t X+ E_T^{\rm miss}$ final states. For further details see the main text.}
\end{center}
\end{figure}

The upper~(lower) panel in~Figure~\ref{fig:summaryportal} summarises the most important constraints on the marginal~(derivative) Higgs portal introduced in~(\ref{eq:LphiH}). The~solid black contours correspond to the current best limit on ${\rm BR} \left (h \to {\rm inv} \right) $ as given in~(\ref{eq:BRhinv}) while the dashed black lines represent the expected HL-LHC 95\%~CL limit ${\rm BR} \left (h \to {\rm inv} \right) < 2.5 \cdot 10^{-2}$~\cite{Cepeda:2019klc}. The purple region in the upper plot is disfavoured by the 90\%~CL bounds of XENON1T~\cite{XENON:2018voc} on  $\sigma_{\rm SI}$. The~vertical orange shaded bands indicate the DM mass ranges that are excluded at 95\%~CL by the $\gamma$-ray observations of dwarf spheroidal galaxies~(dSphs) of the Fermi-LAT and DES collaborations reported in~\cite{Fermi-LAT:2016uux}.~The used experimental bounds assume DM annihilation via $\phi \phi \to b \bar b$ and that $\Omega_{\rm DM} h^2=0.12$. Compared to $\phi \phi \to b \bar b$, the constraints that follow from the latest Fermi-LAT search for monochromatic photons~\cite{Fermi-LAT:2015kyq} lead to weaker constraints. These limits are hence not shown in the figure. In the parameter space below (above) the red curve, the marginal (derivative) Higgs portal model predicts $\Omega_{\rm DM} h^2 > 0.12$, i.e.~larger values of the DM relic density compared to the PLANCK measurement~\cite{Planck:2018vyg}.  The green regions  correspond to the 95\%~CL exclusion limits found in~\cite{Ruhdorfer:2019utl} from a  study of off-shell invisible Higgs production in the~${\rm VBF} + E_T^{\rm miss}$ channel. Finally, the blue domains represent the 95\%~CL constraints obtained by the combined $t \bar t + E_T^{\rm miss}$ and $t W+ E_T^{\rm miss}$~($t X+ E_T^{\rm miss}$) analysis strategy discussed in~\cite{Haisch:2021ugv}. The latter two types of collider limits assume an integrated luminosity of $3 \, {\rm ab}^{-1}$ collected at the HL-LHC.  

From the upper panel in~Figure~\ref{fig:summaryportal} it is evident that  the constraints on the Wilson coefficient~$c_m$ of the marginal Higgs portal from searches for Higgs to invisible decays at the LHC  are more stringent than the DD bounds for DM masses $m_\phi \lesssim 5 \, {\rm GeV}$, while in the range   $5 \, {\rm GeV} \lesssim m_\phi < m_h/2$ they are roughly comparable in strength. In the case~$m_\phi > m_h/2$, the bounds that follow from~$\sigma_{\rm SI}$ are however by more than two orders of magnitude stronger than those that  mono-$X$ searches at the HL-LHC are expected to set. Off-shell invisible Higgs production in the~VBF channel~\cite{Ruhdorfer:2019utl} is likely  the best  probe of the marginal~Higgs portal at the LHC if $m_\phi > m_h/2$. Notice however  that the study~\cite{Ruhdorfer:2019utl} assumes a  systematic uncertainty  of~$1\%$, while the  shown  $tX+E_T^{\rm miss}$ exclusion limits are based on  a  systematic uncertainty  of~$15\%$~\cite{Haisch:2021ugv}. Assuming a reduction of  background uncertainties in  $tX+E_T^{\rm miss}$ down to~$5\%$ would bring the ${\rm VBF} + E_T^{\rm miss}$ and $tX+E_T^{\rm miss}$ exclusion limits closer together. Combining the two  mono-$X$ channels as done in the case of  the LHC searches for the invisible Higgs boson decays~(cf.~for instance~\cite{CMS:2016dhk,CMS:2019bke,CMS:2018yfx,ATLAS:2018bnv,ATLAS:2019cid,ATLAS:2020kdi}) can be expected to improve the~HL-LHC reach. The potential of the high-energy option of the LHC, the   future circular hadron-hadron collider,  the compact linear collider and  a muon collider  in constraining the marginal Higgs portal through ${\rm VBF} + E_T^{\rm miss}$ off-shell Higgs production has been studied  recently in~\cite{Ruhdorfer:2019utl}.  For earlier analyses see also~\cite{Matsumoto:2010bh,Kanemura:2011nm,Chacko:2013lna,Craig:2014lda,Ko:2016xwd}.

For what concerns  the derivative Higgs portal model, the lower panel in~Figure~\ref{fig:summaryportal} shows that in the Higgs on-shell region, i.e.~for $m_\phi < m_h/2$, HL-LHC measurements of invisible Higgs decays  exclude large parts of the parameter space that lead to $\Omega_{\rm DM} h^2 = 0.12$. Only a narrow corridor around the $125 \, {\rm GeV}$ Higgs-boson resonance survives the DM relic density constraint, which is however excluded by DM ID measurements. Given that~$\sigma_{\rm SI}$   is momentum suppressed, the stringent limits from   DM~DD  experiments do not put constraints on the derivative Higgs portal model. This highlights the need to test such models with $m_\phi > m_h/2$ using mono-$X$ searches at the~HL-LHC, however only if  $\Omega_{\rm DM} h^2 < 0.12$. The best tests of the derivative Higgs portal model in the Higgs off-shell region seem again to be  searches for invisible decays of the $125 \, {\rm GeV}$ Higgs boson produced in the VBF channel. This conclusion however  depends once more on the actual size of systematic uncertainties of the relevant mono-$X$ channels under HL-LHC conditions. 

Our discussion so far  was phrased within an EFT, but concrete examples of UV complete models where the two Higgs portal interactions~(\ref{eq:LphiH}) dominate in the low-energy limit have been constructed. For instance, the pNGB DM models  in~\cite{Frigerio:2012uc,Marzocca:2014msa,Balkin:2017aep} lead to a sizeable marginal Higgs portal coupling  $c_m$, while the constructions in~\cite{Balkin:2018tma,Xing:2020uaf} manage to suppress this coupling, making the derivative Higgs portal coupling $c_d$ the leading interaction between the dark and the visible sector. As shown above, in the latter category of models, only DM production at the LHC is able to directly probe pNGB DM models. If~the DM candidate can be produced as a real particle, searches for invisible Higgs boson decays play a key role in such explorations, while  DM masses above the Higgs threshold can be tested by studying mono-$X$ signatures such as ${\rm VBF} +E_T^{\rm miss}$ or  $tX +E_T^{\rm miss}$. Dedicated experimental searches and/or interpretations  by ATLAS and CMS  of the relevant mono-$X$ signatures in the pNGB DM context do not exist at present. 

\subsection{Further considerations}
\label{subsec:further}

\begin{figure}[!t]
\begin{center}
\includegraphics[width=0.65\linewidth]{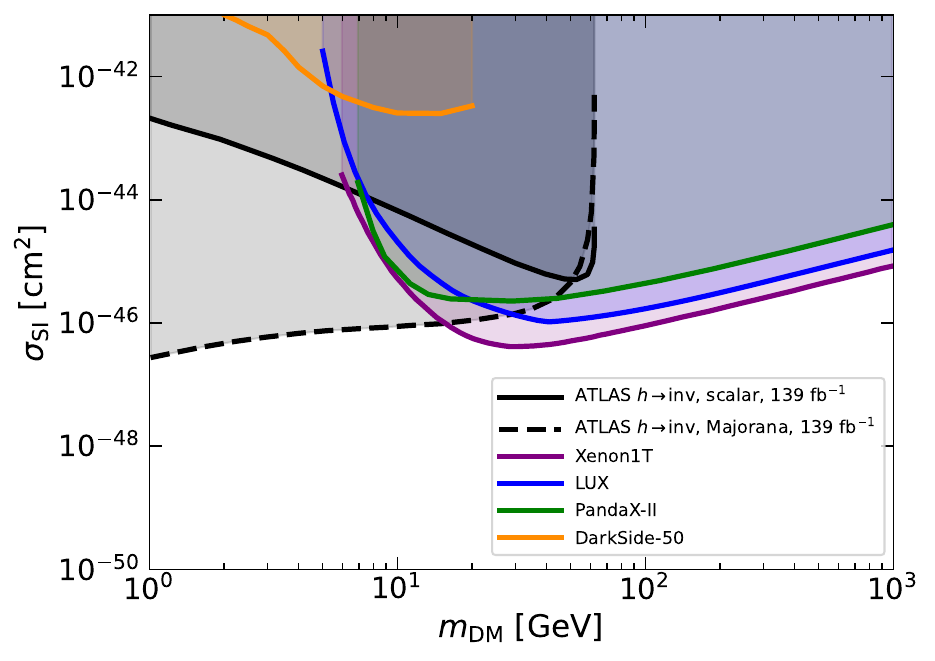} 
\vspace{2mm} 
\caption{\label{fig:ATLASsummary} Comparison of the upper limits at 90\%~CL from DD experiments~\cite{LUX:2016ggv,PandaX-II:2017hlx,XENON:2018voc,DarkSide:2018bpj} on the SI~DM-nucleon cross section to the exclusion limits that derive from~\cite{ATLAS:2020kdi}. The interpretation of the ATLAS results assumes Higgs portal scenarios where the $125 \, {\rm GeV}$ Higgs boson decays to DM which can either be a scalar or a Majorana fermion. The regions above the contours are excluded. }
\end{center}
\end{figure}

As promised we now return to a brief discussion of Higgs portal models with interactions not encoded in~(\ref{eq:LphiH}) such as models with fermionic DM. In~Figure~\ref{fig:ATLASsummary} we show a comparison between the upper limits at 90\%~CL from DD experiments~\cite{LUX:2016ggv,PandaX-II:2017hlx,XENON:2018voc,DarkSide:2018bpj} on the SI~DM-nucleon cross section and the exclusion limits that derive from the latest ATLAS measurement of Higgs to invisible decays~\cite{ATLAS:2020kdi}. At 90\%~CL this measurement leads to ${\rm BR} \left ( h \to {\rm inv} \right ) < 0.09$~---~the corresponding 95\%~CL bound is given in~(\ref{eq:BRhinv}). The~translation of the ${\rm BR} \left ( h \to {\rm inv} \right )$ bound into a limit on $\sigma_{\rm SI}$ relies on an EFT approach under the assumption that the $125 \, {\rm GeV}$ Higgs boson decays to a pair of DM particles are kinematically possible and that the DM particle is either a scalar or a Majorana fermion. In the scalar case, the EFT approach boils down to extract limits on the Wilson coefficient~$c_m$ from~(\ref{eq:hDMDMwidth}) and to insert the obtained values into~(\ref{eq:sigmaSI}) to derive bounds on $\sigma_{\rm SI}$. Notice that the limit on the Wilson coefficient~$c_m$ does only marginal change when going from DM masses of $10 \, {\rm GeV}$ down to~$1 \, {\rm GeV}$ (see the upper panel in~Figure~\ref{fig:summaryportal}) while the bound on the SI~DM-nucleon cross section worsens notable. While this is puzzeling at first, one has to realise that in the scalar case $\sigma_{\rm SI}$ scales with $1/(m_{\phi} + m_{N})^2$ $\big($cf.~(\ref{eq:sigmaSI})$\big)$ which implies that for constant~$c_m$ and the DM mass $m_{\phi}$ sufficiently larger than $m_N$, one has $\sigma_{\rm SI} \propto 1/m_{\phi}^2$. This feature leads to a deterioration of the limit on $\sigma_{\rm SI}$ with decreasing DM masses although the bound on $c_m$ in fact even slightly improves.

In order to understand the behaviour of the exclusion limit for the Higgs portal model with Majorana DM $\chi$ as shown in~Figure~\ref{fig:ATLASsummary} we first provide expressions for the relevant effective interactions, the partial Higgs decay width for $h \to \chi \bar \chi$ and the SI~DM-nucleon cross section. The Higgs portal takes the following form
\begin{equation} \label{eq:LchiH}
{\cal L}_{\chi H} = \frac{c_\chi}{\Lambda} \hspace{0.25mm} \chi \bar \chi  \hspace{0.25mm} ( H^\dagger H ) \,, 
\end{equation}
where $c_\chi$ is a Wilson coefficient assumed to be real and the scale $\Lambda$ suppresses the interaction because the operator is of dimension five. The corresponding partial Higgs decay width reads 
\begin{equation}  \label{eq:hxx}
\Gamma \left ( h \to \chi \bar \chi \right ) =  \left ( 1 - \frac{4 m_\chi^2}{m_h^2} \right)^{3/2}  \hspace{0.5mm}  \frac{c_\chi^2 \hspace{0.125mm}  v^2 \hspace{0.125mm}  m_h }{4 \hspace{0.125mm}\pi  \hspace{0.125mm}  \Lambda^2}\,,  
\end{equation}
while the SI~DM-nucleon cross section is given by
\begin{equation}  \label{eq:SIchiN}
\sigma_{\rm SI}^{\chi  N} = \frac{4 \hspace{0.25mm} c_\chi^2 \hspace{0.125mm}  m_\chi^2 \hspace{0.125mm}  m_N^4 \hspace{0.125mm}  f_N^2}{\pi \hspace{0.125mm}m_h^4 \hspace{0.125mm}  \left ( m_\chi + m_N \right )^2 \hspace{0.125mm}  \Lambda^2} \,.
\end{equation}

\clearpage 

We add that for a complex $c_\chi$ the SI DM-nucleon cross section can be strongly suppressed (cf.~for instance~\cite{GAMBIT:2018eea}). From~(\ref{eq:hxx}) it follows that for $m_\chi \to 0$ measurements of ${\rm BR} \left (h \to {\rm inv} \right)$ simply lead to constant bound on the combination $|c_\chi|/\Lambda$ of parameters. The SI~DM-nucleon cross section~(\ref{eq:SIchiN}) however scales as $m_\chi^2/(m_{\chi} + m_{N})^2$ where compared to~(\ref{eq:sigmaSI}) the additional factor of $m_\chi^2$ appears due to dimensional reasons. It follows that decreasing the DM mass will lead to a steady improvement of the limit on $\sigma_{\rm SI}$ from ${\rm BR} \left (h \to {\rm inv} \right)$ (see~Figure~\ref{fig:ATLASsummary}) even though $|c_\chi|/\Lambda$ stays essentially constant.

The above discussion suggests that from the point of view of collider physics interpreting the limits on ${\rm BR} \left ( h \to {\rm inv} \right )$ in terms of exclusions on $\sigma_{\rm SI}$ is not the optimal choice. A~better way to interpret and to compare the ${\rm BR} \left ( h \to {\rm inv} \right )$ results to the bounds on the SI~DM-nucleon cross section obtained by DD experiments would in our opinion consist in showing limits on the effective interaction strength of the Higgs portal models (i.e.~$|c_m|$, $\Lambda/|c_d|$, $\Lambda/|c_\chi|$, etc.). In the case of the marginal and derivative Higgs portal this has been done in~Figure~\ref{fig:summaryportal} and this is also the standard presentation in most of the theoretical literature~\cite{Silveira:1985rk,Veltman:1989vw,Schabinger:2005ei,Patt:2006fw,OConnell:2006rsp,Kim:2006af,Barger:2007im,Kanemura:2010sh,Djouadi:2011aa,Mambrini:2011ik,Djouadi:2012zc,Alanne:2017oqj,Arcadi:2017kky,Balkin:2018tma,Ruhdorfer:2019utl,Arcadi:2019lka,Das:2020ozo,Arcadi:2021mag,Lebedev:2021xey} on Higgs portal models.

\section{Portals with extended Higgs sectors} 
\label{sec:extendedhiggsportals}

In the spin-0 simplified models with a fermionic singlet DM candidate $\chi$, which were recommended in the articles~\cite{Abdallah:2015ter,Abercrombie:2015wmb} as benchmarks for DM searches at the LHC, the interactions between the mediator and the SM fermions are not invariant under the $SU(2)_L\times U(1)_Y$ gauge symmetry.  This feature leads to unitarity violation at high energies~\cite{Englert:2016joy,Haisch:2018bby}~---~for discussions of unitarity violation in the case of spin-1 simplified DM models see~\cite{Bell:2015sza,Kahlhoefer:2015bea,Haisch:2016usn}. The simplest way to restore gauge invariance in spin-0 simplified~DM models is to introduce a singlet scalar that provides the portal to the dark sector and mixes with the $125 \, {\rm GeV}$ Higgs~\cite{Kim:2008pp,Kim:2009ke,Baek:2011aa,Lopez-Honorez:2012tov,Baek:2012uj,Fairbairn:2013uta,Carpenter:2013xra}. Compared to the spin-0 simplified DM models, this model   leads to additional $E_T^{\rm miss}$ signatures such as Higgs to invisible decays,  mono-$W$, mono-$Z$ and VBF + $E_T^{\rm miss}$ production at tree-level ~---~see~\cite{Albert:2016osu} for a detailed discussion of the collider phenomenology~---~but the stringent constraints on the scalar-Higgs mixing that arises from the Higgs measurements at the LHC~(cf.~\cite{ATLAS-CONF-2020-027,CMS:2020gsy} for the latest global Higgs analyses by ATLAS and CMS)  lead to a suppression of all $E_T^{\rm miss}$ signals. This suppression still allows the LHC to test the model via~(\ref{eq:BRhinv}) if $m_\chi < m_h/2$. However, in the case $m_\chi > m_h/2$, the LHC coverage of the parameter space of the singlet fermionic DM model is  very limited, in particular, if one compares the collider limits to the strong bounds that result from DM~DD experiments. We therefore do not discuss the singlet fermionic DM model any further. 

Restoring gauge invariance in spin-0 simplified~DM models and simultaneously satisfying the stringent LHC constraints from the Higgs boson measurements is possible if the SM Higgs sector it extended.   Two-Higgs-doublet models~(2HDMs)~\cite{Gunion:1989we,Branco:2011iw}  that contain  two Higgs doublets $H_1$ and $H_2$ instead of just $H$  are one of the natural choices for such extensions.  The tree-level 2HDM scalar potential can be written as 
\begin{equation} \label{eq:VH}
\begin{split}
V_H& =\mu_1H_1^{\dag}H_1+\mu_2H_2^{\dag}H_2+\left(\mu_3H_1^{\dag}H_2+\mathrm{h.c.}\right)+\lambda_1\left(H_1^{\dag}H_1\right)^2+\lambda_2\left(H_2^{\dag}H_2\right)^2 \\[2mm]
        & \phantom{xx} +\lambda_3\left(H_1^{\dag}H_1\right)\left(H_2^{\dag}H_2\right)+\lambda_4\left(H_1^{\dag}H_2\right)\left(H_2^{\dag}H_1\right)+\left[\lambda_5\left(H_1^{\dag}H_2\right)^2+\mathrm{h.c.}\right] \,,
\end{split}        
\end{equation}
where  we have imposed a $\mathbb{Z}_2$ symmetry under which $H_1\to H_1$ and $H_2\to -H_2$  but allowed for this discrete symmetry to be broken softly by $\mu_3H_1^{\dag}H_2+\mathrm{h.c.}$ The $\mathbb{Z}_2$ symmetry is the minimal condition necessary to guarantee the absence of flavour-changing neutral currents~(FCNCs) at tree level~\cite{Glashow:1976nt,Paschos:1976ay} and such a symmetry is realised in many well-motivated complete UV theories in the form of supersymmetry~(SUSY), a $U(1)$ symmetry or a discrete symmetry acting on the Higgs doublets.  In order to avoid possible issues with electric dipole moments, it is commonly assumed that all parameters in the scalar potential~(\ref{eq:VH}) are real. In such as case the CP eigenstates that arise after spontaneous symmetry breaking from $V_H$ can be identified with the mass eigenstates, giving rise to  two CP-even scalars $h$ and $H$, one CP-odd  pseudoscalar $A$ and one charged scalar~$H^\pm$.  Besides the masses of the five Higgs bosons, the 2HDM parameter space involves  the angles $\alpha$ and~$\beta$. The former angle describes the mixing of the two CP-even Higgs bosons  while the latter encodes the ratio of the VEVs $v_1$ and $v_2$ of the two Higgs doublets $\tan\beta=v_2/v_1$ with $v = \sqrt{v_1^2 + v_2^2}$. 

In the context of the inert doublet model~\cite{Deshpande:1977rw,Barbieri:2006dq,Cao:2007rm} the scalar potential~(\ref{eq:VH}) alone already allows for interesting DM and collider phenomenology~\cite{Dolle:2009ft,Miao:2010rg,Gustafsson:2012aj,Belanger:2015kga,Ilnicka:2015jba,Poulose:2016lvz,Datta:2016nfz,Hashemi:2016wup,Belyaev:2016lok,Dutta:2017lny,Wan:2018eaz,Kalinowski:2018ylg,Kalinowski:2018kdn,Dercks:2018wch,Kalinowski:2020rmb}. The DM particle in the inert doublet model  is  the lightest neutral component of the second Higgs~doublet making it  a spin-0 state. To accommodate the possibility  of having a spin-1/2 singlet DM candidate (like in the case of the simplified DM models~\cite{Abdallah:2015ter,Abercrombie:2015wmb}) considering only~(\ref{eq:VH}) is therefore not enough and one generically needs besides the DM particle an additional mediator. If one insists that this mediator has spin-0 and one does not want to violate CP, the additional mediator can either be a scalar or pseudoscalar.  Dedicated studies of the collider aspects and DM properties  in the  latter type of next-generation simplified DM models have been presented in~\cite{Ipek:2014gua,No:2015xqa,Goncalves:2016iyg,Bell:2016ekl,Bauer:2017ota,Tunney:2017yfp,Bell:2017rgi,Arcadi:2017wqi,Pani:2017qyd,Bauer:2017fsw,Bell:2018zra,Abe:2018emu,Abe:2018bpo,CidVidal:2018eel,Ertas:2019dew,Abe:2019wjw,Arcadi:2020gge,Butterworth:2020vnb,Robens:2021lov}. Below we review in detail two of the models discussed in these articles.

\subsection{2HDM+$a$ model}
 \label{sec:2HDMa}
 
The 2HDM+$a$ model~\cite{Ipek:2014gua,No:2015xqa,Goncalves:2016iyg,Bauer:2017ota} is the simplest gauge-invariant and renormalisable extension of the simplified pseudoscalar DM model~\cite{Abdallah:2015ter,Abercrombie:2015wmb}. It includes a Dirac fermion~$\chi$, which transforms as a singlet under the SM gauge group and therefore provides a DM candidate, and a CP-odd mediator $P$ that furnishes the dominant portal between the dark and the visible sector. Since the DM~DD constraints are weaker for models with pseudoscalar mediators compared to models with scalar mediators, the observed DM relic abundance can be reproduced in large regions of parameter space. These features allow for a host of $E_T^{\rm miss}$ signatures at colliders which  can be consistently compared and combined, making the 2HDM+$a$  model~one of the main pillars of the LHC DM search programme~\cite{Abe:2018bpo,ATLAS:2017hoo,CMS:2018zjv,ATL-PHYS-PUB-2018-027,ATLAS:2019wdu,CMS:2020ulv,ATLAS:2020yzc,ATLAS:2021jbf,ATLAS:2021shl,ATLAS-CONF-2021-029,ATLAS-CONF-2021-036}. 

\subsubsection{Theory}
\label{sec:2HDMatheory}

If the Dirac DM field $\chi$ and the pseudosalar $P$ are taken to transform under the $\mathbb{Z}_2$ symmetry as   $\chi \to -\chi$ and $P \to P$, the only  renormalisable  DM-mediator coupling that is allowed by  symmetry is
\begin{equation} \label{eq:2HDMaLchi}
{\cal L}_\chi = - i y_\chi P \bar \chi \gamma_5 \chi \,.
\end{equation}
Here it is assumed that the dark-sector Yukawa coupling $y_\chi$ is real in order not to violate~CP. Besides~(\ref{eq:VH}) the scalar potential in the 2HDM+$a$ model contains the following terms 
\begin{equation}  \label{eq:VHP}
V_{HP} = P \left ( i b_P H_1^\dagger H_2 + {\rm h.c.} \right ) + P^2 \left ( \lambda_{P1} H_1^\dagger H_1 +   \lambda_{P2} H_2^\dagger H_2 \right ) \,, 
\end{equation}
that connect  singlets to doublets. The parameters $b_P$, $\lambda_{P1}$ and $\lambda_{P2}$ are taken to be real to not violate CP. Notice that the first term in $V_{HP}$ breaks the  $\mathbb{Z}_2$ symmetry softly. The~singlet potential reads 
\begin{equation}  \label{eq:VP}
V_{P} = \frac{1}{2} m_P^2 P^2 \,. 
\end{equation}
A quartic term $P^4$ is not  included in~(\ref{eq:VP}) since it does not lead to any substantial modification of the LHC phenomenology, in particular such an addition would have no relevant effects in any  of the $E_T^{\rm miss}$ observables discussed below. 

Including the mass of the DM particle, the  Lagrangian of the 2HDM+$a$ model contains~14 free parameters in addition to the SM ones. After rotation to the mass eigenbasis, these~14 parameters can be traded for seven physical masses, three mixing angles and four couplings:
\begin{equation} \label{eq:2HDMainput}
\begin{Bmatrix}
\mu_1,\mu_2,\mu_3,b_P,m_P,m_{\chi},\\
y_{\chi},\lambda_1,\lambda_2,\lambda_3,\lambda_4,\lambda_5,\\
\lambda_{P1},\lambda_{P2}
\end{Bmatrix} \ \
\Longleftrightarrow
\ \
\begin{Bmatrix}
v,m_h,m_A,m_H,m_{H^{\pm}},m_a,m_{\chi},\\
\cos(\beta-\alpha),\tan\beta,\sin\theta,\\
y_{\chi},\lambda_3,\lambda_{P1},\lambda_{P2}
\end{Bmatrix} \,.
\end{equation}
Here $\sin\theta$ represents the mixing of the two CP-odd weak spin-0 eigenstates and the additional CP-odd mediator $a$ is mostly composed of $P$ for $\sin \theta \simeq 0$. The parameters appearing on the right-hand side of~(\ref{eq:2HDMainput}) are used as input in the analyses of the 2HDM+$a$ model. Since the VEV $v$ and the Higgs mass $m_h$ are already fixed by observations there are~12 input parameters. 

Some of the 2HDM+$a$ parameters are constrained by Higgs physics,  electroweak~(EW) precision observables (EWPOs), vacuum stability considerations, flavour physics and LHC searches for additional spin-0 bosons. The mixing angle $\alpha$ between the CP-even scalars~$h$ and $H$ is for instance constrained by Higgs coupling strength measurements~\cite{ATLAS-CONF-2020-027,CMS:2020gsy}. For~arbitrary values of~$\tan \beta$ only parameter choices with $\cos (\beta - \alpha) \simeq 0$ are experimentally allowed. In order to satisfy the constraints from Higgs physics, the existing experimental 2HDM+$a$ analyses~\cite{ATLAS:2017hoo,CMS:2018zjv,ATL-PHYS-PUB-2018-027,ATLAS:2019wdu,CMS:2020ulv,ATLAS:2020yzc,ATLAS:2021jbf,ATLAS:2021shl,ATLAS-CONF-2021-029,ATLAS-CONF-2021-036} have concentrated on the so-called alignment limit of the~2HDM where $\cos (\beta - \alpha) = 0$~\cite{Gunion:2002zf}, treating $\tan \beta$ as a free parameter. 

The measurements of the EWPOs constrain the differences between the masses of the additional scalar and pseudoscalar particles $m_H$, $m_A$, $m_{H^\pm}$ and $m_a$, because the exchange of spin-0 states modifies the propagators of the EW bosons starting at the one-loop level. In~\cite{Bauer:2017ota} it has been shown that the sum of the potentials~(\ref{eq:VH}) and~(\ref{eq:VHP}) has a custodial symmetry if   $\cos (\beta - \alpha) = 0$ and $m_A = m_H = m_{H^\pm}$. For such parameter choices the EWPOs are satisfied for any value of $\sin \theta$ and  $m_a$, which renders the choice  $m_A = m_H = m_{H^\pm}$ a~good starting point to explore the 2HDM+$a$ parameter space. 

The requirement that the scalar potential $V_H+V_{HP}+V_P$ of the 2HDM+$a$ model is bounded from below  restricts the possible choices of the spin-0 boson masses, mixing angles and quartic couplings. Assuming that $\lambda_{P1},\lambda_{P2} > 0$ and  $m_A = m_H = m_{H^\pm}$ one can show~\cite{Abe:2018bpo} that there are two bounded from below conditions:  
\begin{equation} \label{eq:2HDMaBFB}
\lambda_3 > \frac{m_h^2}{v^2} \simeq 0.26 \,, \qquad 
\lambda_3 > \frac{m_A^2 - m_a^2}{v^2} \sin^2 \theta -  \frac{m_h^2}{v^2} \cot^2 \left ( 2 \beta \right ) \,. 
\end{equation}
These inequalities suggest that $\lambda_3$ has to be sufficiently large, in particular, if the  mass splitting of the two pseudoscalars and/or  $\sin\theta$ are large. Since the relations~(\ref{eq:2HDMaBFB}) are modified in models with more general scalar potentials including dimension-six operators~\cite{Bauer:2017fsw,Haisch:2018djm} the bounded from below requirements are not directly imposed in the ATLAS and CMS interpretations of the $E_T^{\rm miss}$ searches~\cite{ATLAS:2017hoo,CMS:2018zjv,ATL-PHYS-PUB-2018-027,ATLAS:2019wdu,CMS:2020ulv,ATLAS:2020yzc,ATLAS:2021jbf,ATLAS:2021shl,ATLAS-CONF-2021-029,ATLAS-CONF-2021-036}. Instead the choice~$\lambda_3 = 3$ is employed which generically allows for heavy Higgses above $1 \, {\rm TeV}$, while keeping~$\lambda_3$ small enough to stay in the perturbative regime. 

The quartic couplings $\lambda_3$, $\lambda_{P1}$ and $\lambda_{P2}$ affect the decay pattern of the heavy CP-odd and CP-even 2HDM scalars. In the alignment limit and assuming that $m_A = m_H = m_{H^\pm}$ the  $Aah$ and $Haa$ couplings take the following form~\cite{Bauer:2017ota}
\begin{eqnarray} \label{eq:2HDMacouplings1}
\begin{split}
g_{Aah}&=\frac{1}{2m_Av}\bigg [ \hspace{0.25mm} m_A^2-m_a^2+ m_h^2 -2\lambda_3v^2 +2 \left (\lambda_{P1}\cos^2\beta+\lambda_{P2}\sin^2\beta \right )v^2 \hspace{0.25mm} \bigg]\sin \left ( 2 \theta \right )  \,, \hspace{2mm} \\[2mm]
g_{Haa}&=\frac{1}{m_Hv}\bigg [ \hspace{0.25mm} 2 \cot \left (2 \beta \right ) \left ( m_h^2 -  \lambda_3 v^2 \right ) \sin^2 \theta + \sin \left (2 \beta \right ) \left (\lambda_{P1} - \lambda_{P2} \right  ) \hspace{0.25mm} v^2 \cos^2 \theta \hspace{0.25mm} \bigg ] \,.
\end{split}
\end{eqnarray}
These expressions imply that parameter choices of the form $\lambda_3 = \lambda_{P1} = \lambda_{P2}$ are well suited to keep the total widths $\Gamma_A$ and $\Gamma_H$ better under control in the limit of heavy Higgs masses, since such choices lead to cancellations in~(\ref{eq:2HDMacouplings1}). 

Indirect constraints on the charged Higgs-boson mass $m_{H^\pm}$  arise from $Z \to b \bar b$, $B \to X_s \gamma$, $B_s \to \mu^+ \mu^-$ and  $B$-meson mixing~---~see~\cite{Bauer:2017ota,Abe:2018bpo,Robens:2021lov} for details and relevant references. These constraints are more model-dependent than the bounds that derive from the EWPOs because they depend on the choice of the Yukawa sector of the 2HDM model.  For instance, in the case of the 2HDM of type-II the inclusive $B \to X_s \gamma$ decay sets a 95\%~CL lower limit $m_{H^\pm} > 800 \, {\rm GeV}$~\cite{Misiak:2017bgg,Misiak:2020vlo} that for $\tan \beta  \gtrsim 2$ is practically independent of the specific choice of $\tan \beta$. In other 2HDM realisations such as a fermiophobic 2HDM model of type-I~(see~\cite{Fox:2017uwr,Haisch:2017gql} for  recent detailed discussions) all flavour bounds can however be significantly relaxed, allowing for EW-scale values of $m_{H^\pm}$. Furthermore, since the FCNC constraints arise from loop corrections they can in principle be weakened by the presence of additional particles that are too heavy to be produced at the LHC. This makes the bounds from flavour indicative, and   the analyses~\cite{ATLAS:2017hoo,CMS:2018zjv,ATL-PHYS-PUB-2018-027,ATLAS:2019wdu,CMS:2020ulv,ATLAS:2020yzc,ATLAS:2021jbf,ATLAS:2021shl,ATLAS-CONF-2021-029,ATLAS-CONF-2021-036} have  not directly imposed them on the parameter space of the 2HDM+$a$ model. 

Direct searches for heavy Higgs bosons can also be used to explore and to constrain the parameter space of the 2HDM+$a$ model. Discussions of the existing LHC constraints can be found in~\cite{Bauer:2017ota,Abe:2018bpo,Butterworth:2020vnb}. The most interesting signatures arise if one deviates from the alignment limit and/or gives up the assumption  $m_A = m_H = m_{H^\pm}$. 
In particular searches for  final states involving tops and/or $W$ bosons~\cite{Pani:2017qyd,ATL-PHYS-PUB-2018-027,Haisch:2018djm,CidVidal:2018eel,Butterworth:2020vnb} provide   interesting avenues to test the 2HDM+$a$ model, and future LHC searches should consider  such channels and interpret the obtained results in the context of the 2HDM+$a$ model. In~Section~\ref{eq:2HDMasummary} we will give an example of a non-$E_T^{\rm miss}$ that already now provides relevant constraints on the 2HDM+$a$  parameter space. 

\subsubsection{Anatomy of $E_T^{\rm miss}$ signatures}
\label{sec:2hdmaanatomy}

\begin{figure}[t!]
\centering
\includegraphics[width=.75\linewidth]{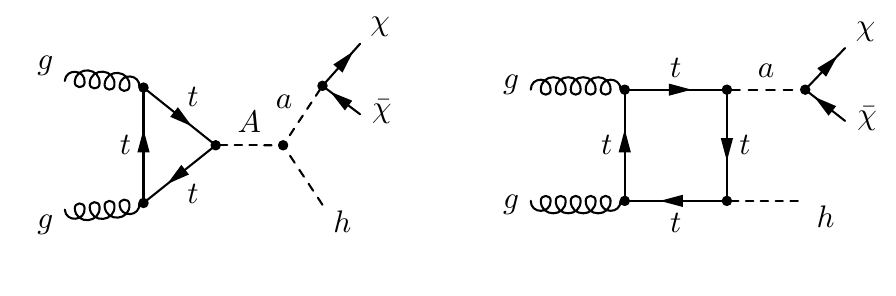}

\vspace{-2mm}

\includegraphics[width=.75\linewidth]{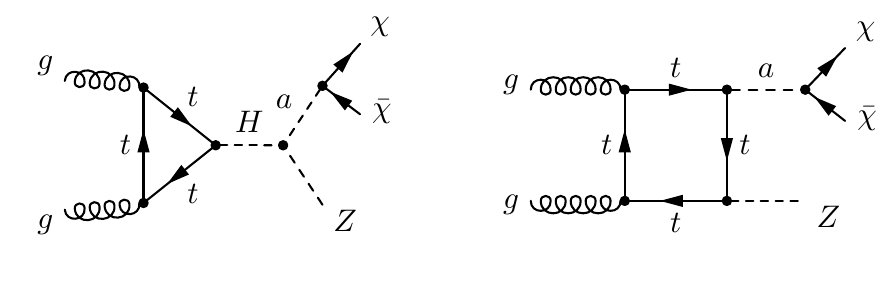}

\vspace{-1mm}

\includegraphics[width=.75\linewidth]{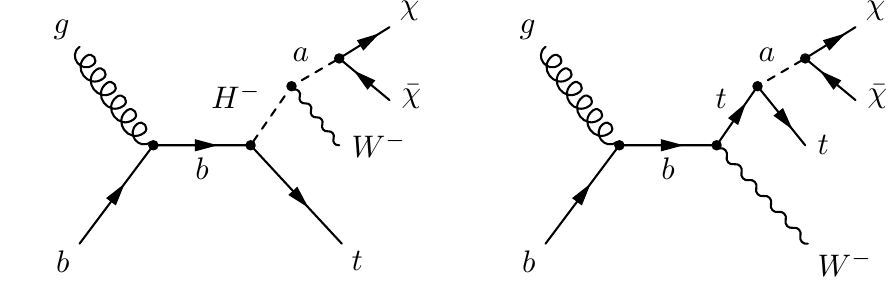}

\vspace{2mm}
\caption{\label{fig:2HDMadiagrams} Representative 2HDM+$a$  Feynman diagrams that give rise to a mono-Higgs signal~(upper row), a mono-$Z$ signature~(middle row) and $tW+E_T^{\rm miss}$ signal~(lower row) in ggF production. For further explanations see the main text.}
\end{figure}

\begin{figure}[t!]
\centering
\vspace{4mm}
\includegraphics[width=.75\linewidth]{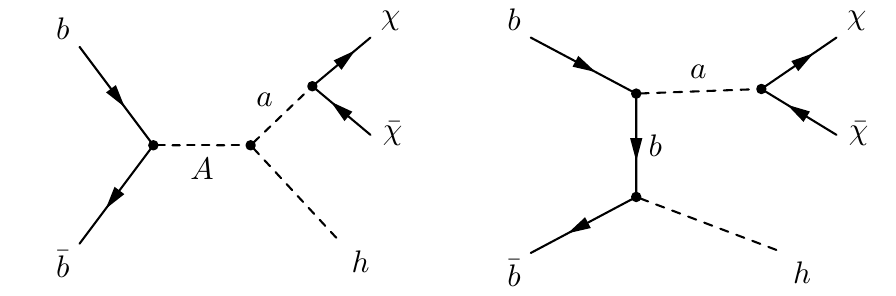}

\vspace{4mm}

\includegraphics[width=.75\linewidth]{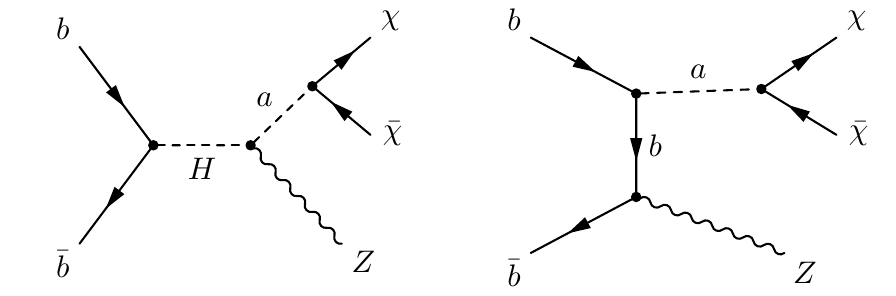}
\vspace{2mm}
\caption{\label{fig:2HDMabbdiagrams} Examples of  2HDM+$a$  Feynman diagrams that give rise to a mono-Higgs signal~(upper row) and  a mono-$Z$ signature~(lower row) in  $b \bar b$-fusion. Consult the  text for additional details.}
\end{figure}

The 2HDM+$a$ model provides a wide variety of $E_T^{\rm miss}$ signatures that can be searched for at the LHC. As pointed out in the works~\cite{Goncalves:2016iyg,Bauer:2017ota,Pani:2017qyd} the most interesting signals are $h+E_T^{\rm miss}$~(i.e.~mono-Higgs), $Z+E_T^{\rm miss}$~(i.e.~mono-$Z$) and $tW+E_T^{\rm miss}$ production, because these signals can be resonantly enhanced. Examples of representative  diagrams that lead to the discussed $E_T^{\rm miss}$ signals in ggF production are displayed in~Figure~\ref{fig:2HDMadiagrams}. Diagrams with bottom-quark loops and graphs in which the internal~$a$ is replaced by an $A$ also exist. In the case of the mono-Higgs signature it is evident from the~figure that for $m_A > m_h + m_a$ the triangle graph shown on the left-hand side in the upper row allows for resonant $h+E_T^{\rm miss}$ production. Similar resonance enhancements arise from the graphs on the left-hand side for the mono-$Z$~(middle row) and $tW+E_T^{\rm miss}$ channel~(lower row) if $m_H > m_Z + m_a$ and $m_{H^\pm} > m_W + m_a$, respectively. Notice that resonant mono-Higgs, mono-$Z$ and $tW+E_T^{\rm miss}$ production is not possible in the simplified pseudoscalar DM model~\cite{Abdallah:2015ter,Abercrombie:2015wmb} because the mediator couples only to fermions at tree level. As a result only the Feynman diagrams displayed on the right-hand side of~Figure~\ref{fig:2HDMadiagrams} are present in this model. Notice that the appearance  of the resonance contributions due to an on-shell~$A$, $H$ and $H^\pm$ not only enhances the mono-Higgs, mono-$Z$ and $tW+E_T^{\rm miss}$ compared to the simplified pseudoscalar DM model, but also leads to a quite different kinematics in these channels~\cite{Goncalves:2016iyg,Bauer:2017ota,Pani:2017qyd,Abe:2018bpo}. Besides ggF production, important contributions to the mono-Higgs and mono-$Z$ channels can arise from $b \bar b$-fusion production in 2HDM realisations such as type-II  models  in which the bottom-quark Yukawa coupling is enhanced by $\tan \beta$. The corresponding Feynman diagrams are shown in~Figure~\ref{fig:2HDMabbdiagrams}. As for the ggF mono-Higgs and mono-$Z$ signals there are also resonant (left column) and non-resonant contributions (right column) in the case of $b \bar b$-fusion. 

An additional $E_T^{\rm miss}$ signature that allows to put constraints on the parameter space of the  2HDM+$a$ model are invisible Higgs decays~\cite{Bauer:2017ota}. The relevant decay channel  is $h \to aa$ followed by $a \to \chi \bar \chi$.  The decay of the $125 \, {\rm GeV}$ Higgs to a pair of pseudoscalars $a$ is proportional to the square of the $haa$ coupling. For $\cos (\beta - \alpha) = 0$ and $m_A = m_H = m_{H^\pm}$ this coupling takes the form 
\begin{equation}  \label{eq:2HDMacouplings2} 
\begin{split}
g_{haa} & = \frac{1}{m_h v} \bigg [ \left ( 2 m_A^2-2 m_a^2+ m_h^2 -2\lambda_3v^2  \right ) \sin^2 \theta  \\[2mm] 
& \phantom{xxxxxxx} - 2 \left (\lambda_{P1}\cos^2\beta+\lambda_{P2}\sin^2\beta \right ) \hspace{0.25mm} v^2 \cos^2 \theta \hspace{0.25mm} \bigg] \,. 
\end{split} 
\end{equation}
Ignoring fine-tuned solutions for which $|g_{haa}| \ll 1$, one can show that for sufficiently light DM masses the bound~(\ref{eq:BRhinv}) leads to a lower limit of $m_a \gtrsim 100 \, {\rm GeV}$. We add that the direct measurements of the total Higgs width~\cite{CMS:2017dib,ATLAS:2018tdk} that impose  $\Gamma_h < 1.1 \, {\rm GeV}$ at 95\%~CL furthermore imply that $m_a \gtrsim m_h/2 \simeq 62.5 \, {\rm GeV}$. This  bound also holds for heavy~DM, because  the processes $h \to aa$ with $a  \to  f \bar f$, where  $f$ denotes all the kinematically accessible SM fermions, always provides a sizeable non-SM contribution to~$\Gamma_h$ unless~(\ref{eq:2HDMacouplings2}) is  artificially small.

In addition to the $E_T^{\rm miss}$ signatures discussed so far, the 2HDM+$a$ model also leads to mono-jet, $t \bar t + E_T^{\rm miss}$ and $b \bar b + E_T^{\rm miss}$ signals. In contrast to the case of the  simplified pseudoscalar DM model where these mono-$X$ channels provide the leading constraints (see~\cite{Haisch:2018bby,CMS:2017dcx,ATLAS:2017hoo,Haisch:2018hbm,CMS:2019zzl,ATLAS:2020xzu,ATLAS:2021yij,ATLAS:2021hza,ATLAS:2021kxv,CMS:2021far} for the latest theoretical and experimental analyses)  this is not the case in the 2HDM+$a$,  because the mono-jet, $t \bar t + E_T^{\rm miss}$ and $b \bar b + E_T^{\rm miss}$ channels do not receive resonantly enhanced contributions. Reinterpreting the existing spin-0 simplified DM model searches in the 2HDM+$a$ context is relatively straightforward~\cite{Bauer:2017ota,Abe:2018bpo}. 

\subsubsection{Summary plots}
\label{eq:2HDMasummary}

In the articles~\cite{Abe:2018bpo,ATLAS:2017hoo,CMS:2018zjv,ATL-PHYS-PUB-2018-027,ATLAS:2019wdu,CMS:2020ulv,ATLAS:2020yzc,ATLAS:2021jbf,ATLAS:2021shl,ATLAS-CONF-2021-029,ATLAS-CONF-2021-036} various 2HMD+$a$ benchmark scans have been recommended and studied.  The following parameters are common to all existing benchmark scans
\begin{equation} \label{eq:bench1}
m_A = m_H = m_{H^\pm} \,, \quad \cos \left ( \beta - \alpha \right ) = 0 \,, \quad  y_\chi = 1 \,, \quad \lambda_3=\lambda_{P1} = \lambda_{P2} =3 \,,
\end{equation}
and we have motivated these choices in Section~\ref{sec:2HDMatheory}. Unless the relevant parameter is scanned over the plots shown below also employ 
\begin{equation} \label{eq:bench2}
\tan \beta = 1 \,, \qquad m_\chi = 10 \, {\rm GeV} \,,
\end{equation}
where the latter choice ensures a sizeable ${\rm BR} (a \to \chi \bar \chi)$ for the pseudoscalar masses of interest,~i.e.~$m_a \gtrsim 100 \, {\rm GeV}$.  The Yukawa sector of the 2HDM is chosen to be of type-II. In~the alignment limit the couplings of the neutral non-SM scalars to the top-quark  and bottom-quark  therefore scale like 
\begin{equation} \label{eq:2HDMfermions}
\begin{split}
& g_{Ht\bar{t}} \propto y_t \cot \beta \,, \qquad g_{Hb\bar{b}} \propto y_b \tan\beta \,, \\[2mm]
& \hspace{-7.5mm} g_{At\bar{t}} \propto y_t  \cot \beta \cos \theta \,, \qquad g_{Ab\bar{b}} \propto y_b \tan\beta \cos \theta \,, \\[2mm]
& \hspace{-7.5mm} g_{at\bar{t}} \propto y_t  \cot \beta \sin \theta \,, \qquad \phantom{i} g_{ab\bar{b}} \propto y_b \tan\beta \sin \theta \,,
\end{split}
\end{equation}
where $y_t = \sqrt{2} m_t/v \simeq 0.94$ and  $y_b = \sqrt{2} m_b/v \simeq 0.02$  are the top-quark and bottom-quark Yukawa coupling, respectively. These expressions imply that for the benchmark choice~(\ref{eq:bench2}) of $\tan \beta$  the top-quark loop diagrams in~Figure~\ref{fig:2HDMadiagrams} provide the dominant contributions to the mono-Higgs and mono-$Z$, while for sufficiently large values of $\tan \beta$ the $b \bar b$-induced graphs of~Figure~\ref{fig:2HDMabbdiagrams} can give a relevant or even the dominant contribution. 

The mono-$X$ searches that provide at present the bulk of the sensitivity to the 2HDM+$a$ model are the $h+E_T^{\mathrm{miss}}$ channel with either $h\to b\bar{b}$ $\big($$h\hspace{0.25mm} (b\bar{b}) + E_T^{\rm miss}$$\big)$~\cite{ATLAS:2021shl,CMS:2018zjv} or~$h\to \gamma\gamma$~$\big($$h\hspace{0.25mm} (\gamma \gamma) + E_T^{\rm miss}$$\big)$~\cite{ATLAS:2021jbf} decays, the $Z+E_T^{\mathrm{miss}}$ signal followed by $Z\to \ell^+\ell^-$$\big($$Z\hspace{0.25mm} (\ell^+ \ell^-) + E_T^{\rm miss}$$\big)$~\cite{CMS:2020ulv,ATLAS-CONF-2021-029} and $tW+E_T^{\mathrm{miss}}$~production~\cite{ATLAS:2020yzc}. Thus we will focus on  these searches below, disregarding other mono-$X$ searches such as mono-jets or $t \bar t + E_T^{\rm miss}$.  To evade the limits from invisible Higgs decays (cf.~Section~\ref{sec:2hdmaanatomy}), we consider only $m_a$ values satisfying $m_a > 100 \, {\rm GeV}$ when studying the 2HDM+$a$ parameter space.  In addition,  we will show that the recent search for charged Higgs bosons in the channel $H^{\pm}\to tb$~\cite{ATLAS:2021upq} allows to set stringent constraints on $m_{H^{\pm}}$, which under the assumption of degenerated 2HDM spin-0 masses~(\ref{eq:bench1}), translate into bounds on~$m_A$ or $m_H$. As we will see, the obtained limits are to first approximation independent of $m_a$, and as a result the $H^\pm \to tb$ search provides complementary constraints with respect to the mono-$X$  signatures, because these searches depend on the precise choice for $m_a$. Other non-$E_T^{\rm miss}$ searches  such as for instance four-top production~\cite{Abe:2018bpo,ATL-PHYS-PUB-2018-027} are not considered. 

\paragraph{\bf Scans in the $\bm{m_A} \hspace{0.25mm}$--$\hspace{0.5mm} \bm{m_a}$ plane}

\begin{figure}[t!]
\centering
\includegraphics[width=.45\linewidth]{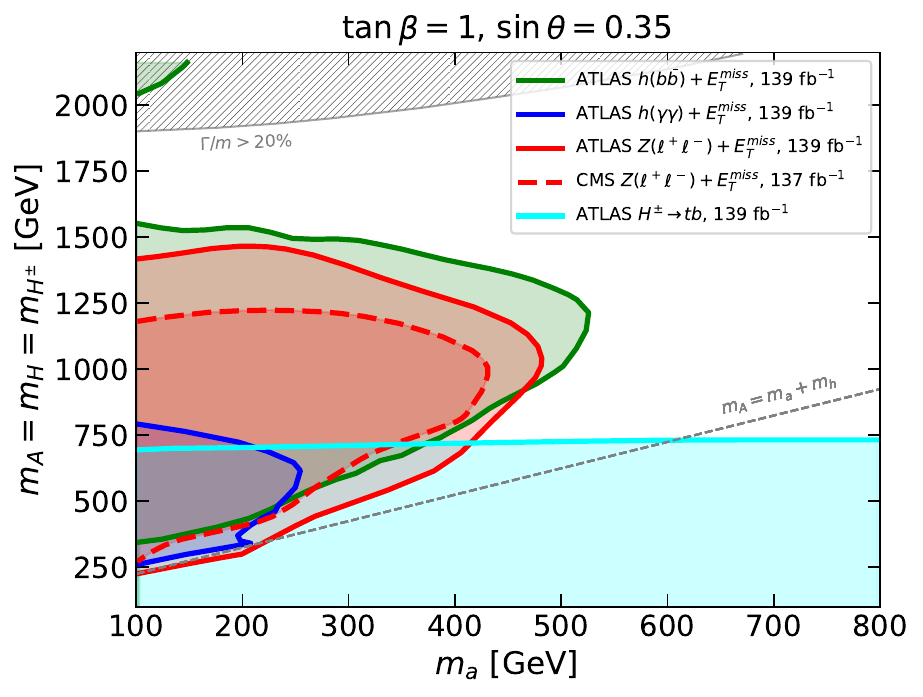} \qquad 
\includegraphics[width=.45\linewidth]{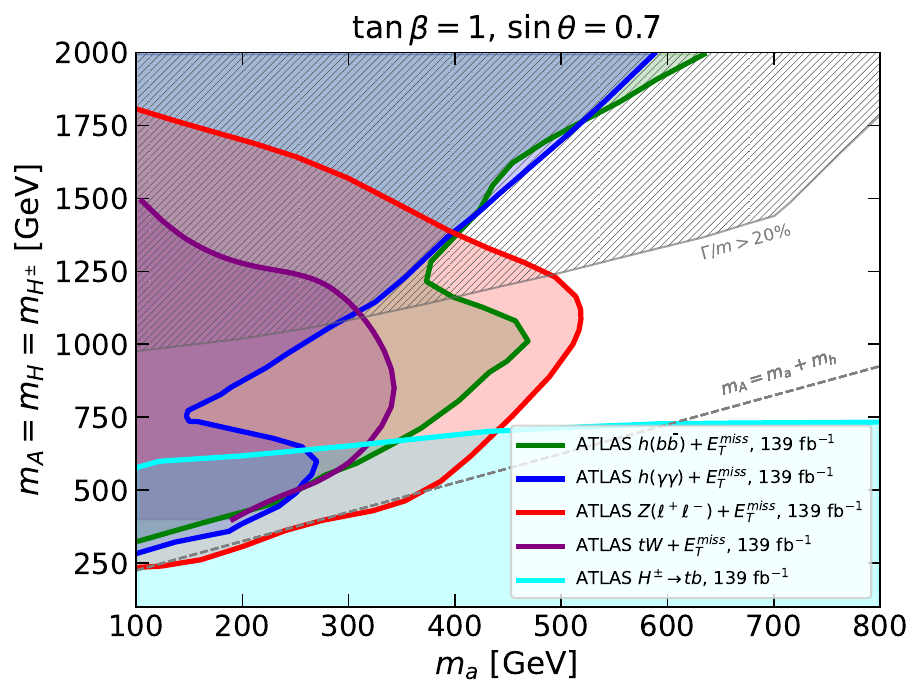}
\vspace{2mm}
 \caption{Observed exclusions at 95\%~CL for the 2HDM+$a$ model in the $m_A \hspace{0.25mm}$--$\hspace{0.5mm} m_a$  plane. The left and right panel corresponds to $\sin\theta=0.35$ and $\sin\theta=0.7$, respectively. The other parameters are chosen as in~(\ref{eq:bench1}) and~(\ref{eq:bench2}).  The coloured exclusions arise from~\cite{ATLAS:2021shl,ATLAS:2021jbf,ATLAS:2020yzc,CMS:2020ulv,ATLAS-CONF-2021-029,ATLAS:2021upq}. The hatched grey regions correspond to the parameter space where any of the additional Higgs bosons has a relative width of more than 20\%. This is indicated by the label $\Gamma/m > 20\%$. The dashed grey line corresponds to the equality $m_A =  m_a + m_h$. See text for further details. \label{fig:2hdma_2D_exclusion}}
\end{figure}

An assortment of 95\%~CL exclusions in  the  $m_A \hspace{0.25mm}$--$\hspace{0.5mm} m_a$ plane are shown in~Figure~\ref{fig:2hdma_2D_exclusion} for $\sin\theta=0.35$~(left panel) and $\sin\theta=0.7$~(right panel). At present the  $h\hspace{0.25mm} (b\bar{b})+E_T^{\mathrm{miss}}$ and the  $Z\hspace{0.25mm} (\ell^+\ell^-)+E_T^{\mathrm{miss}}$ signatures allow to constrain the largest parts of the $m_A \hspace{0.25mm}$--$\hspace{0.5mm} m_a$ plane in both $\sin \theta$ scenarios. In fact, in the high-mass region the $h \hspace{0.25mm}  (b\bar{b})+E_T^{\mathrm{miss}}$ is the most sensitive strategy in the case of $\sin \theta = 0.35$, while it becomes less sensitive than the $Z \hspace{0.25mm} (\ell^+ \ell^-)+E_T^{\mathrm{miss}}$ search close to $m_A=m_a+m_h$~(dashed grey line), where the $E_T^{\mathrm{miss}}$ spectrum becomes softer and the acceptance is reduced due to the high threshold of the $E_T^{\mathrm{miss}}$ triggers. The $Z \hspace{0.25mm} (\ell^+ \ell^-)+E_T^{\mathrm{miss}}$ search on the other hand relies on lepton triggers, making it more sensitive to the parameter space close to  $m_A=m_a+m_h$. One also observes that for $\sin\theta =0.7$, which corresponds to close-to-maximal mixing in the pseudoscalar sector, the $Z \hspace{0.25mm} (\ell^+ \ell^-)+E_T^{\mathrm{miss}}$ and $tW+E_T^{\mathrm{miss}}$ signatures have increased sensitivity compared to the case of low $\sin \theta$. This feature is readily understood by noticing that the couplings relevant for resonant production scale with $\sin \theta$. In the alignment limit with $m_A = m_H = m_{H^\pm}$ one has explicitly:
\begin{equation} \label{eq:2HDMacouplings3} 
\begin{split}
g_{HaZ} & = \frac{1}{m_Hv}\left[(m_A^2-m_a^2-m_Z^2)^2-4m_a^2m_Z^2\right]\sin\theta \,, \\[2mm]
g_{H^{\pm}aW^{\pm}} & = \frac{m_W^2}{m_{H^{\pm}}v}\left[m_W^2 -4m_A^2 \right]\sin\theta \,.
\end{split}
\end{equation}

In the case of the mono-Higgs exclusions two features that are visible in the plot on the left-hand side of~Figure~\ref{fig:2hdma_2D_exclusion} deserve an explanation. To understand the dips in the  $h +E_T^{\mathrm{miss}}$ constraints one first has to realise that~(\ref{eq:2HDMacouplings1}) and~(\ref{eq:2HDMacouplings2}) imply that for an on-shell $A$ the real part of the sum of the  $gg \to A \to  h a  \to  h + \chi \bar \chi$  and $gg \to a  \to  h a \to  h + \chi \bar \chi$ amplitudes is proportional to 
\begin{equation} \label{eq:magicformula}
\frac{\left ( 2 m_A^2 - 2 m_a^2 + m_h^2 \right ) \sin^2 \theta - 6 v^2}{m_A^2-m_a^2} \,.
\end{equation}
Here we have employed the parameter choices~(\ref{eq:bench1}). For fixed values of $m_a$ and $\sin \theta$, the numerator in~(\ref{eq:magicformula}) is  not sign-definite being negative~(positive) for sufficiently small~(large)~$m_A$. Close to the zero point of~(\ref{eq:magicformula}) the resonant contribution to mono-Higgs production is hence strongly suppressed,  leading to a weakening of the  constraints which in such a case arise almost entirely  from the non-resonant contributions~---~see the right Feynman diagram in the upper row of~Figure~\ref{fig:2HDMadiagrams}. 

\begin{figure}[t!]
\centering
\includegraphics[width=.45\linewidth]{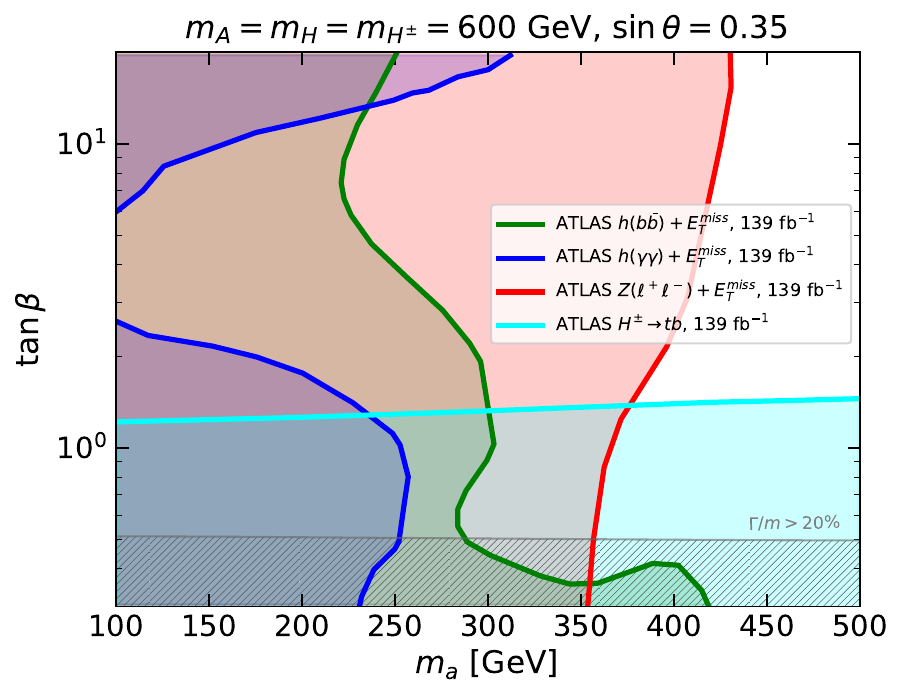}  \qquad 
\includegraphics[width=.45\linewidth]{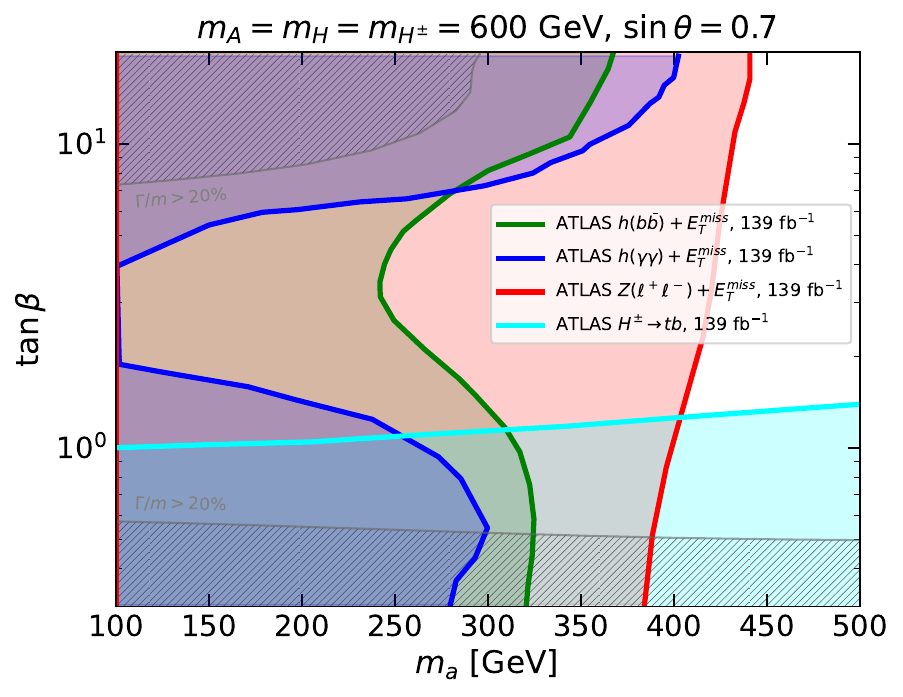}

\vspace{4mm}

\includegraphics[width=.45\linewidth]{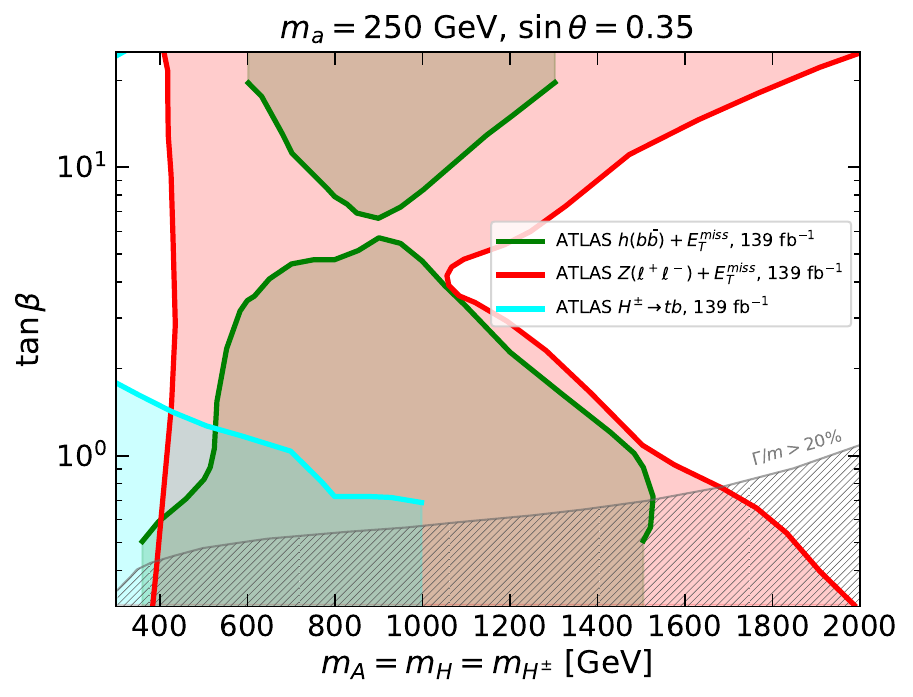}  \qquad 
\includegraphics[width=.45\linewidth]{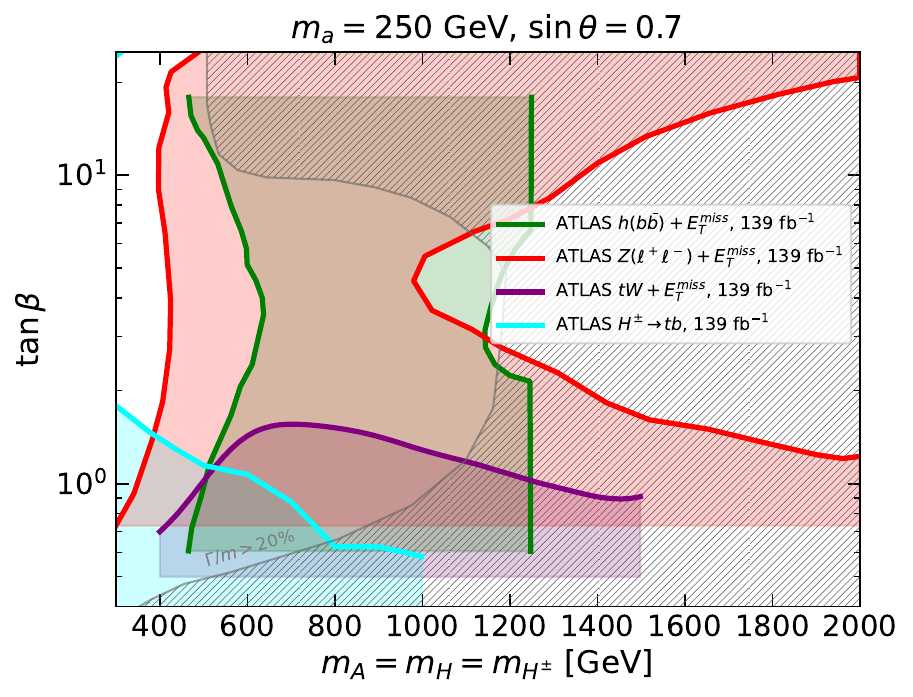}
\vspace{4mm}
 \caption{As~Figure~\ref{fig:2hdma_2D_exclusion}  but for the $m_a \hspace{0.25mm}$--$\hspace{0.5mm} \tan \beta$ plane~(upper panels) and $m_A \hspace{0.25mm}$--$\hspace{0.5mm} \tan \beta$  plane~(lower panels).  The parameters choices not given in~(\ref{eq:bench1}) are indicated in the headlines of the plots. \label{fig:2hdma_tanb_exclusion}}
\end{figure}

The enhanced  sensitivity  of the mono-Higgs searches at high $m_A$  is due to the quadratic dependence of~(\ref{eq:2HDMacouplings2}) on the 2HDM pseudoscalar mass. We~also note that for large $m_A\hspace{0.25mm}$--$\hspace{0.5mm}m_a$ mass splittings, the widths of all non-SM Higgs  bosons become large with the effect being more pronounced for increasing $\sin\theta$~---~cf.~(\ref{eq:2HDMacouplings1}),~(\ref{eq:2HDMacouplings2}) and~(\ref{eq:2HDMacouplings3}). While off-shell effects are taken into account in~Figure~\ref{fig:2hdma_2D_exclusion},  possible modifications of the line shape of the intermediate Higgs bosons~\cite{Seymour:1995np,Passarino:2010qk,Goria:2011wa} are ignored.  The latter effects have been studied in~\cite{Anastasiou:2011pi,Anastasiou:2012hx}, where it has been shown that for a heavy Higgs boson different treatments of its propagator can lead to notable difference in the inclusive production cross sections compared to the case of a Breit-Wigner with a fixed width, as used here.  The hatched grey regions in the two plots of~Figure~\ref{fig:2hdma_2D_exclusion} correspond to the parameter space where any of the additional Higgs bosons has a relative width of more than 20\%. The mono-$X$ exclusions in these regions carry some (hard to quantify) model dependence related to the precise treatment of the widths of the internal Higgs bosons. 

Notice finally that in the case of the benchmark scenario~(\ref{eq:bench1})  the constraints from $H^\pm \to tb$ are to a very good approximation independent of $m_a$, and we find that charged Higgs boson masses $m_{H^\pm} \lesssim 700 \, {\rm GeV}$ ($m_{H^\pm} \lesssim 600 \, {\rm GeV}$) are strongly disfavoured for $\sin \theta = 0.35$~($\sin \theta = 0.7$). Searches for $H^\pm \to tb$ hence cover an area largely complementary to the results of the mono-$X$ searches. We also add that the obtained limits are almost as strong as the indirect bounds that follow from $B \to X_s \gamma$ (see Section~\ref{sec:2HDMatheory}). With more luminosity to be collected at the LHC one can expect the direct charged Higgs searches to become competitive with the flavour bounds. 

\paragraph{\bf Scans in $\bm{\tan \beta}$}

\begin{figure}[t!]
\centering
\includegraphics[width=.45\linewidth]{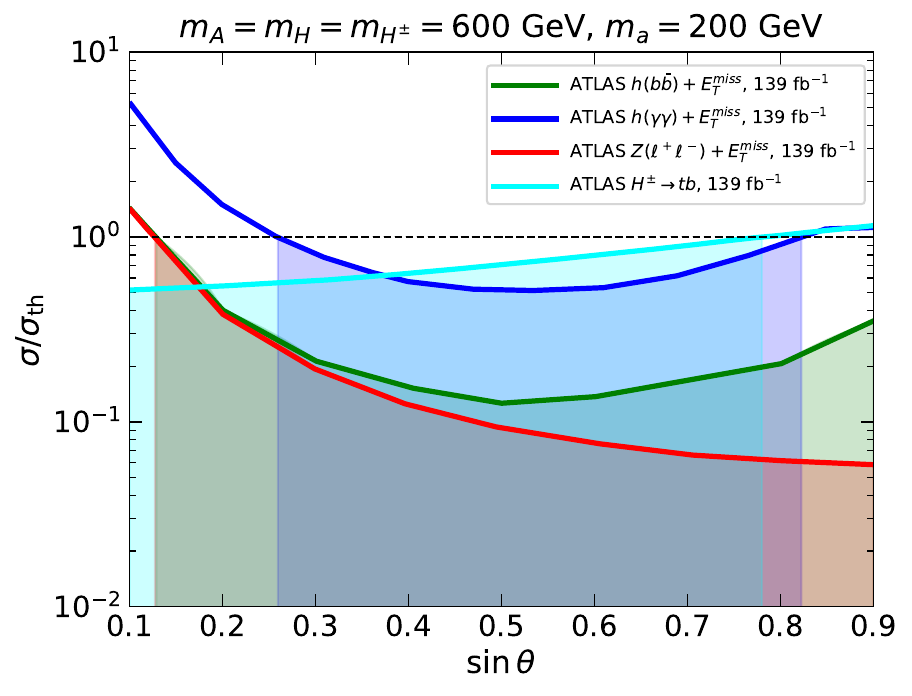} \qquad 
\includegraphics[width=.45\linewidth]{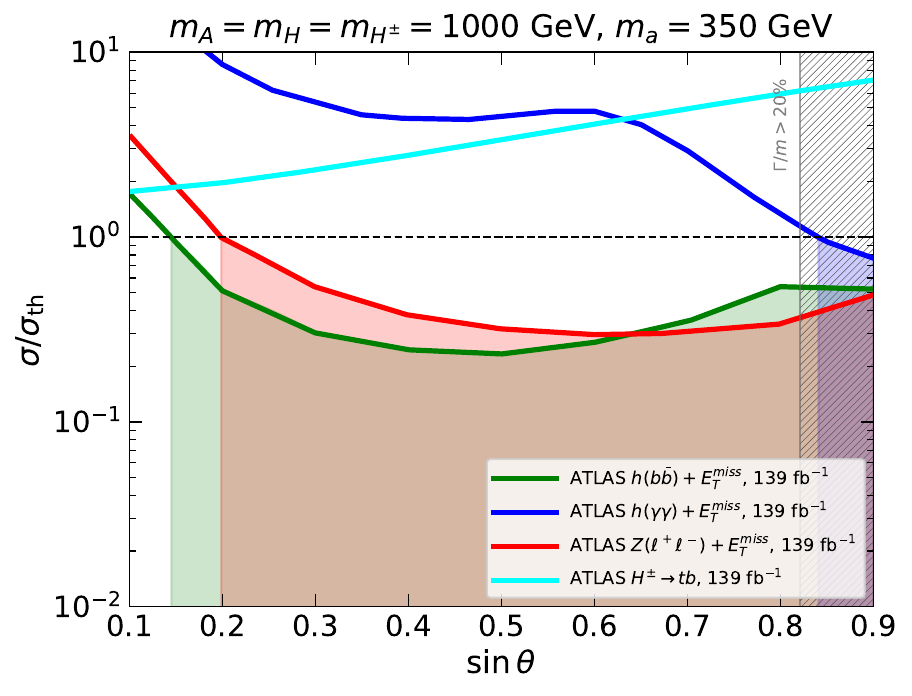}
\vspace{2mm}
\caption{Exclusion limits at 95\%~CL for the 2HDM+$a$ model as a function of $\sin\theta$ in terms of the excluded cross section ($\sigma$) over the cross section predicted by the model ($\sigma_{\rm th}$). The limits are taken from~\cite{ATLAS:2021shl,ATLAS:2021jbf,ATLAS-CONF-2021-029}. The left and right panel shows results for $m_A = 600 \, {\rm GeV}$, $m_a = 200 \,  {\rm GeV}$ and $m_A = 1000 \, {\rm GeV}$, $m_a = 350 \,  {\rm GeV}$, respectively. The other parameters are chosen as in~(\ref{eq:bench1}) and~(\ref{eq:bench2}). In~the plot on the right-hand side the  parameter space where any of the non-SM Higgs bosons has a relative width of more than 20\% is indicated by the  hatched grey region and the label~$\Gamma/m > 20\%$. 
   \label{fig:2hdma_sinp_exclusion}}
\end{figure}

The 95\%~CL limits that the  searches~\cite{ATLAS:2021shl,ATLAS:2021jbf,ATLAS:2020yzc,CMS:2020ulv,ATLAS-CONF-2021-029,ATLAS:2021upq} set in the  $m_a \hspace{0.25mm}$--$\hspace{0.5mm} \tan \beta$ and $m_A \hspace{0.25mm}$--$\hspace{0.5mm} \tan \beta$  planes are displayed  in~Figure~\ref{fig:2hdma_tanb_exclusion}.  For $\tan\beta\gtrsim 5$, the $b \bar b$-induced production becomes dominant for both the mono-Higgs  and mono-$Z$ signatures, since the couplings~(\ref{eq:2HDMfermions}) of the  bottom quark to the neutral Higgs bosons are proportional to $\tan\beta$ in the case of a 2HDM model of type-II. In fact, the sensitivity of the $h+E_T^{\mathrm{miss}}$ searches at high $\tan\beta$ solely stems from $b \bar b$-fusion production, while at low $\tan\beta$ the sensitivity comes entirely  from ggF production. Production via $b \bar b$-fusion leads to final states with more bottom-quarks jets~($b$-jets), therefore dedicated selections with additional $b$-jets can be employed to increase the sensitivity in this region. In particular,  the $h \hspace{0.25mm} (b\bar{b})+E_T^{\mathrm{miss}}$ exclusion at high $\tan\beta$ and  low $\sin\theta$ has been shown to come almost exclusively from a selection with additional $b$-jets~\cite{ATLAS:2021shl}. Notice that the $H^{\pm}\to tb$ search is sensitivity to both low and high values of $\tan\beta$, since the associated coupling involves both  $y_t \cot \beta$ and $y_b \tan\beta$ tems. Parameter regions with $\tan\beta\lesssim 0.3$ are incompatible with the requirement of having a perturbative top-quark Yukawa coupling~\cite{Branco:2011iw} and therefore this region is not displayed in the exclusion plots, while flavour observables disfavour very high values of $\tan\beta$~---~see for instance~\cite{Abe:2018bpo,Bauer:2017ota,Robens:2021lov}. 

\paragraph{\bf Scans in $\bm{\sin \theta}$}

In~Figure~\ref{fig:2hdma_sinp_exclusion} we show the 95\%~CL exclusion limits as a function of $\sin\theta$ that follow from~\cite{ATLAS:2021shl,ATLAS:2021jbf,ATLAS-CONF-2021-029} for two sets of pseudoscalar masses $m_A$ and $m_a$. The sensitivity of all searches improves as $\sin\theta$ increases from zero, since for $\sin\theta=0$ the DM mediator $a$ decouples. The shape of the exclusion curves is due to the interplay between the production cross section and the acceptance, in particular at intermediate values of $\sin\theta$ values, the $E_T^{\mathrm{miss}}$ spectrum becomes harder, leading to an increased signal acceptance. The improvement of the sensitivity of the $h \hspace{0.25mm} (\gamma\gamma)+E_T^{\mathrm{miss}}$ search at high $m_A$ and $\sin\theta$  is due to the scaling of the $g_{haa}$ coupling~(\ref{eq:2HDMacouplings2}), as discussed previously.

\paragraph{\textbf{DM mass scan}}

\begin{figure}[t!]
\centering
\includegraphics[width=.65\linewidth]{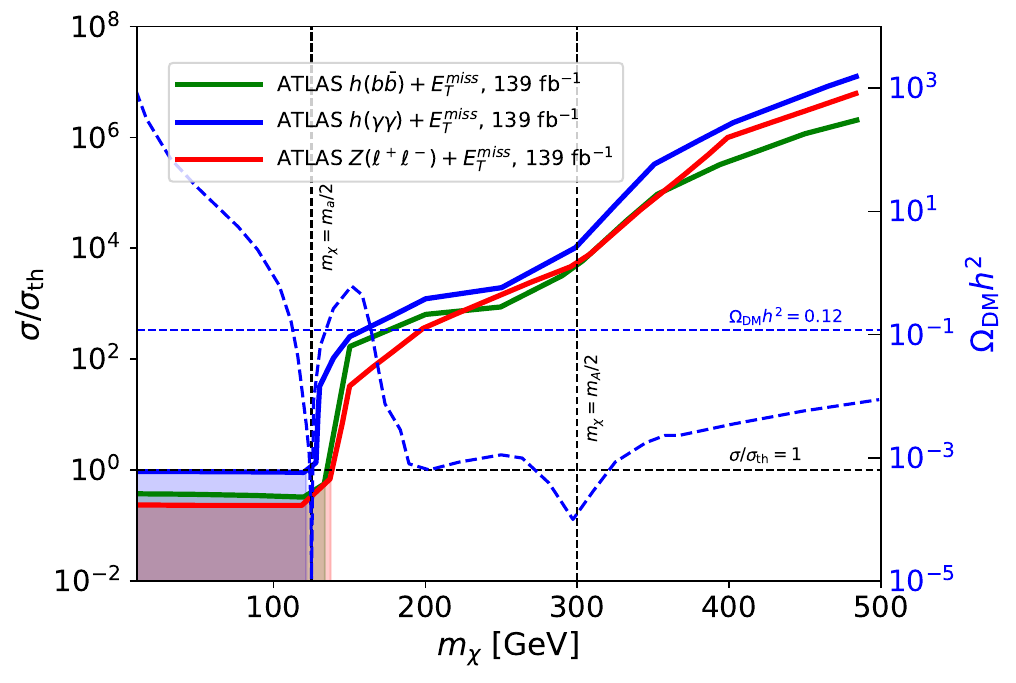}
\vspace{2mm}
 \caption{As~Figure~\ref{fig:2hdma_sinp_exclusion} but as a function of $m_{\chi}$.  The parameters choices not given in~(\ref{eq:bench1}) and~(\ref{eq:bench2}) are  $m_A =  600 \, {\rm GeV}$, $m_a = 250 \, {\rm GeV}$ and $\sin \theta = 0.35$. The solid lines correspond to the  limits arising from the mono-$X$ searches~\cite{ATLAS:2021shl,ATLAS:2021jbf,ATLAS-CONF-2021-029}, while the dashed blue line corresponds to the calculated relic density for the studied 2HDM+$a$ benchmark model. \label{fig:2hdma_mdm}}
\end{figure}

Mono-$X$ searches can also provide constraints on the mass $m_\chi$ of the DM candidate.~Figure~\ref{fig:2hdma_mdm} illustrates these constraints at 95\%~CL for~(\ref{eq:bench1}),~(\ref{eq:bench2}), $m_A = 600 \, {\rm GeV}$, $m_a = 250 \, {\rm GeV}$ and $\sin \theta = 0.35$. In the region $m_{\chi}<m_{a}/2$, where the $a\to \chi\bar{\chi}$ decay is kinematically allowed, a variation of $m_{\chi}$ has a negligible impact on the production cross sections and the $E_T^{\mathrm{miss}}$ spectrum, and in consequence the sensitivity of the $X+E_T^{\mathrm{miss}}$ searches is independent of $m_{\chi}$. For higher masses, the production cross sections decrease significantly and the $E_T^{\mathrm{miss}}$ becomes softer, significantly worsening the sensitivity. For this particular benchmark the existing searches can exclude DM masses below $m_{\chi}\lesssim 140 \, {\rm GeV}$. The DM relic density calculation for this benchmark is also shown in~Figure~\ref{fig:2hdma_mdm} . This calculation is performed using {\tt MadDM}~\cite{Backovic:2015tpt} and relies on the simplified assumption that the DM relic density is solely determined by the interactions predicted in the model. This assumption would be violated in the presence of additional hidden degrees of freedom or interactions. As a result, overproduction or underproduction of DM should not be interpreted as an argument for excluding certain parameter ranges of the model~\cite{Albert:2017onk,Abe:2018bpo}.

\paragraph{\textbf{Outlook}}

The above discussion shows that the 2HDM+$a$ model predicts a rich phenomenology of processes resulting in a diverse range of final-state signatures with and without~$E_T^{\rm miss}$.  The searches that are considered in this review  to constrain the 2HDM+$a$ parameter space include the mono-Higgs, mono-$Z$, $tW+E_T^{\rm miss}$ and $H^\pm \to tb$ channels~---~see also the recent ATLAS note~\cite{ATLAS-CONF-2021-036}. The HL-LHC prospects of the $tW+E_T^{\rm miss}$ and four-top channel have been discussed in~\cite{CidVidal:2018eel,ATL-PHYS-PUB-2018-027}. All existing 2HDM+$a$  collider studies~\cite{Goncalves:2016iyg,Bauer:2017ota,Pani:2017qyd,Abe:2018bpo,ATLAS:2017hoo,CMS:2018zjv,ATL-PHYS-PUB-2018-027,ATLAS:2019wdu,CMS:2020ulv,ATLAS:2020yzc,ATLAS:2021jbf,ATLAS:2021shl,ATLAS-CONF-2021-029,ATLAS-CONF-2021-036,Arcadi:2020gge,Butterworth:2020vnb,Robens:2021lov} have assumed that the Yukawa sector of the model is of~type-II. In this class of models the  bounds from FCNC processes and LHC  searches for heavy Higgses are strong, pushing the masses of the additional 2HDM Higgses above the $500 \, {\rm GeV}$ range.  It is well-known~(cf.~for example~\cite{Fox:2017uwr,Haisch:2017gql}) that in fermiophobic 2HDM models of type-I the constraints on the additional Higgs boson can be significantly  relaxed, thereby allowing for new scalars and pseudoscalars with   masses of order of the EW scale. The fermiophobic nature of the Higgs bosons leads to unconventional production mechanism and also the decays of the non-SM spin-0 particles can have unfamiliar patterns. The mono-$X$ phenomenology in fermiophobic 2HDM+$a$ models awaits explorations. 

\subsection{2HDM+$s$ model} 
\label{sec:2HDMs}

Instead of mixing an additional CP-odd singlet $P$ with the 2HDM pseudoscalar~$A$, as done in~(\ref{eq:VHP}), it is also possible to mix a scalar singlet $S$ with the CP-even spin-$0$ states $h$ and $H$. This gives rise to the class of 2HDM+$s$ models~\cite{Bell:2016ekl,Bell:2017rgi}, which are  the natural, gauge invariant extension of the simplified DM models with a single scalar mediator~\cite{Abdallah:2015ter,Abercrombie:2015wmb}.  Like in the case of the 2HDM+$a$ model, the presence of non-SM Higgs bosons in the 2HDM+$s$ model can lead to novel $E_T^{\rm miss}$ that are not captured by a DM simplified model with just a single scalar mediator. While the CP nature of the mediator in the 2HDM+$s$ model typically leads to large SI~DM-nucleon interactions that are problematic in view of the stringent DM~DD constraints, we will give an example of a 2HDM+$s$ model realisation to which the DM~DD searches are blind. Detailed discussions of the 2HDM+$s$ model have been presented in~\cite{Bell:2016ekl,Bell:2017rgi,Arcadi:2020gge} and we will summarise the most important findings of these articles below.  

\subsubsection{Theory}

The construction of the DM-mediator interactions and the full scalar potential of the 2HDM+$s$ model proceeds like in the case of the 2HDM+$a$ model with minor modifications. First, the mediator $S$ couples to the  dark scalar current $\chi \bar \chi$ and not to the pseudoscalar current $\bar \chi \gamma_5 \chi$.  Second,  the coupling of the $S$ to the term $(H_1^\dagger H_2 + {\rm h.c.})$ that appears in~(\ref{eq:VHP}) is taken to be purely real so that only CP-even components can mix. Alternatively, one can assume that the singlet $S$ develops a VEV, and in this way  obtain a mixing between $S$ and the CP-even scalars of the 2HDM doublets~\cite{Bell:2016ekl,Bell:2017rgi}. In view of the stringent constraints from the measurement of the Higgs signal strengths at the LHC, it is furthermore natural to work in a generalised alignment limit~\cite{Bell:2016ekl} where only the weak eigenstates $H$ and $S$ mix, giving rise to the mass eigenstates $S_1$ and~$S_2$. In~the works~\cite{Bell:2016ekl,Bell:2017rgi,Arcadi:2020gge} the mixing angle $\theta$ between $H$ and $S$ has been defined such that for $\sin \theta = 0$ the state $S_2$ is a pure $S$ state. We will employ this definition hereafter as well. 

Before discussing the  constraints on the  2HDM+$s$ model that arise from mono-$X$ searches, let us also spend some words on the restrictions on the parameter space that stem from DM~DD experiments. Expressed in terms of mass eigenstates the DM-mediator interactions in the 2HDM+$s$ model take the form  
\begin{equation} \label{eq:LDM2HDMs}
{\cal L}_\chi = - y_\chi \hspace{0.25mm} \left ( \sin \theta  \hspace{0.5mm} S_1 + \cos \theta  \hspace{0.5mm} S_2 \right ) \hspace{0.5mm}  \chi \bar \chi \,.
\end{equation}
The couplings between $S_1$ and $S_2$ and the SM quarks depend on the choice of the 2HDM Yukawa sector. In the case of a type-II model one has for instance
\begin{equation} \label{eq:LY2HDMs}
{\cal L}_Y = -  \frac{\cos \theta  \hspace{0.5mm} S_1 - \sin \theta  \hspace{0.5mm} S_2}{\sqrt{2}} \left [ \sum_{q=u,c,t} y_q \cot \beta \hspace{0.5mm}  \bar q q - \sum_{q=d,s,b} y_q \tan \beta  \hspace{0.5mm}  \bar q q \right ] \,.
\end{equation}

The SI~DM-nucleon cross section takes the general form
\begin{equation} \label{eq:SI1}
\sigma_{\rm SI} = \left ( \frac{m_N m_\chi}{m_N + m_\chi} \right )^2  \, \frac{c_N^2}{\pi} \,,
\end{equation}
where $c_N$ is the Wilson coefficient of the dimension-six nucleon operator $O_N  = \chi \bar \chi \bar N N$ that can be found by integrating out the mediators $S_1$ and $S_2$.  In the case of~(\ref{eq:LDM2HDMs}) and~(\ref{eq:LY2HDMs}) this coefficient reads~\cite{Bell:2016ekl,Bell:2017rgi,Arcadi:2020gge}
\begin{equation} \label{eq:cN}
\begin{split}
c_N  & =  \frac{m_N}{v} \, \frac{y_\chi \sin \left ( 2 \theta \right )}{2} \left ( \frac{1}{m_{S_1}^2} -  \frac{1}{m_{S_2}^2} \right ) \\[2mm]
& \phantom{xx} \times  \left [ \cot \beta \hspace{0.5mm} f_{T_u}^N - \tan \beta \sum_{q=d,s} f_{T_q}^N + \frac{4 \cot \beta -2 \tan \beta }{27} \,  f_{T_G}^N  \right ] \,, 
\end{split}
\end{equation}
where $f_{T_u}^N \simeq 0.019$, $f_{T_d}^N \simeq 0.045$ and $f_{T_s}^N \simeq 0.043$~\cite{Alarcon:2011zs,Alarcon:2012nr,Junnarkar:2013ac,Hoferichter:2015dsa} are  the quark-nucleon matrix elements and  $f_{T_G}^N  = 1- \sum_{q=u,d,s} f_{T_q}^N \simeq 0.89$ is the effective gluon-nucleon coupling. Besides the obvious possibilities to suppress the Wilson coefficient~$c_N$,~i.e.~$y_\chi \to 0$, $\theta \to 0$ or $m_{S_1} \to m_{S_2}$, in the case of the 2HDM of type-II there is a fourth possibility to achieve that $c_N \simeq 0$. The trick is to choose $\tan \beta$ such that the square bracket in~(\ref{eq:cN}) vanishes.  Numerically, this happens for $\tan \beta \simeq 1$, and this cancellation is only possible because the up- and down-type contributions to~$c_N$ interfere destructively in the case of the 2HDM of type-II.  Note that in a scenario where a singlet scalar  mixes with the $125 \, {\rm GeV}$ Higgs to provide the DM portal~\cite{Kim:2008pp,Kim:2009ke,Baek:2011aa,Lopez-Honorez:2012tov,Baek:2012uj,Fairbairn:2013uta,Carpenter:2013xra} or a  2HDM model of type-I, interference effects between different quarks are not possible as all fermion couplings to the mediator are proportional to the SM Yukawa couplings with the same proportionality coefficient.  The above discussion shows that the Yukawa freedom introduced by the 2HDM is an important feature of the 2HDM+$s$ model, because it can be used to mitigate the stringent DM~DD constraints. As a result, it is possible to obtain the correct relic density without having excessive contributions to $\sigma_{\rm SI}$. See~\cite{Bell:2017rgi} for a detailed discussion of this point. 

\subsubsection{LHC constraints}

\begin{figure}[t!]
\centering
\includegraphics[width=.75\linewidth]{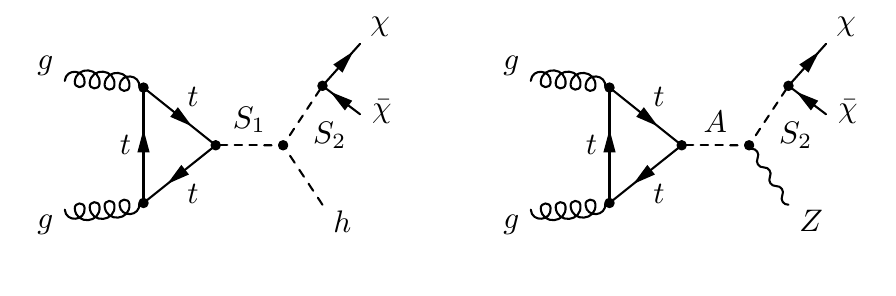}
\vspace{0mm}
\caption{\label{fig:2HDMsdiagrams}  Feynman diagrams that give rise to a resonant mono-Higgs signal~(left) and a mono-$Z$ signature~(right) in ggF production in the 2HDM+$s$ model.}
\end{figure}

While ATLAS and CMS  have not published interpretations of the DM searches in the context of the 2HDM+$s$ model, theoretical reinterpretations of LHC searches have been performed.  In the~article~\cite{Arcadi:2020gge} the ATLAS and CMS searches for the $h \hspace{0.25mm} (b\bar{b})+E_T^{\mathrm{miss}}$~\cite{ATLAS:2017uis,CMS:2018zjv} and the $Z \hspace{0.25mm} (\ell^+ \ell^-)+E_T^{\mathrm{miss}}$~\cite{ATLAS:2017nyv,CMS:2017nxf} final states were examined, and it was found that like in the case of the 2HDM+$a$ model these signatures provide the best coverage of the parameter space of the model. This is again related to the fact that the mono-Higgs and mono-$Z$ signatures receive resonant contributions. The~corresponding ggF production channels are displayed in~Figure~\ref{fig:2HDMsdiagrams}.  Notice that resonant contributions to the $tW + E_T^{\rm miss}$ signal also exist in the 2HDM+$s$ model.  The~relevant graph is obtained from the left one in the lower row  in~Figure~\ref{fig:2HDMadiagrams} by replacing the $H^- a W^-$ vertex by a $H^- S_2 W^-$ vertex. The~$tW + E_T^{\rm miss}$ signature has so far not been studied in the context to the 2HDM+$s$ model. 

The analysis performed in~\cite{Arcadi:2020gge} employs the generalised alignment limit and uses the following benchmark parameter choices
\begin{equation} \label{eq:bench2HDMs}
\begin{split}
& m_A = m_{S_1} = m_{H^\pm} \,,  \quad   m_\chi = 10 \, {\rm GeV} \,, \quad \sin \theta = 0.3 \,, \\[2mm]
& \phantom{xx} y_\chi = 1 \,, \quad \lambda_3= \frac{m_h^2}{v^2} \simeq 0.26  \,, \quad \lambda_{S1} = \lambda_{S2} =0 \,,
\end{split}
\end{equation}
where $\lambda_{S1}$ and $\lambda_{S2}$ are the analogues of  $\lambda_{P1}$ and $\lambda_{P2}$ introduced in~(\ref{eq:VHP}). The Yukawa sector of the 2HDM is taken to be of type-II. While these parameter choices have common features with the 2HDM+$a$ benchmark~(\ref{eq:bench1}), the low value of $\lambda_3$ and the vanishing quartic couplings~$\lambda_{S_1}$ and  $\lambda_{S_2}$ lead to a modified hierachy of the mono-$X$ signatures with respect to the 2HDM+$a$ model.  This happens because the $S_1S_2h$ coupling like~$g_{Aah}$ in~(\ref{eq:2HDMacouplings1})  involves~a~term proportional to $\lambda_{S_1}\cos^2\beta+\lambda_{S_2}\sin^2\beta$, while the $AS_2Z$ coupling similar to  $g_{HaZ}$~in~(\ref{eq:2HDMacouplings3})  does not depend on the quartic couplings. The parameter choices~(\ref{eq:bench2HDMs})  thus favours a mono-$Z$ signal over a mono-Higgs signature, while in the case of the 2HDM+$a$ model the opposite is the case because of the sizeable quartic couplings employed in the benchmark~(\ref{eq:bench1}). 

\begin{figure}[!t]
\centering
	\includegraphics[width=0.35\textwidth]{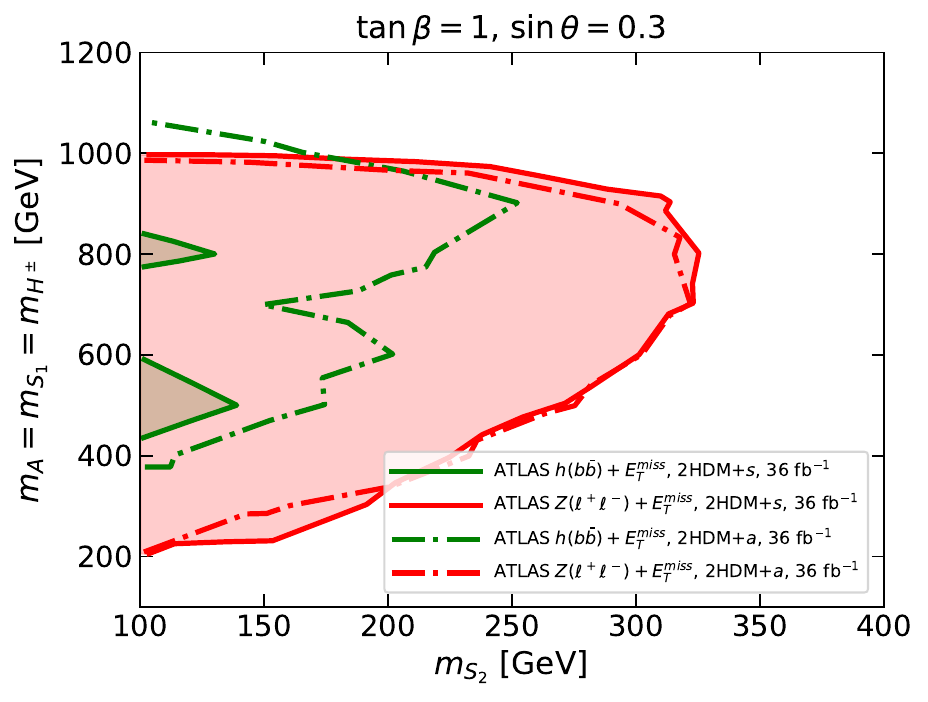} \quad 
	\includegraphics[width=0.34\textwidth]{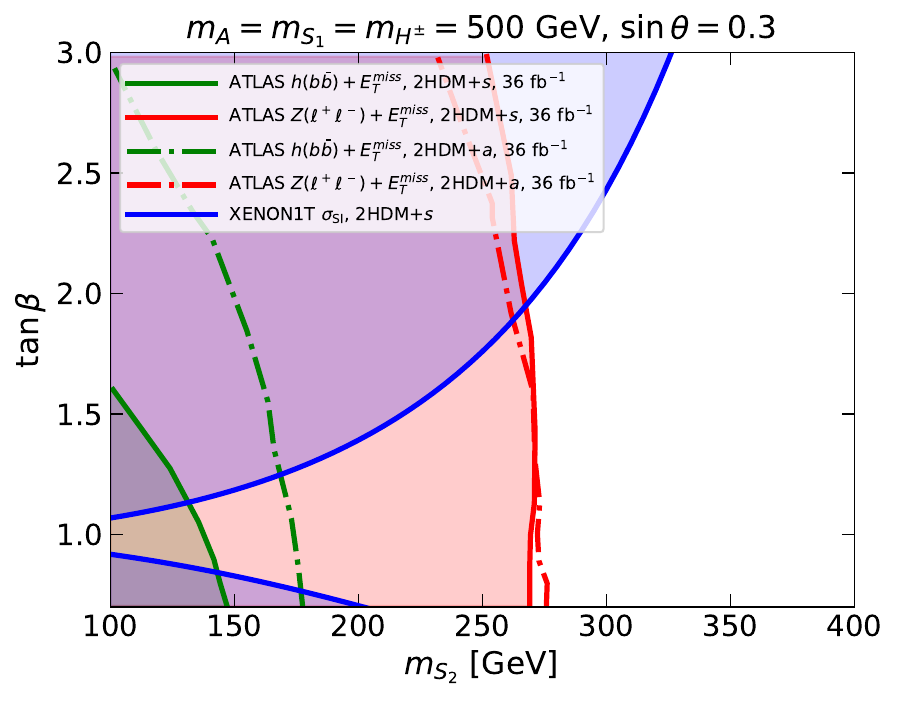}
\vspace{2mm}
\caption{98\%~CL constraints on the 2HDM+$s$ model in the $m_{S_2}\hspace{0.25mm} $--$\hspace{0.5mm} m_A$ plane~(left panel) and the $m_{S_2}\hspace{0.25mm} $--$ \hspace{0.5mm} \tan\beta$ plane~(right panel) displayed as solid lines. For comparison also the constraints in the 2HDM+$a$ model are shown as dash-dotted lines. All LHC exclusions use the benchmark parameters~(\ref{eq:bench2HDMs}) and are taken from the publication~\cite{Arcadi:2020gge}. The solid blue contour in the right plot corresponds to the parameter space that is excluded at 90\%~CL by the XENON1T upper limit on the SI~DM-nucleon cross section~\cite{XENON:2018voc}. For further details consult the main text.}\label{fig:2hdmsexclusion}
\end{figure}

In~Figure~\ref{fig:2hdmsexclusion} we summarise the most relevant constraints on the 2HDM+$s$ model in the $m_{S_2}\hspace{0.25mm} $--$\hspace{0.5mm} m_A$ plane~(left panel) and the $m_{S_2}\hspace{0.25mm} $--$ \hspace{0.5mm} \tan\beta$ plane~(right panel). The shown exclusions have been obtained in~\cite{Arcadi:2020gge} by recasting the ATLAS searches for $h \hspace{0.25mm} (b\bar{b})+E_T^{\mathrm{miss}}$~\cite{ATLAS:2017uis} and $Z \hspace{0.25mm} (\ell^+ \ell^-)+E_T^{\mathrm{miss}}$~\cite{ATLAS:2017nyv}. As a result of the choice of the quartic couplings in~(\ref{eq:bench2HDMs}), the mono-$Z$ search provides compared to the mono-Higgs search far better constraints in both two-dimensional parameter planes. For comparison, also the exclusions in the 2HDM+$a$ model using the same input are shown in~Figure~\ref{fig:2hdmsexclusion}. One observes that the mono-$Z$ constraints are similar in shape and reach in both models. This feature is readily understood by noting that the relevant processes that lead to the mono-$Z$ signal in the two models are $gg \to A \to S_2Z$ and $gg \to H \to aZ$. However, for a scalar and pseudoscalar of the same mass and with the same couplings to quarks one has approximately $\sigma \left (gg \to A \right )/\sigma \left (gg \to H \right ) \simeq 2$ and ${\rm BR} \left (A \to S_2Z \right )/{\rm BR} \left (H \to aZ \right ) \simeq 1/2$~\cite{Arcadi:2020gge}, which leads to the same signal strength in both cases. For the mono-Higgs signal, the relevant resonant processes are instead $gg \to S_1 \to S_2 h$ and $gg \to A \to a h$. In this case one has however that $\sigma \left (gg \to S_1 \right )/\sigma \left (gg \to A \right ) \simeq 1/2$ and ${\rm BR} \left (S_1 \to S_2 h \right )/{\rm BR} \left (A \to ah \right ) \simeq 1$~\cite{Arcadi:2020gge}, and as a result for the same parameters the 2HDM+$a$ model predicts a higher mono-Higgs cross section than the 2HDM+$s$ model. This results in stronger exclusion bounds in the former than in the latter case, as evident from~Figure~\ref{fig:2hdmsexclusion}.

Since the 2HDM+$s$ interactions generically  lead to large SI DM-nucleon cross sections, we show in the two panels of Figure~\ref{fig:2hdmsexclusion} also the restrictions on the parameter space that stem from DM~DD experiments. The shown results employ (\ref{eq:SI1}), (\ref{eq:cN}) and (\ref{eq:bench2HDMs}). In~the case of the scan in the $m_{S_2}\hspace{0.25mm} $--$\hspace{0.5mm} m_A$ plane (left panel) the latest XENON1T results~\cite{XENON:2018voc} do not lead to any constraint because of the parameter choice $\tan \beta = 1$. The situation is  different for the $m_{S_2}\hspace{0.25mm} $--$ \hspace{0.5mm} \tan\beta$ plane  (right panel)  where $\tan \beta$ values sufficiently different from one are excluded at 90\%~CL by the XENON1T measurement, as indicated by the solid blue line. Notice that the DD constraint become weaker for increasing mass~$m_{S_2}$. All these features can be understood from the structure of the coefficient $c_N$ as given in~(\ref{eq:cN}).  The scans in Figure~\ref{fig:2hdmsexclusion} demonstrate that, in contrast to naive expectation, in the 2HDM+$s$ model of type-II it is possible to have interesting LHC signatures that a compatible with the severe limits imposed by DM~DD experiments if $\tan \beta = {\cal O} (1)$. 

The above comparison of the mono-$X$ phenomenology in the 2HDM+$s$ and 2HDM+$a$ models furthermore stresses the complementarity of  mono-Higgs and mono-$Z$ searches in exploring DM two-mediator models (see also~\cite{Bauer:2017fsw} for a related discussion). Unfortunately, the 2HDM+$s$ model has largely escaped the  attention of both the theoretical and experimental community. Neither recommendations a la~\cite{Abe:2018bpo} exist for the 2HDM+$s$ model nor is there a single 2HDM+$s$  interpretation by ATLAS and CMS of the plethora of LHC DM searches. In view of the fact that the articles~\cite{Bell:2016ekl,Bell:2017rgi,Arcadi:2020gge} give a detailed account of the relevant $E_T^{\rm miss}$ signatures, the DM~DD and ID bounds and the relic density calculation, one would hope that combined analyses such as~\cite{ATLAS-CONF-2021-036} will be done in the future in the case of the 2HDM+$s$ model as well. 

\section{Portals with extended Higgs and gauge sectors} 
\label{sec:extendedhiggsgaugeportal}

In Section~\ref{sec:extendedhiggsportals} we have discussed two-mediator models where both portal particles have spin-0. Constructing renormalisable interactions between the dark and the visible sector  that involve a spin-0 as well as a spin-1 state is however also possible. Given that the interactions between DM and SM particles are severely constrained experimentally, such construction typically involve extensions of both the Higgs and gauge sector of the~SM. The resulting theories fall into the class of dark $Z^\prime$ or dark Higgs models, which typically have a rich collider and DM phenomenology. In the following we will discuss two of these models in more detail. The choice of models is  motivated by the fact that ATLAS and CMS have already searched for the discussed DM models. 

\subsection{2HDM+$Z^\prime$ model}  
\label{sec:2HDMZp}

The dark $Z^\prime$ model that has received by far the most attention at the LHC~(see for example~\cite{ATLAS:2017uis,CMS:2017prz,CMS:2018nlv,CMS:2018zjv,CMS:2018ljc,CMS:2019ykj,ATLAS:2021shl,ATLAS:2021jbf}) is the so-called 2HDM+$Z^\prime$ models introduced in~\cite{Berlin:2014cfa}. In this model only the  right-handed up-type quarks are assumed to be charged under the $U(1)_{Z^\prime}$ symmetry, while all the other SM fermion fields are chosen as SM singlets. This choice allows LHC production of the $Z^\prime$ boson in $u \bar u$- and $c \bar c$-fusion, but avoids the stringent constraints from searches for dilepton resonances since the charged leptons carry no $U(1)_{Z^\prime}$  charge. The Higgs sector of the  2HDM+$Z^\prime$ model is taken to be a 2HDM of type-II. To incorporate DM interactions, it is assumed that the heavy CP-odd pseudoscalar~$A$ that arises  from the 2HDM  possesses a large coupling to DM particles, such that the corresponding branching ratio is close to one. Models in which the DM candidate $\chi$ is a spin-1/2 or a spin-0 particle have been sketched in~\cite{Berlin:2014cfa}.  In the first case $\chi$ is a Majorana fermion that arises from singlet-doublet DM~\cite{Mahbubani:2005pt,Enberg:2007rp,Cohen:2011ec,Cheung:2013dua}, while in the second case $\chi$ is the lightest component of a complex scalar field. To avoid the stringent constraints from invisible Higgs boson decays and/or DM direct detection experiments the fundamental parameters in the underlying models that give rise to $\chi$ need to be tuned, but after tuning it is possible to obtain large branching ratios of $A$ to DM for $\chi$ masses in the ballpark of $100 \, {\rm GeV}$ in both cases. For concreteness we will assume hereafter that the DM candidate in the  2HDM+$Z^\prime$  model is a Majorana fermion that couples to the $A$ with the coupling strength $g_\chi$. The same assumption is made in the ATLAS and CMS analyses~\cite{ATLAS:2017uis,CMS:2017prz,CMS:2018nlv,CMS:2018zjv,CMS:2018ljc,CMS:2019ykj,ATLAS:2021shl,ATLAS:2021jbf}. In addition we will focus our discussion on the LHC phenomenology of the 2HDM+$Z^\prime$  model, ignoring the DM phenomenology because it is more model dependent. 

\subsubsection{Theory}

The tight  constraints from the Higgs coupling measurements at the LHC are avoided most easily by working in the alignment limit of the  2HDM+$Z^\prime$ model. In this limit the partial decay widths of the $Z^\prime$ boson that are relevant for the discussion hereafter take the following form 
\begin{eqnarray} \label{eq:Zdarkdecays}
\begin{split}
\Gamma\left (Z^\prime\to q \bar q \right )&= \frac{g_{Z^\prime}^2 m_{Z^\prime}}{32 \pi} \left ( 1 - 4 x_{q/Z^\prime} \right )^{1/2}  \left ( 1 - x_{q/Z^\prime} \right ) \,, \\[2mm]
\Gamma\left (Z^\prime\to Ah \right )&= \frac{g_{Z^\prime}^2  \sin^2 (2\beta) \hspace{0.5mm} m_{Z^\prime}}{768 \pi} \left (x_{A/Z^\prime}^2 + ( 1 - x_{h/Z^\prime} )^2 - 2 x_{A/Z^\prime} (1+x_{h/Z^\prime})  \right )^{3/2} \, ,\\[2mm]
\Gamma\left (Z^\prime\to Zh \right )&=  \frac{g_{Z^\prime}^2 \sin^4 \beta \hspace{0.75mm} m_{Z^\prime}}{192 \pi}   \left (x_{Z/Z^\prime}^2 + ( 1 - x_{h/Z^\prime} )^2 - 2 x_{Z/Z^\prime} (1+x_{h/Z^\prime})  \right )^{1/2} \\[1mm]
& \phantom{xx} \times \left (x_{Z/Z^\prime}^2 + ( 1 - x_{h/Z^\prime} )^2 - 2 x_{Z/Z^\prime} (1+x_{h/Z^\prime})  + 12 x_{Z/Z^\prime} \right )\, .
\end{split}
\end{eqnarray}
Here $g_{Z^\prime}$ denotes the $U(1)_{Z^\prime}$ gauge coupling,  we have used the abbreviations $x_{i/j} = m_i^2/m_j^2$ and  have neglected contributions that are suppressed by the mass ratio $m_Z^2/m_{Z^\prime}^2$ of the $Z$-boson and the $Z^{\prime}$-boson mass. Such terms arise from $Z$--$Z^{\prime}$ mixing, but are numerically subdominant in the partial decay rates of the $Z^\prime$ boson. Notice that the $Z^\prime$ boson can only decay to up-quark and charm-quark pairs as well as to top-quark pairs if $m_{Z^\prime} > 2 m_t$ with $m_t \simeq 173 \, {\rm GeV}$ denoting the top-quark mass, but not to the other SM fermions because these matter fields are assumed to be $U(1)_{Z^\prime}$ singlets. 

If the $A$ is sufficiently heavy, the coupling $g_\chi$ is sizeable and $\tan \beta$ is small, important  decays of the heavy pseudoscalar are to DM and top-quark pairs. In the alignment limit the corresponding partial decay rates are given by 
\begin{equation} \label{eq:Adecays}
\begin{split}
\Gamma \left ( A \to \chi \bar \chi \right ) & =  \frac{g_\chi^2 m_A}{8 \pi} \, \left ( 1 - 4 x_{\chi/A} \right )^{1/2} \,, \\[2mm]
\Gamma \left ( A \to   t \bar t \right ) & = \frac{3 y_t^2 \cot^2 \beta \hspace{0.75mm}  m_A}{16 \pi} \, \left ( 1 - 4 x_{t/A} \right )^{1/2}\,. 
\end{split}
\end{equation}
Notice that the partial decay width $A \to Zh$ vanishes in the alignment limit, but depending on the precise value of $\tan \beta$ and/or the masses of the heavy 2HDM spin-0 states the decays $A \to b \bar b$, $A \to \tau^+ \tau^-$, $A \to H Z$ and $A \to H^\pm W^\mp$ can  be relevant in this limit.

\begin{figure}[t!]
\centering
\includegraphics[width=.85\linewidth]{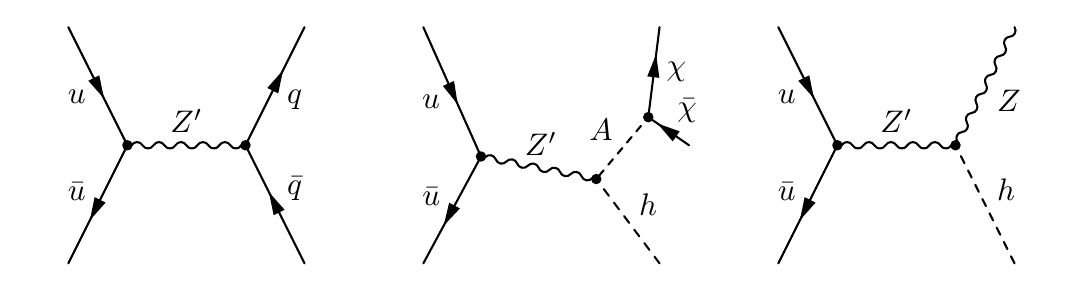}
\vspace{4mm}
\caption{\label{fig:2HDMZp} Examples of Feynman diagrams that lead to dijet or $t \bar t$ production~(left) depending on whether the final-states fermions are  up, charm or top quarks, a mono-Higgs signal~(middle) and resonant $Zh$ production~(right) in the 2HDM+$Z^\prime$ model.  $Z^\prime$-boson production in $c \bar c$-fusion is also possible but not shown explicitly.  For further details consult the main text. 
}
\end{figure}

From the expressions for the partial decay widths~(\ref{eq:Zdarkdecays}) and~(\ref{eq:Adecays})  it follows that if a heavy $Z^\prime$ boson is produced on-shell in the LHC collisions this gives rise to a dijet or a $t \bar t$ final state, a mono-Higgs signature or  resonant $Zh$ production. Examples of the corresponding Feynman diagrams are shown in~Figure~\ref{fig:2HDMZp}. In our numerical study of the 2HDM+$Z^\prime$ model we will consider all four collider signatures that arise from the exchange of $Z^\prime$ bosons, and study only 2HDM+$Z^\prime$ benchmarks in which the dominant $A$ decay modes are described by the expressions~(\ref{eq:Adecays}). 

Other laboratory constraints on the 2HDM+$Z^\prime$ model arise from the measurements of the EWPO and flavour physics. In the context of the EWPO the most important constraint turns out to be the bound on the amount of custodial symmetry breaking as measured by the Peskin-Takeuchi parameter $T$. In the 2HDM+$Z^\prime$ model, one obtains at tree level a strictly positive correction to $T$ of the form  
\begin{equation} \label{eq:T2HDMZp}
T = \frac{\sin^2 \left ( 2 \theta_w \right) \,  g_{Z^\prime}^2 \sin^4 \beta}{16 \pi \alpha^2}  \frac{x_{Z/Z^\prime}}{1-x_{Z/Z^\prime}} \,, 
\end{equation}
where $\sin \theta_w$ denotes the sine of the weak mixing angle $\theta_w$ with $\sin^2 \theta_w \simeq 0.23$, respectively, and $\alpha \simeq 1/128$  is the electromagnetic coupling constant in the $\overline{\rm MS}$ scheme at the mass of the $Z$ boson. A simultaneous fit to the EWPO gives~\cite{Zyla:2020zbs}
\begin{equation} \label{eq:Texp}
T = 0.03 \pm 0.12 \,,
\end{equation}
and this constraint leads to a lower bound on the mass $m_{Z^\prime}$ of the $Z^\prime$ boson for fixed values of $g_{Z^\prime} $ and $\tan \beta$. Constraints on the masses of the extra spin-0 states of the 2HDM in general arise from the $T$ parameter once loop corrections are included. However, such effects are absent if all non-SM Higgs  bosons are taken to be mass degenerate,~i.e.~if $m_A = m_H = m_{H^\pm}$ is assumed, because in this case the 2HDM Higgs potential~(\ref{eq:VH}) is custodially invariant (see the related discussion in Section~\ref{sec:2HDMa}). Below we will only consider this limit. 

As explained in Section~\ref{sec:2HDMa} indirect constraints on the charged Higgs-boson mass arise from flavour physics with  $B \to X_s \gamma$ providing a particularly strong 95\%~CL lower limit of $m_{H^\pm} > 800 \, {\rm GeV}$ in the case of the 2HDM of type-II. Direct searches for heavy Higgs bosons can also be used to explore the parameter space of the 2HDM+$Z^\prime$ model but  turn out to be less restrictive than the $Z^\prime$-boson search strategies discussed below. 

\subsubsection{Summary plots}

The existing LHC analyses~\cite{ATLAS:2017uis,CMS:2017prz,CMS:2018nlv,CMS:2018zjv,CMS:2018ljc,CMS:2019ykj,ATLAS:2021shl,ATLAS:2021jbf}  that consider the 2HDM-$Z^\prime$ model employ the alignment limit and adopt the following parameter choices:
\begin{equation}  \label{eq:2HDMZpbench}
g_\chi = 1 \,, \quad g_{Z^\prime} =0.8 \,,  \quad \tan \beta =1 \,, \quad 
m_{\chi} =100 \, {\rm GeV} \,, \quad m_{A} =m_{H}=m_{H^{\pm}} \,.
\end{equation}
This benchmark has been established in~\cite{Abercrombie:2015wmb} and we will also focus on it in what follows. The phenomenological impact  that modifications of~(\ref{eq:2HDMZpbench}) have will however also be discussed briefly. Before analysing  the relevant LHC constraints, we remark that a combination of~(\ref{eq:T2HDMZp}) and~(\ref{eq:Texp}) leads for~(\ref{eq:2HDMZpbench}) to the 95\%~CL bound 
\begin{equation} \label{eq:Tconstraint}
m_{Z^\prime} > 1080 \, {\rm GeV} \,.
\end{equation}
Notice that this bound can be relaxed by decreasing the values of $g_{Z^\prime}$ and/or $\tan \beta$ compared to~(\ref{eq:2HDMZpbench}) but such parameter choices generically also reduce the signal strengths of the LHC signatures. The $T$ parameter therefore always provides a relevant constraint in the  2HDM-$Z^\prime$ model in the parameter space that is testable at the LHC. 

Dedicated searches for the 2HDM-$Z^\prime$ model have been carried out by ATLAS and CMS in mono-Higgs final states with the Higgs decaying to bottom-quark~\cite{ATLAS:2017uis,CMS:2017prz,CMS:2018zjv,CMS:2018ljc,ATLAS:2021shl},  photon~\cite{CMS:2018nlv,ATLAS:2021jbf} or tau pairs~\cite{CMS:2018nlv} as well as a combination of all relevant Higgs decay channels~\cite{CMS:2019ykj}. The dominant contribution to the mono-Higgs signal arises in the 2HDM+$Z^\prime$ model from the graph shown in the middle of~Figure~\ref{fig:2HDMZp} with subleading effects stemming from $c \bar c$-fusion and $pp \to Z^\prime \to Zh$ followed by $Z \to \nu \bar \nu$. 

Mono-Higgs searches are however not the only relevant collider constraints that need to be considered if one wants to obtain a global picture of the allowed parameter space in the  2HDM-$Z^\prime$ model. That this is the case can simply be seen by evaluating the partial decay width of the $Z^\prime$ boson. Using~(\ref{eq:2HDMZpbench}) together with $m_{Z^\prime} = 2 \, {\rm TeV}$ and $m_A = 800 \, {\rm GeV}$ and employing exact formulas for all possible $Z^\prime$ decay channels gives 
\begin{equation} \label{eq:BRZp}
\begin{split}
& \sum_{q = u, c}  {\rm BR} \left ( Z^\prime \to q \bar q \right ) = 62.8\% \,, \qquad 
 {\rm BR} \left ( Z^\prime \to t \bar t \right ) = 30.7\% \,, \\[2mm]
& \hspace{4.5mm} {\rm BR} \left ( Z^\prime \to Ah \right ) = 0.8\% \,, \qquad 
{\rm BR} \left ( Z^\prime \to Zh \right ) = 1.3\% \,, 
\end{split}
\end{equation}
in good agreement with the approximations~(\ref{eq:Zdarkdecays}).  One furthermore has 
\begin{equation} \label{eq:BRA}
 {\rm BR} \left (A \to \chi  \bar \chi \right ) = 41.8\% \,, \qquad 
{\rm BR} \left ( A \to t \bar t  \right ) = 57.7\% \,. 
\end{equation}
The numbers~(\ref{eq:BRZp}) and~(\ref{eq:BRA}) for the branching ratios of the $Z^\prime$ and the $A$ indicate that besides the mono-Higgs channel, also LHC searches for dijet final states~(see~\cite{ATLAS:2019fgd,CMS:2019gwf} for the latest ATLAS and CMS results), $t \bar t$ resonances (see for example~\cite{CMS:2018rkg,ATLAS:2020lks}) and resonant $Zh$ production~\cite{CMS:2017sjn,ATLAS:2017ptz,CMS:2018hir,CMS:2019kca,ATLAS:2020qiz,ATLAS-CONF-2020-043} can be expected to lead to relevant constraints in the $m_{Z^\prime}\hspace{0.25mm} $--$\hspace{0.5mm}  m_A$ plane for  benchmark parameter choices like~(\ref{eq:2HDMZpbench}).  To our knowledge, ATLAS and CMS interpretations of the existing dijet, $t \bar t$ and $Zh$ resonance searches in the  2HDM-$Z^\prime$ model do not exist, but recasts are straightforward as we will show below. 

\begin{figure}[!t]
\begin{center}
\includegraphics[width=.65\linewidth]{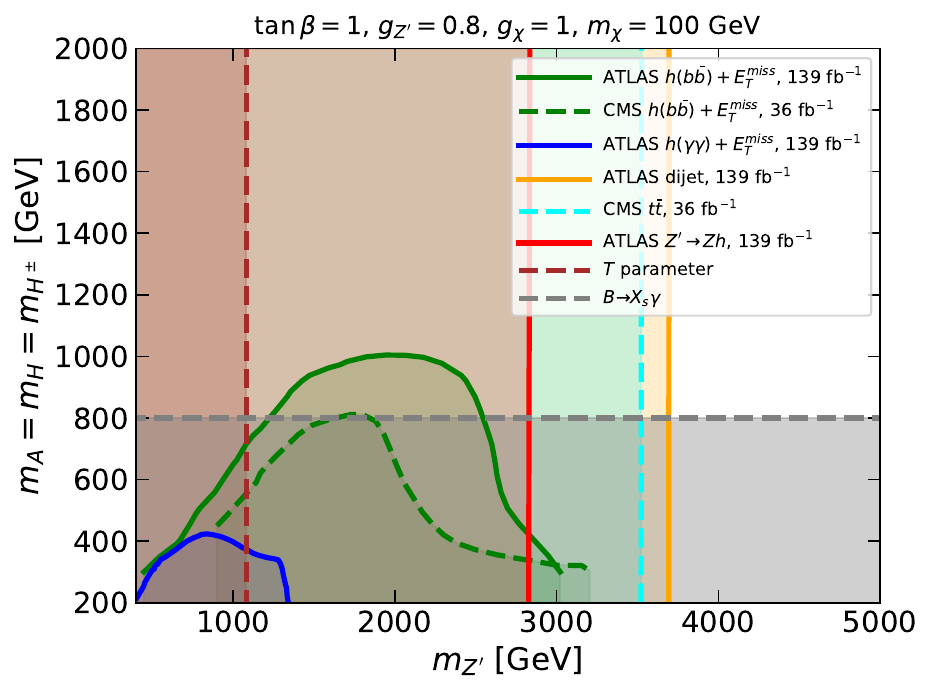} 
\vspace{2mm} 
\caption{\label{fig:2HDMZpMM}   Observed exclusions in the $m_{Z^\prime}\hspace{0.25mm} $--$\hspace{0.5mm}  m_A$   plane  at 95\%~CL. The shown excluded regions  correspond to the 2HDM-$Z^\prime$  benchmark model~(\ref{eq:2HDMZpbench}). The solid blue, solid green and dashed green domains are taken from~\cite{ATLAS:2021jbf},~\cite{ATLAS:2021shl} and~\cite{CMS:2018ljc}, respectively.  The solid orange, dashed cyan and solid red exclusions have instead been obtained from a recast of the dijet search~\cite{ATLAS:2019fgd}, the $t \bar t$ search~\cite{CMS:2018rkg} and the search for resonant $Zh$ production~\cite{ATLAS-CONF-2020-043}. For comparison also the indirect constraints arising from  the $T$ parameter~(dashed brown  line) and $B\to X_s \gamma$~\cite{Misiak:2017bgg,Misiak:2020vlo}~(dashed grey line) are shown. See text for further explanations.}
\end{center}
\end{figure}

In~Figure~\ref{fig:2HDMZpMM} we summarise the relevant 95\%~CL constraints on the 2HDM-$Z^\prime$ model in the $m_{Z^\prime}\hspace{0.25mm} $--$\hspace{0.5mm}  m_A$   plane. The shown exclusions correspond to the benchmark parameter choices~(\ref{eq:2HDMZpbench}).  The  blue and green exclusions represent the latest mono-Higgs constraints by ATLAS~\cite{,ATLAS:2021shl} and CMS~\cite{CMS:2018ljc}. The searches $h\hspace{0.25mm} (b\bar{b}) + E_T^{\rm miss}$ provide stronger constraints than  $h\hspace{0.25mm} (\gamma\gamma) + E_T^{\rm miss}$ because  ${\rm BR} \left (h \to b \bar b \right)/{\rm BR} \left (h \to \gamma \gamma \right) \simeq 250$. The sensitivity decreases with increasing $m_A$ due to other decay channels like $A\to t\bar{t}$ becoming kinematically accessible, thereby reducing  BR$(A\to\chi\bar{\chi})$. It should be noted that while the ATLAS search~(solid green line) uses the Higgs invariant mass as a discriminant to separate signal from background events, the CMS search~(dashed green line) employs the transverse mass of the $b\bar{b}+E_T^{\mathrm{miss}}$ system, which is better suited for resonant signatures. As a consequence the CMS search is competitive with the ATLAS one despite using not $139 \, {\rm fb}^{-1}$ but only $36\, {\rm fb}^{-1}$ of integrated luminosity collected at $13 \, {\rm TeV}$. 

The solid orange, dashed cyan  and solid red vertical lines indicate the 95\%~CL limits on $m_{Z^\prime}$ that follow from our recast of the dijet search~\cite{ATLAS:2019fgd}, the~$t \bar t$~search~\cite{CMS:2018rkg} and  the resonant $Zh$ search~\cite{ATLAS-CONF-2020-043}, respectively. These bounds are essentially independent of the precise value of the mass of the $A$, and for $m_A \ll m_{Z^\prime}$ read
\begin{equation} \label{eq:threebounds}
m_{Z^\prime } > 3690 \, {\rm GeV} \,,  \qquad 
m_{Z^\prime } > 3520 \, {\rm GeV} \,,  \qquad 
m_{Z^\prime } > 2820 \, {\rm GeV} \,.
\end{equation}
Using~\cite{CMS:2019gwf} and~\cite{ATLAS:2020lks} would lead to a slightly  weaker  dijet and $t \bar t$ limit, respectively. Notice that the exclusions~(\ref{eq:threebounds}) are significantly more stringent than~the limit~(\ref{eq:Tconstraint}) which follows from the $T$ parameter. The latter limit is indicated by a dashed brown  line in~Figure~\ref{fig:2HDMZpMM}. In fact, the obtained dijet and $t \bar t$ constraint are so strong that they exceed the maximal mass reach of the existing mono-Higgs  searches that reads 
\begin{equation}
m_{Z^\prime } > 3190 \, {\rm GeV} \,,
\end{equation}
and is obtained for light pseudoscalar masses. As can be seen from the dashed grey horizontal line such small values of $m_A$ are under the assumption $m_A = m_H = m_{H^\pm}$ in conflict with $B \to X_s \gamma$ if the Yukawa sector of the 2HDM-$Z^\prime$ model is taken to be of type~II as done in~\cite{Berlin:2014cfa} and all existing LHC analyses~\cite{ATLAS:2017uis,CMS:2017prz,CMS:2018nlv,CMS:2018zjv,CMS:2018ljc,CMS:2019ykj,ATLAS:2021shl,ATLAS:2021jbf}. Notice that the recent ATLAS $Zh$ search~\cite{ATLAS-CONF-2020-043} considers  the combination of the $\ell^+ \ell^- b \bar b$ and $\nu \bar \nu b \bar b$ final states to set limits. In our recast we have only included contributions  with a $Z^\prime Zh$ vertex that lead to both of these final states (cf.~the right diagram in~Figure~\ref{fig:2HDMZp}). Contributions with a  $Z^\prime Ah$ vertex where the pseudoscalar decays via $A \to \chi \bar \chi$ have instead been  neglected (cf.~the middle diagram in~Figure~\ref{fig:2HDMZp}). While such diagrams mimic the $\nu \bar \nu b \bar b$ final state they lead to a different $E_T^{\rm miss}$ distribution than the former contributions with $Z \to \nu \bar \nu$ and thus to a different experimental acceptance. Neglecting diagrams with an internal $A$ in the reinterpretation is however always a conservative approach, because this contribution necessarily leads to a larger signal strength. 

\begin{figure}[!t]
\begin{center}
\includegraphics[width=.65\linewidth]{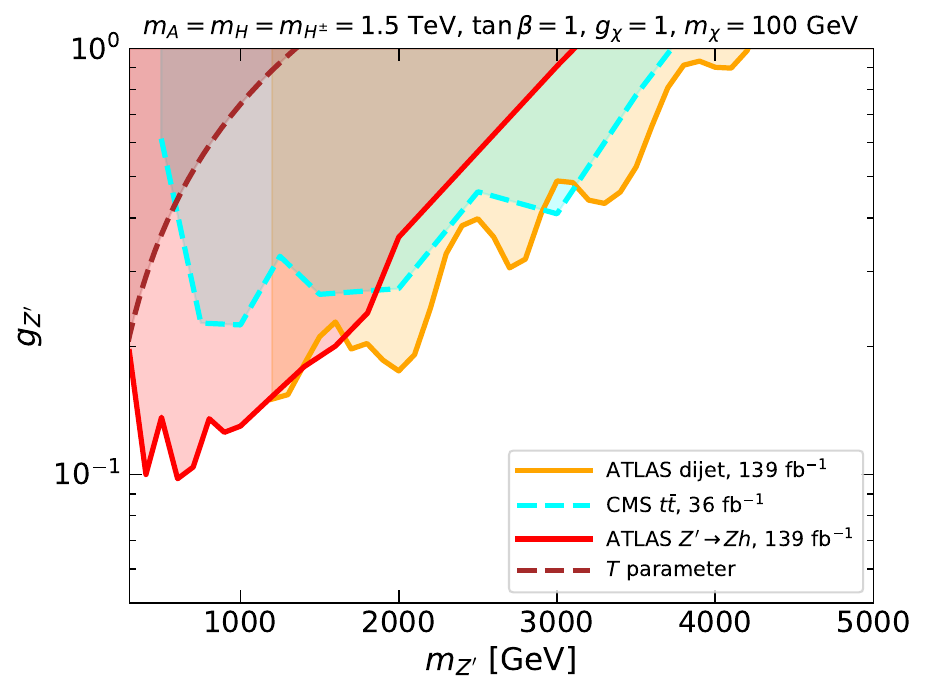} 
\vspace{2mm} 
\caption{\label{fig:2HDMZpMg}  95\%~CL  exclusions in the $m_{Z^\prime}\hspace{0.25mm} $--$\hspace{0.5mm}  g_{Z^\prime}$   plane. The  2HDM-$Z^\prime$  parameter choices are indicated in the headline of the~figure.   The solid orange, dashed cyan and solid red  domains correspond to the limits obtained from the dijet search~\cite{ATLAS:2019fgd}, the~$t \bar t$ resonance search~\cite{CMS:2018rkg} and  the search for resonant $Zh$ production~\cite{ATLAS-CONF-2020-043}, respectively. The parameter space excluded by the $T$~parameter~is also displayed as a dashed brown  line. Further details can be found in the main text.}
\end{center}
\end{figure}

In order to highlight the complementary of  dijet, $t \bar t$  and $Zh$ searches in probing the 2HDM-$Z^\prime$ parameter space we show in~Figure~\ref{fig:2HDMZpMg} the 95\%~CL  exclusions in the $m_{Z^\prime}\hspace{0.25mm} $--$\hspace{0.5mm}  g_{Z^\prime}$   plane that follow from our recasts of~\cite{ATLAS:2019fgd},~\cite{CMS:2018rkg} and~\cite{ATLAS-CONF-2020-043}. We employ the parameters given in~(\ref{eq:2HDMZpbench}) with $m_A = 1.5 \, {\rm TeV}$ while keeping the $U(1)_{Z^\prime}$ coupling $g_{Z^\prime}$ as a free parameter. One observes that the used dijet ($t \bar t$) search allows to set  better constraints on $g_{Z^\prime}$ than the used $Zh$ search in the mass region $m_{Z^\prime} \gtrsim 1.5 \ {\rm TeV}$ ($m_{Z^\prime} \gtrsim 2 \ {\rm TeV}$) for lower $Z^\prime$-boson  masses the situation is reversed. For $m_{Z^\prime} \lesssim 1 \, {\rm TeV}$  low-mass dijet searches such as~\cite{ATLAS:2018hbc,ATLAS:2018qto,CMS:2018mgb,ATLAS:2019itm,CMS:2019emo,CMS:2019mcu} allow to set further relevant constraints on the 2HDM-$Z^\prime$ model. These constraints are however not included in the~figure. Notice that the shown LHC constraints are stronger than the indirect constraint from the $T$ parameter~(dashed brown line) in the entire mass range shown. We finally add that for the parameter choices employed in~Figure~\ref{fig:2HDMZpMg} the existing   mono-Higgs searches do not provide any constraint because the  $A$ is too heavy. 

The above discussion should have made clear that mono-Higgs searches are not the only way to probe the parameter space of the 2HDM-$Z^\prime$ model at the LHC. In fact, as we have argued stringent constraints arise in general from searches that look for resonant $Z^\prime$-boson production in final states such as dijets,  $t \bar t$ or $Zh$. Similar to what has been shown in~\cite{ATLAS:2019wdu} for the case spin-1 single-mediator DM simplified models, each $Z^\prime$-boson resonance search  is sensitive to complementary regions of the $m_{Z^\prime}\hspace{0.25mm} $--$\hspace{0.5mm}  g_{Z^\prime}$ parameter space, and only by combining the whole suite of searches one is able to exploit the full LHC potential. ATLAS and CMS interpretations of dijet, $t \bar t$  and $Zh$  resonance searches in the 2HDM-$Z^\prime$ model do no exist but could be easily  added to the exotics search canon. 

\subsection{2MDM model} 
\label{sec:darkHiggs}

While simplified models with a single spin-1 boson mediating the DM-SM interactions may lead to unitarity violation~\cite{Bell:2015sza,Kahlhoefer:2015bea,Haisch:2016usn}, it has been shown that the introduction of an extra scalar mediator can cure this problem~\cite{Kahlhoefer:2015bea}. Furthermore, when this scalar field acquires a VEV it gives rise to a dark Higgs mechanism, generating masses for all the particles in the dark sector~\cite{Bell:2016uhg} and opening up new channels for DM annihilation, which can allow for the relic density constraints to be met~\cite{Duerr:2016tmh,Bell:2016fqf}. A concrete model implementation of this general idea is discussed in what follows. 

\subsubsection{Theory} 
\label{subsec:darkHiggs_th}

The 2MDM model introduced in~\cite{Duerr:2016tmh} includes an additional $U(1)_{Z^\prime}$ gauge symmetry with a new gauge boson $Z^\prime$ and a complex scalar field $S$ that is a singlet under the SM gauge group. DM is taken to be a Majorana fermion $\chi$ in order to evade the constraints from DM~DD experiments. The terms of the Lagrangian that contain the interactions relevant for the discussion  below can be written as 
\begin{equation} \label{eq:DH2mDM1}
\begin{split}
\mathcal{L}& = -\frac{g^\prime }{2} \hspace{0.25mm} q_{\chi} \hspace{0.25mm} Z^{\prime \mu} \hspace{0.25mm} \bar{\chi}\gamma_{\mu}\gamma_5 \chi-\frac{y_\chi}{2} \hspace{0.25mm} \bar{\chi}  \hspace{0.25mm} \left (P_L S+P_R S^\ast \right )  \hspace{0.25mm}\chi \\[2mm]
& \phantom{xx} + 4 g^{\prime 2}\hspace{0.25mm}  q_\chi^2 \hspace{0.25mm}  Z^{\prime \mu}Z^\prime_{\mu}S^{\dag}S-g^\prime \hspace{0.25mm}  q_q \hspace{0.25mm} Z^{\prime \mu}\sum_{q} \bar{q}\gamma_{\mu}q \,.
\end{split}
\end{equation}
Here  $g^\prime$ is the $U(1)_{Z^\prime}$ gauge coupling, $y_\chi$ is a Yukawa coupling, $q_{\chi}$ and $q_q$ denote the $U(1)_{Z^\prime}$ quantum numbers of the DM and the quark fields, respectively, and $P_L$ and $P_R$ project onto left- and right-handed fields.  The $U(1)_{Z^\prime}$ charge of $S$ is fixed to be $q_S = - 2 q_\chi$ so that the Yukawa term does not break the symmetry explicitly. In addition it has been assumed that the SM Higgs doublet carries no $U(1)_{Z^\prime}$ charge to avoid mass mixing between the $Z$ and the $Z^\prime$ boson which is severely constrained by the measurements of the EWPOs. In such a case there is also no $Z^\prime Z h$ vertex at tree level, and as a result searches for resonant $Zh$ production do not provide any constraint. Note that the $Z^\prime$ boson has flavour-diagonal and flavour-universal vector couplings to quarks, while it  does not couple to leptons. The former feature automatically avoids FCNCs, while the latter feature  is crucial in view of the stringent bounds from dilepton searches. Such a charge assignment can be obtained  by gauging baryon number~\cite{Pais:1973mi}, and realistic models that implement this mechanism have been discussed for example in~\cite{Duerr:2013dza,FileviezPerez:2014lnj,Duerr:2013lka,Duerr:2014wra,Ohmer:2015lxa,Duerr:2015vna}. Models of this type generically have additional fermionic degrees of freedom to cancel gauge anomalies~(cf.~\cite{Ekstedt:2016wyi,Ismail:2016tod,Ellis:2017tkh} for recent discussions) but it is tacitly assumed that the only new fermion that plays an important role at LHC energies  is the DM candidate $\chi$. 

The scalar $S$ is assumed to acquire a VEV $\langle S \rangle = 1/\sqrt{2} \hspace{0.5mm} (0, w)^T$, spontaneously breaking the $U(1)_{Z^\prime}$ and thereby giving masses to the $Z^\prime$ boson and the DM particles: 
\begin{equation}
 m_{Z^\prime}=2g^\prime q_{\chi} w \,,   \qquad \, m_\chi = \frac{y_{\chi}w}{\sqrt{2}} \,.
\end{equation}
Introducing now the coupling strengths $g_\chi = g^\prime q_\chi$ and $g_q= g^\prime q_q$, the Lagrangian~(\ref{eq:DH2mDM1}) can be rewritten as
\begin{equation} \label{eq:DH2mDM2}
\begin{split}
\mathcal{L}& = -\frac{g_\chi}{2}  \hspace{0.25mm} Z^{\prime \mu} \hspace{0.25mm} \bar{\chi}\gamma_{\mu}\gamma_5 \chi- \frac{g_\chi m_\chi}{m_{Z^\prime}} \hspace{0.25mm} s \chi \bar \chi + 2 g_\chi^2 \hspace{0.25mm}    Z^{\prime \mu}Z^\prime_{\mu} \left ( \frac{m_{Z^\prime}}{g_\chi} \hspace{0.25mm} s+ s^2 \right ) -g_q \hspace{0.25mm}Z^{\prime \mu}\sum_{q} \bar{q}\gamma_{\mu}q \,,
\end{split}
\end{equation}
where $s$ is the scalar field excitation around the VEV of $S$. Besides the interaction terms given in~(\ref{eq:DH2mDM2}) the full 2MDM model  contains further interactions.  First, there is the possibility of kinetic mixing between the $U(1)_Y$ and the $U(1)_{Z^\prime}$ gauge bosons. The kinetic mixing is constrained to be very small by the measurements of EWPOs (see for example~\cite{Duerr:2016tmh}) and therefore ignored in the following. Second, the dark Higgs and the $125 \, {\rm GeV}$ Higgs can mix with the amount of mixing conventionally parameterised by $\sin \theta$. The good agreement between the Higgs signal strength measurements at the LHC and the SM expectations~(cf.~\cite{ATLAS-CONF-2020-027,CMS:2020gsy}) require $|\!\sin \theta| < 0.25$ at 95\%~CL if  no additional Higgs decay channels are open. For the values of $\sin \theta$ considered below, this bound plays no role, and we will also show that  additional Higgs decay channels are not an issue. 

\subsubsection{Experimental constraints} 
\label{subsec:darkHiggs_exp}

The LHC searches~\cite{ATLAS:2019ivx,Aad:2020sef,CMS-PAS-EXO-20-013}  have provided interpretations  in the 2MDM model for
\begin{equation} \label{eq:DH2MDMbench}
g_\chi =1 \,, \qquad g_q = 0.25 \,, \qquad \sin \theta = 0.01 \,, \qquad m_\chi = 200 \, {\rm GeV} \,, \quad 
\end{equation}
presenting exclusion limits in the $m_{Z^\prime} \hspace{0.25mm}$--$ \hspace{0.5mm} m_s$ plane. The used value of $\sin \theta$ clearly satisfies the aforementioned limit on $\sin \theta$ imposed by Higgs physics. Furthermore, the decay channels $h \to \chi \bar \chi$ and  $h \to Z^\prime Z^\prime$ are kinematically closed. Since the mass $m_s$ is scanned over in the works~\cite{ATLAS:2019ivx,Aad:2020sef,CMS-PAS-EXO-20-013}, on-shell Higgs decays  of the form $h \to ss$ are in principle allowed if $m_s < m_h/2$. In the limit $\sin \theta \ll 1$, the corresponding partial decay width can be approximated  by 
\begin{equation}
\Gamma \left ( h \to s s  \right ) = \frac{g_\chi^2 \sin^2 \theta  \hspace{0.5mm} m_h^3}{8 \pi m_{Z^\prime}^2} \, \left (  1 - 4 x_{s/h} \right )^{1/2}  \left ( 1 + 2  x_{s/h} \right ) ^2 \,. 
\end{equation}
For the parameters~(\ref{eq:DH2MDMbench}) and taking $m_{Z^\prime} = 1 \, {\rm TeV}$ and $m_s \ll m_{Z^\prime}$, one finds numerically $\Gamma \left ( h \to s s  \right ) = 1.9 \cdot 10^{-3} \, \Gamma_h^{\rm SM}$. Such a small modification easily passes the bounds from the direct measurements of the total Higgs width~\cite{CMS:2017dib,ATLAS:2018tdk} as well as the numerous limits that stem from searches for $h \to 4 f$ (see for example~\cite{Haisch:2018kqx} for a recent detailed discussion of these bounds).  It follows that Higgs physics leads to no restrictions on the 2MDM model for the parameter choices~(\ref{eq:DH2MDMbench}). 

\begin{figure}[!t]
\begin{center}
\includegraphics[width=0.45\textwidth]{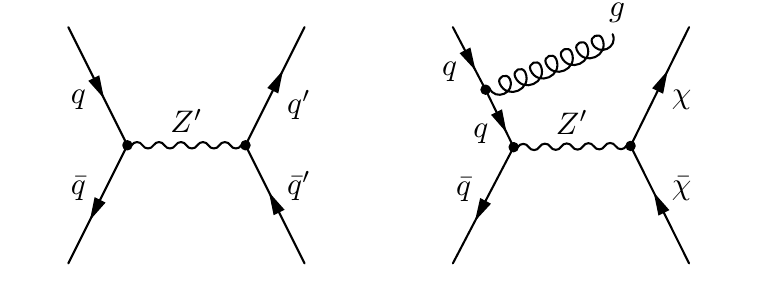} 
\vspace{2mm} 
\caption{\label{fig:DH2MDM1}  Representative Feynman diagrams that lead to dijet or heavy-quark pair production~(left) and a mono-jet signal~(right) in the 2MDM model.   }
\end{center}
\end{figure}

In~Figure~\ref{fig:DH2MDM1} we show two example  diagrams that give rise to relevant LHC signatures in the 2MDM model.  These diagrams only involve the exchange of the $Z^\prime$ boson and lead to a signature with two light- or heavy-quark jets in the final state  or to a classic mono-jet signature where the jet arises as initial-state radiation~(ISR). These topologies can hence be probed for instance   via   dijet, $t \bar t$ or $j + E_T^{\rm miss}$ searches. Since all these signatures result from $s$-channel $Z^\prime$-boson exchange, an important ingredient that will determine the signal strength in a certain channel is the branching ratio of the $Z^\prime$ boson into the corresponding final state. The necessary partial decay widths  take the following form 
\begin{equation}
\begin{split}
\Gamma \left ( Z^\prime \to \chi \bar \chi \right ) & =   \frac{g_\chi^2 \hspace{0.25mm} m_{Z^\prime}}{24 \pi} \left ( 1 - 4 x_{\chi/Z^\prime} \right )^{3/2}\,,  \\[2mm]
\Gamma \left ( Z^\prime \to q \bar q \right ) & =   \frac{g_q^2 \hspace{0.25mm} m_{Z^\prime}}{4 \pi} \left (  1 - 4 x_{q/Z^\prime} \right )^{1/2}  \left ( 1 + 2  x_{q/Z^\prime} \right )  \,.
\end{split}
\end{equation}
Using  the input parameters given in~(\ref{eq:DH2MDMbench}) together with $m_{Z^\prime} = 1 \, {\rm TeV}$, these expressions lead to ${\rm BR} \left ( Z^\prime \to  \chi \bar \chi  \right ) = 25.5\%$, ${\rm BR} \left ( Z^\prime \to  q \bar q \right ) = {\rm BR} \left ( Z^\prime \to  t \bar t \right )  = 12.4 \%$ and a relative width of $\Gamma_{Z^\prime}/m_{Z^\prime} = 4.0\%$ These branching ratios  are rather similar to the values of the branching ratios that are predicted in the spin-1 simplified DM benchmark models with $g_\chi = 1$, $g_q = 0.25$ and $g_\ell = 0$. As a result, one expects to find the same hierarchy of sensitivities as in the  spin-1 simplified DM case~---~see~\cite{ATL-PHYS-PUB-2021-006,SummaryPlotsEXO13TeV} for the latest DM summary plots for $s$-channel mediators by ATLAS and CMS~---~with the non-$E_T^{\rm miss}$ searches providing better bounds on $m_{Z^\prime}$ than the $E_T^{\rm miss}$ searches. We will see below that this naive expectation is in fact correct. 

\begin{figure}[!t]
\begin{center}
\includegraphics[width=0.45\textwidth]{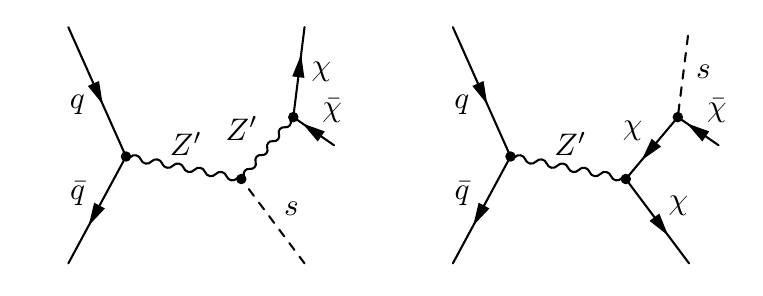} 
\vspace{2mm} 
\caption{\label{fig:DH2MDM2}  Examples of Feynman diagrams that lead to  $s+E_T^{\rm miss}$ production  in the 2MDM model. Depending on its mass the dominant decay modes of the mediator $s$ are to  pairs of bottom-quarks,  $W$, $Z$ or Higgs bosons. This results in a $b \bar b + E_T^{\rm miss}$, $W^+ W^- + E_T^{\rm miss}$, $ZZ + E_T^{\rm miss}$ and  $hh + E_T^{\rm miss}$ signal, respectively.  Further explanations can be found in the text.}
\end{center}
\end{figure}

The 2MDM model however also gives rise to signatures not present in the spin-1 simplified DM models. These novel signatures are illustrated in ~Figure~\ref{fig:DH2MDM2}. Besides the $Z^\prime$ boson the displayed diagrams contain a dark Higgs $s$ that can either be emitted in a $Z^\prime Z^\prime s$ vertex (left graph) or a $\chi \bar \chi s$ vertex (right graph). The dark Higgs can decay into SM and DM particles.  Assuming that the decay channel $s \to \chi \bar \chi$ is kinematically inaccessible, the relevant partial decay widths are given in the limit $\sin \theta \ll 1$ by 
\begin{equation} \label{eq:DH2Mbr}
\begin{split}
\Gamma \left ( s \to  b \bar b \right ) & =   \frac{3   \hspace{0.25mm}  y_b^2  \hspace{0.25mm}  \sin^2 \theta  \hspace{0.5mm} m_s}{16 \pi} \left ( 1 - 4 x_{b/s} \right )^{3/2}\,,  \\[2mm]
\Gamma \left ( s \to W^+ W^- \right ) & =   \frac{ \sin^2 \theta  \hspace{0.5mm}  m_s^3}{16 \pi  \hspace{0.25mm}  v^2}  \hspace{0.25mm}  \left ( 1 - 4 x_{W/s} \right )^{1/2}  \left ( 1 - 4  x_{W/s} + 12 x_{W/s}^2 \right ) \,,   \\[2mm]
\Gamma \left ( s \to Z Z \right ) & =   \frac{ \sin^2 \theta  \hspace{0.5mm}  m_s^3}{32 \pi  \hspace{0.25mm}  v^2}  \hspace{0.25mm}  \left ( 1 - 4 x_{Z/s} \right )^{1/2}  \left ( 1 - 4  x_{Z/s} + 12 x_{Z/s}^2 \right ) \,, \\[2mm]
\Gamma \left ( s \to h h  \right ) & =    \frac{ \sin^2 \theta  \hspace{0.5mm}  m_s^3}{32 \pi  \hspace{0.25mm}  v^2}  \hspace{0.25mm}  \left ( 1 - 4 x_{h/s} \right )^{1/2}  \left ( 1 + 2  x_{h/s} \right )^2   \,. 
\end{split}
\end{equation}
Notice that the first three expressions  have the same functional form than the corresponding partial decay widths of the SM Higgs boson (see for instance~\cite{Djouadi:2005gi})  which is expected because the dark Higgs obtains its SM coupling solely by mixing with the $125 \, {\rm GeV}$ Higgs. In consequence, the decay pattern of the mediator $s$ resemble to first approximation those of a SM Higgs boson with mass $m_s$. In the  mass range $50 \, {\rm GeV} \lesssim m_s \lesssim 140 \, {\rm GeV}$ the dominant decay mode is $s \to b \bar b$ with branching ratios of around  80\% to 40\%. In~the mass range $140 \, {\rm GeV} \lesssim m_s \lesssim 300 \, {\rm GeV}$ the dominant decay modes are to $W^+ W^-$, $ZZ$ and $hh$ with ${\rm BR} \left ( s \to W^+ W^- \right ) \simeq 100\%$ at $m_s \simeq 170 \, {\rm GeV}$ and ${\rm BR} \left ( s \to W^+ W^- \right ) \simeq  2  \hspace{0.25mm}   {\rm BR} \left ( s \to ZZ \right )  \simeq  2 \hspace{0.25mm} {\rm BR} \left ( s \to hh \right )  \simeq 50\%$ at the upper end of the considered mass range. The latter feature is expected from the $SU(2)_L \times U(1)_Y$ symmetric~limit. For $m_s$ values above the top-quark threshold also the decay $s \to t \bar t$ is relevant. In the following numerical analysis, we restrict ourselves to the mass range $50 \, {\rm GeV} < m_s < 350 \, {\rm GeV}$, and thus the dominant final states  that arise from the graphs  in~Figure~\ref{fig:DH2MDM2} are $b \bar b + E_T^{\rm miss}$, $VV + E_T^{\rm miss}$ with $VV = W^+W^-, ZZ$ and  $hh + E_T^{\rm miss}$. 

We now turn to the numerical analysis of the constraints that the various LHC searches impose on the parameter space of the 2MDM model. We start by considering the exclusions that can derived from the signatures  that result from the  diagrams displayed in~Figure~\ref{fig:DH2MDM1}. For the benchmark parameter choices~(\ref{eq:DH2MDMbench}) we find by reinterpreting the 
 dijet~\cite{ATLAS:2019fgd}, $t \bar t$~\cite{CMS:2018rkg}  and mono-jet  search~\cite{ATLAS:2021kxv} the following 95\%~CL limits on the $Z^\prime$-boson mass: 
\begin{equation} \label{eq:DH2MDMlimits} 
m_{Z^\prime} > 3680 \, {\rm GeV} \,, \qquad  m_{Z^\prime} > 3190 \, {\rm GeV} \,, \qquad  m_{Z^\prime} > 1520 \, {\rm GeV} \,. 
\end{equation}
The dijet~\cite{CMS:2019gwf}, $t \bar t$~\cite{ATLAS:2020lks} and mono-jet~\cite{CMS:2021far} searches have comparable sensitivities to the searches leading to~(\ref{eq:DH2MDMlimits}). We~emphasise that our recast of the ATLAS mono-jet search includes only $j + E_T^{\rm miss}$ events with ISR jets (see the right diagram in~Figure~\ref{fig:DH2MDM1}) while we have not considered contributions that follow from $s\hspace{0.5mm} (VV)+E_T^{\rm miss}$ production with the EW gauge bosons decaying fully hadronically (see~Figure~\ref{fig:DH2MDM2}). We will comment on this simplification below.  As anticipated, the limits~(\ref{eq:DH2MDMlimits}) have the same hierarchy as the bounds obtained in the standard benchmark considered in spin-1 simplified DM models~(cf.~\cite{ATL-PHYS-PUB-2021-006,SummaryPlotsEXO13TeV}). 

For the $s+E_T^{\rm miss}$ signatures, it is important to realise that the dark Higgs is produced with high momentum due to the large $m_{Z^\prime}\hspace{0.25mm}$--$\hspace{0.5mm}m_s$ mass splitting. As a result, the dark Higgs decay products are strongly collimated.  According to~\eqref{eq:DH2Mbr}, the dark Higgs decays predominantly to a pair of $b$-jets for $m_s\lesssim 140 \, {\rm GeV}$. This decay mode is targeted by a reinterpretation~\cite{ATLAS:2019ivx} of the ATLAS $h \hspace{0.25mm} (b\bar{b})+E_T^{\mathrm{miss}}$ search that considers the invariant mass distribution of the $b\bar b$ pair down to $50 \, {\rm GeV}$, allowing to probe $m_s>50 \, {\rm GeV}$. For $m_s\gtrsim140 \, {\rm GeV}$, the $s\to VV$ decay mode becomes important. ATLAS targeted this decay mode by exploring fully hadronic final states where both $V$ bosons decay into a quark pair each~\cite{Aad:2020sef}. The challenging multi-prong decay $s \to V \hspace{0.25mm} (q \bar q) V \hspace{0.25mm} (q \bar q)$ is reconstructed with a novel technique~\cite{ATL-PHYS-PUB-2018-012} aimed at resolving the dense topology from boosted $VV$ pairs using reclustered jets in the calorimeter and tracking information. Recently, CMS has probed the same dark Higgs mass range in the $W^+W^-+E_T^{\mathrm{miss}}$ final state where each $W$ boson decays leptonically~\cite{CMS-PAS-EXO-20-013}. The signal was extracted from a two-dimensional fit to the dilepton invariant mass and the transverse mass of the trailing lepton plus $E_T^{\rm miss}$ system. 

\begin{figure}[t!]
\centering
\includegraphics[width=.65\linewidth]{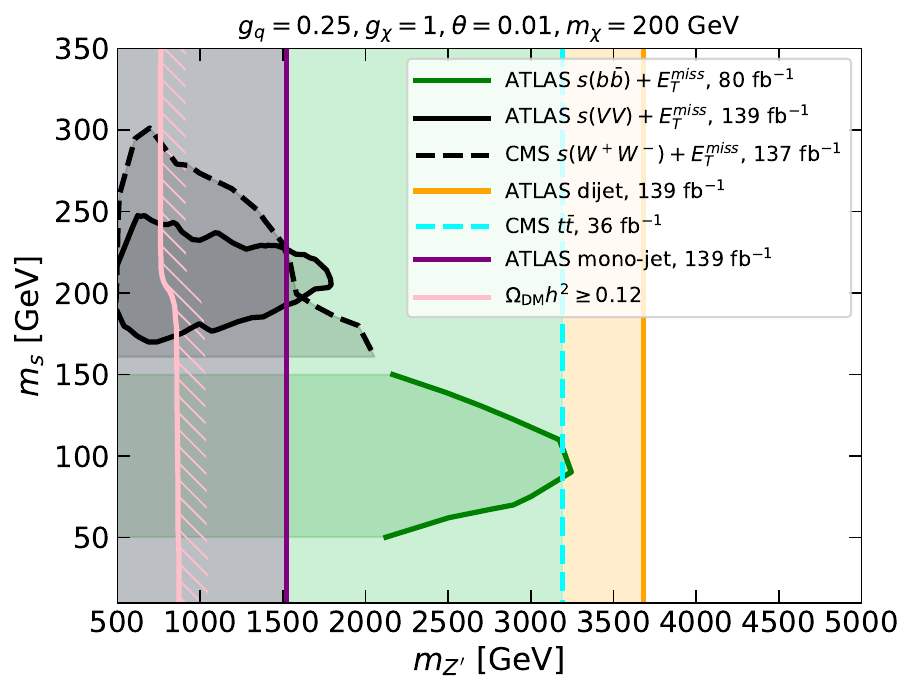}
\vspace{2mm}
 \caption{Most relevant 95\%~CL exclusion contours in the $m_{Z^\prime}\hspace{0.25mm}$--$\hspace{0.5mm}m_s$ plane in the 2MDM model. The~shown results correspond to the benchmark scenario~(\ref{eq:DH2MDMbench}) as indicated in the headline of the plot. The solid black line corresponds to the analysis in \cite{Aad:2020sef} with hadronic decays of $W$ or $Z$ bosons, while the dashed black line corresponds to the analysis in \cite{CMS-PAS-EXO-20-013} with leptonic decays of $W$ bosons. The solid green line is taken from \cite{ATLAS:2019ivx}. The solid orange, dashed cyan and solid purple vertical lines follow from a reinterpretation of the dijet~\cite{ATLAS:2019fgd}, $t \bar t$~\cite{CMS:2018rkg} and mono-jet search~\cite{ATLAS:2021kxv}, respectively. The~pink hatching indicates the direction for which the DM relic density is larger than observed. Further details are discussed in the main text.}
\label{fig:monoS}
\end{figure}

The constraints on the $m_{Z^\prime}\hspace{0.25mm}$--$\hspace{0.5mm}m_s$ plane  from the $s+E_T^{\rm miss}$ analyses by the ATLAS collaboration~\cite{ATLAS:2019ivx,Aad:2020sef} are presented in~Figure~\ref{fig:monoS}. The~95\%~CL exclusion limits~(\ref{eq:DH2MDMlimits}) are also shown in the~figure for comparison. Focusing  on the $E_T^{\rm miss}$ searches first one observes that the novel $s \hspace{0.25mm} (VV)+E_T^{\mathrm{miss}}$ and $s \hspace{0.25mm} (b\bar b)+E_T^{\mathrm{miss}}$ search strategies allow to exclude additional parameter space of the 2MDM model with respect to our mono-jet recast of the ATLAS search~\cite{ATLAS:2021kxv}. In this context one however has to remember that the mono-jet limit given in~\eqref{eq:DH2MDMlimits} does not include the contributions of the $s \hspace{0.25mm} (VV) + E_T^{\rm miss}$ and $s \hspace{0.25mm} (b \bar b) + E_T^{\rm miss}$ processes.  The obtained mono-jet bound therefore provides a conservative lower limit on the actual sensitivity of the $j + E_T^{\rm miss}$ search~\cite{ATLAS:2021kxv} for the 2MDM benchmark~(\ref{eq:DH2MDMbench})~---~we estimate that including $s+E_T^{\rm miss}$ contributions would strengthen the mono-jet limits by around 15\%  (5\%) at $m_s = 100 \, {\rm GeV}$ ($m_s = 200 \, {\rm GeV}$).  It is also evident from the~figure that our recast of the dijet~\cite{ATLAS:2019fgd} and $t \bar t$~search~\cite{CMS:2018rkg} has a higher mass reach in $m_{Z^\prime}$ than the mono-$X$ searches. We note that the 2MDM model can also give rise to $hh + E_T^{\rm miss}$ and $t \bar t +  E_T^{\rm miss}$ signatures.  The former signal can be constrained for instance by using the results of~\cite{CMS:2017nin,ATLAS:2018tti} that studies final states with at least three $b$-jets and~$ E_T^{\rm miss}$ (see also~\cite{Blanke:2019hpe} for an exploration of the $hh +E_T^{\rm miss}$ signature in simplified models of hidden sectors).  Estimating the sensitivity of multi $b$-jet plus $E_T^{\rm miss}$ searches to the $hh + E_T^{\rm miss}$ signature in the 2MDM model is however beyond the scope of this review. The latter  signal can  be targeted by standard $t\bar t + E_T^{\rm miss}$ searches (cf.~for example~\cite{ATLAS:2017hoo,ATLAS:2020xzu,ATLAS:2021hza}). Employing~\cite{ATLAS:2021hza} we expect that for~(\ref{eq:DH2MDMbench}) the HL-LHC with $3 \, {\rm ab}^{-1}$ may be able to set the bounds $M_{Z^\prime} \gtrsim 500 \, {\rm GeV}$ for $m_s \simeq 750 \, {\rm GeV}$ and $M_{Z^\prime} \gtrsim 1.5 \, {\rm TeV}$ for $m_s \simeq 400 \, {\rm GeV}$.  We finally add that the 2HDM model can  lead to interesting LLP signatures if the mixing between the  SM and the dark Higgs is switched off, because in such a case the dark Higgs can only decay through EW gauge boson loops.  For details see the recent publication~\cite{Bernreuther:2020xus}. 

\section{Models with exotic  Higgs decays involving LLPs}
 \label{sec:LLP}

BSM scenarios with hidden sectors that are connected to the SM sector through the $125 \, {\rm GeV}$ Higgs boson are being actively explored at colliders. Such scenarios are often characterised by new electrically neutral LLPs. These LLPs typically decay into SM particles, leaving a displaced vertex signature in the detector. The identification of such signatures is non-trivial and often requires dedicated triggering and reconstruction algorithms~---~see~\cite{Lee:2018pag,Alimena:2019zri} for a detailed review of experimental aspects of LLPs at the LHC. Generally speaking, LLP signatures can arise in a multitude of BSM models ranging from SUSY, theories of neutral naturalness,  hidden valley models and composite Higgs theories, to just name a few examples. Depending on the context, the appearance of BSM LLPs can be theoretically motivated for instance by the naturalness problem of the Higgs mass, the DM puzzle, baryogenesis or the smallness of neutrino masses~---~a comprehensive collection of BSM theories with LLPs can be found in~\cite{Curtin:2018mvb}. Below we will consider three hidden sector models in which the LLPs arise from decays of the $125 \, {\rm GeV}$ Higgs boson. The collider phenomenology of axion-like particles (see for example~\cite{Bauer:2017ris,Bauer:2018uxu} for recent comprehensive studies) that might also lead to LLP signatures from Higgs decays will not be discussed.  
 
\subsection{Neutral naturalness}
\label{sec:NN}

The discovery of a light, seemingly elementary Higgs boson has escalated the seriousness of the EW hierarchy problem, while the steadily increasing LHC limits on coloured BSM states exclude more and more of the natural parameter space of the standard solution to the EW hierarchy problem such as SUSY or compositeness. Models of neutral naturalness like twin Higgs~\cite{Chacko:2005pe}, folded SUSY~\cite{Burdman:2006tz}, quirky little Higgs~\cite{Cai:2008au} and orbifold Higgs~\cite{Craig:2014aea} provide compelling alternative solutions to the EW hierarchy problem. In these theories, the large radiative corrections to the Higgs mass associated with the top quark, are cancelled by top partners that carry no colour, thereby relaxing the most stringent LHC constraints that follow from strong production. The cancellation is achieved with discrete symmetries that must be nearly exact in the top sector, but may be approximate for the other partner or mirror states~\cite{Craig:2015pha}. In~fact, since the QCD coupling drives the renormalisation group~(RG) running of the top-quark Yukawa coupling, for the near-exact discrete symmetry in the top sector to be preserved, viable theories of neutral naturalness contain at least one new QCD-like hidden gauge group with a coupling $\hat \alpha_s$ that is comparable in strength to its SM counterpart $ \alpha_s$. Connecting DM with neutral naturalness is possible (see~for instance~\cite{GarciaGarcia:2015fol,Craig:2015xla,GarciaGarcia:2015pnn,Farina:2015uea,Freytsis:2016dgf,Farina:2016ndq,Barbieri:2016zxn,Barbieri:2017opf,Hochberg:2018vdo,Cheng:2018vaj,Terning:2019hgj,Koren:2019iuv}) but more model-dependent than the LLP phenomenology on which we will focus in the following.

\subsubsection{Theory}

In models of neutral naturalness  the coupling between the $125 \, {\rm GeV}$ Higgs boson and the top partners  gives rise to an effective coupling~$h \hat g \hat g$ between the $125 \, {\rm GeV}$ Higgs $h$ and the hidden gluons $\hat g$ at the one-loop level. This is in full analogy to the SM where top-quark loops provide the dominant contribution to the effective $h gg$  coupling. The effective $h \hat g \hat g$ interactions can be parameterised by 
\begin{equation} \label{eq:higgshiddenglue}
{\cal L} = \frac{\hat \alpha_s}{12 \pi} \hspace{0.25mm} \frac{h}{v}  \hspace{0.5mm} \hat G_{\mu \nu}^a   \hat G^{a, \mu \nu}  \hspace{0.25mm} \hat \zeta \,, 
\end{equation}
where  $\hat G_{\mu \nu}^a$ denotes the  field strength tensor of the hidden $SU(3)$ gauge group and $\hat \zeta$ is a model-dependent mixing angle. One generically has $|\hat \zeta| = {\cal O} (v^2/M^2)$ where depending on the model $M$ is either the scale of spontaneous global symmetry breaking or the mass of the top partners. Fine-tuning arguments bound the scale $M$  to lie at or below the TeV scale, implying that  the relevant mixing angles fall into the range $0.1 \lesssim |\hat \zeta| \lesssim 1$. 

The coupling~(\ref{eq:higgshiddenglue}) provides a portal for production of states in the hidden QCD sector. Once produced, states in the hidden sector cascade down to the lightest accessible mirror state, which is typically a bound state of hidden QCD. This state can then  decay back to SM particles with the associated lifetime depending on the exact nature of the  hidden sector. A canonical signal that can appear in models of neutral naturalness without light mirror matter, such as folded SUSY and quirky little Higgs and some realisations of twin Higgs, is the formation of mirror glueballs~\cite{Juknevich:2009ji,Juknevich:2009gg}. 

The lightest mirror glueball is a scalar state with quantum numbers $J^{CP} = 0^{++}$  and its mass $\hat m_0$ is entirely determined by the RG running of the  strong coupling $\hat \alpha_s$ in the hidden sector. In fact,  $\hat m_0$ is related  to the  mass $m_0  \simeq 1.7 \, {\rm GeV}$ of the~$0^{++}$ glueball in~QCD~\cite{Morningstar:1999rf,Chen:2005mg,Meyer:2008tr}  by the following simple rescaling~\cite{Curtin:2015fna}
\begin{equation} \label{eq:magic1}
\hat m_0 = \frac{\hat \Lambda_{\rm Landau}}{ \Lambda_{\rm Landau}} \, m_0 \,, 
\end{equation}
with $ \Lambda_{\rm Landau} \simeq 150 \, {\rm MeV}$  the  Landau pole in QCD,~i.e.~the scale where $1/\alpha_s ( \Lambda_{\rm Landau}  ) = 0$. Without knowing the exact mass spectrum of states in the hidden sector it is not possible to give a precise value of $\hat \Lambda_{\rm Landau}$ and therefore $\hat m_0$. One can however estimate the typical range of $\hat m_0$ values  by studying the following toy model. Let $M_\ast$ be the scale  where the discrete symmetry $\hat \alpha_s (M_\ast) =  \alpha_s (M_\ast)$ between the two strong couplings is broken and denote with $M$ the scale of the lightest top partner. The one-loop beta function in the hidden sector  takes the form 
\begin{equation}
\hat \beta = \left \{ \begin{matrix} 11 - \displaystyle \frac{2}{3}  \hspace{0.125mm} \hat N_f  \,, & M < \mu < M_\ast \,, \\[4mm]
11 \,, & \mu < M  \,,  
\end{matrix}  \right .
\end{equation}
where $ \hat N_f$ denotes the number of flavours that are active for renormalisation scales $\mu$ in the range $M < \mu < M_\ast$. Solving the one-loop RG equation of $\hat \alpha_s$ it is easy to show that the Landau pole in our toy model occurs at 
\begin{equation}  \label{eq:magic2}
\hat \Lambda_{\rm Landau} = M_\ast \, \left ( \frac{M_\ast}{M} \right )^{-\frac{2 \hspace{0.125mm} \hat N_f}{33}} e^{-\frac{2 \pi}{11 \hspace{0.125mm} \alpha_s (M_\ast) }} \,, 
\end{equation}
where 
\begin{equation} \label{eq:magic3}
\alpha_s (M_\ast) = \alpha_s (m_t)  \hspace{0.5mm} \left [ 1 - \frac{7 \alpha_s (m_t)}{2\pi} \ln \left ( \frac{m_t}{M_\ast} \right ) \right ]^{-1}  \,, 
\end{equation}
with $ \alpha_s (m_t) \simeq 0.11$. Assuming now that there is only one light hidden state of mass~$M$, while all other mirror states have a mass $M_\ast$, and taking  $M = v$ and $M_\ast = 1 \, {\rm TeV}$ or $M = 2 \, {\rm TeV}$ and $M_\ast = 20 \, {\rm TeV}$, one finds using~(\ref{eq:magic1}),~(\ref{eq:magic2}) and~(\ref{eq:magic3})  the following approximate range of mirror glueball masses:
\begin{equation} \label{eq:G0mass}
15 \, {\rm GeV} \lesssim \hat m_0 \lesssim 50 \, {\rm GeV} \,.
\end{equation}
While more sophisticated calculations (see for instance~\cite{Curtin:2015fna}) lead to slightly larger mass ranges, they do not change the conclusion that a representative mirror sector gives rise to glueballs that can be pair-produced in the decays of the $125 \, {\rm GeV}$ Higgs.  Glueball production is therefore a smoking gun signature in many theories of neutral naturalness. 

Under the assumption that the $0^{++}$ mirror glueballs are dominantly produced in symmetric two-body Higgs decays, one can estimate the corresponding exclusive Higgs branching ratio. One finds~\cite{Curtin:2015fna}
\begin{equation}
{\rm BR} \left ( h \to 0^{++} 0^{++} \right ) = \left (1 -\frac{4 \hspace{0.125mm} \hat m_0^2}{m_h^2} \right )^{1/2} \left ( \frac{\hat \alpha_s (m_h)}{\alpha_s (m_h)} \, \hat \zeta \right )^2  \,  \kappa (\hat m_0) \, {\rm BR} \left ( h \to gg \right )_{\rm SM}  \,, 
\end{equation}
where ${\rm BR} \left ( h \to gg \right )_{\rm SM} \simeq 8.2\%$ is the Higgs to digluon branching ratio in the SM. The~parameter $\kappa (\hat m_0)$  encodes our ignorance about the hadronisation of the lightest mirror glueball and the mixing effects of excited hidden glueball states with the $125 \, {\rm GeV}$  Higgs. In the article~\cite{Curtin:2015fna} it has been argued that $\kappa ( 15 \, {\rm GeV}) \simeq 0.1$ and $\kappa ( 50 \, {\rm GeV}) \simeq 1$. Using these numbers as well as the estimate~$\hat \zeta \simeq v^2/M^2$, one obtains for  the two benchmarks that led to~(\ref{eq:G0mass}) the following range of  branching ratios:
\begin{equation} \label{eq:BRmagic}
 2 \cdot  10^{-5} \lesssim{ \rm BR} \left ( h \to 0^{++} 0^{++} \right )   \lesssim  1 \cdot 10^{-2} \,.
\end{equation}
In view of  all the approximations and estimates that went into~(\ref{eq:BRmagic}), the given range should only be taken as an indication of the typical values  of the  $ h \to 0^{++} 0^{++}$ branching ratio that arise in theories of neutral naturalness. 

Since the  $0^{++}$ mirror glueball has the same quantum numbers as the SM Higgs  both states mix by  virtue of~(\ref{eq:higgshiddenglue}). Once produced, the  $0^{++}$ mirror glueballs can hence decay to  all kinematically available SM particles $Y$ via an off-shell Higgs,~i.e.~through the process~$0^{++} \to h^\ast \to YY$. The corresponding partial decay widths have been calculated in the work~\cite{Juknevich:2009gg} and in the case of~(\ref{eq:higgshiddenglue}) take the form
\begin{equation} \label{eq:magicgamma}
\Gamma \left ( 0^{++} \to YY \right ) =  \left ( \frac{\hat \alpha_s (\hat m_0) \hspace{0.25mm} \hat f_0\hspace{0.5mm}  \hat \zeta}{6 \hspace{0.125mm}  \pi \hspace{0.125mm} v \left (m_h^2-\hat m_0^2 \right )} \right )^2 \; \Gamma \left ( h^\ast \to YY \right )_{\rm SM}  \,, 
\end{equation}
where $\hat f_0$ is the annihilation matrix element  of the lightest mirror glueball through~(\ref{eq:higgshiddenglue}) and $\Gamma \left ( h^\ast \to YY \right )_{\rm SM} $ denotes the partial decay width for a SM Higgs boson with mass $\hat m_0$. It follows that the decay pattern of the lightest scalar mirror glueball resembles that  of a SM Higgs boson of appropriate mass.  For mirror glueball masses in the range~(\ref{eq:G0mass}) the $0^{++}$  decays around 80\%, 10\% and 10\% of the time to $b \bar b$, $c \bar c$ and $\tau^+ \tau^-$ final states, respectively. Using again $\hat \zeta \simeq v^2/M^2$ as well as~$\hat \alpha_s (\hat m_0) \hat f_0\hspace{0.5mm}  \simeq 0.18 \hspace{0.25mm} \hat m_0^3$~\cite{Meyer:2008tr,Curtin:2015fna} together with~(\ref{eq:magicgamma}) the proper decay length of the lightest mirror glueball can be approximated by 
\begin{equation} \label{eq:meanlifeglue}
c \hspace{0.125mm} \tau_{0^{++}} \simeq 2 \, {\rm m} \, \cdot \, \left ( \frac{15 \, {\rm GeV}}{\hat m_0} \right )^{7}  \left ( \frac{M}{1 \, {\rm TeV}} \right )^{4}  \,.
\end{equation}
For typical values of $\hat m_0$ and $M$ realised in theories neutral naturalness, the proper decay length of the $0^{++}$ mirror glueball ranges from microns to kilometers. In fact, the strong scaling of~(\ref{eq:meanlifeglue}) with both $\hat m_0$ and $M$ suggests that $c \hspace{0.125mm} \tau_{0^{++}}$ can in practice be treated as an almost free parameter in  the framework of neutral naturalness. 

Notice that hidden valley models~\cite{Strassler:2006im,Strassler:2006ri,Han:2007ae}  share many of the phenomenological features discussed above. Like in theories of neutral naturalness also in hidden valley models  a new confining gauge group is added to the SM.  However, the confining gauge group in hidden valley models makes, in full analogy to QCD, hidden hadrons out of hidden quarks. If the hidden sector comprises two light flavours of quarks the spectrum of hidden hadrons contains  a hidden pion $\pi_h$. Given its pseudoscalar nature the hidden pion preferentially decays to heavy SM flavours for example $\pi_h \to b \bar b$. As argued in the articles~\cite{Strassler:2006im,Strassler:2006ri,Han:2007ae}, the typical  $\pi_h$ masses and proper decay lengths  fall into the ballpark of~(\ref{eq:G0mass}) and (\ref{eq:meanlifeglue}), respectively.  Hidden valley models may  also contain hidden Higgses  which partake in the mass generation of the hidden quarks. If one of these hidden Higgs fields mixes with the $125 \, {\rm GeV}$ Higgs boson  it is possible to obtain $h \to \pi_h \pi_h$ branching ratios that are observable at the LHC~\cite{Strassler:2006im,Strassler:2006ri,Han:2007ae}. The phenomenology of the $\pi_h$ is therefore very similar to that of the~$0^{++}$ with the most obvious  difference that the hidden pion is a pseudoscalar whereas~the lightest mirror glueball is a scalar.

While the above considerations  broadly motivate searches for displaced Higgs decays at the LHC, the models presently employed in the interpretation of such searches by the experimental collaborations are more generic than the theories of neutral naturalness or the hidden valley models discussed above. The used simplified  models assume SM production of a Higgs boson and its subsequent decay to a pair of scalar ($s$) or pseudoscalar ($a)$ particles. The~$s$~($a$) is assumed to decay like the SM Higgs, while  its mass $m_s$ ($m_a$), the relevant Higgs branching ratio ${\rm BR} \left ( h \to ss \right )$  $\big($${\rm BR} \left ( h \to aa \right )$$\big)$ and its proper decay length $c  \tau_{s}$ ($c  \tau_{a}$) are treated as free parameters. For what concerns models of neutral naturalness,  this approach is  motivated by~(\ref{eq:G0mass}),~(\ref{eq:BRmagic}) and~(\ref{eq:meanlifeglue}). The scalar (pseudoscalar) case can be thought to cover theories of neutral naturalness (hidden valley models) with the lightest mirror glueball (the hidden pion) being the LLP.  For definiteness we will hereafter refer to the LLP produced in exotic Higgs decays as an $a$ particle.

\subsubsection{Experimental constraints}

The first searches for pair-produced neutral LLPs in the context of Higgs portal models have been performed by the CDF~\cite{CDF:2011dnt} and D\O~\cite{D0:2009mtx} collaborations at the Tevatron. Both searches looked for displaced vertices in their tracking system only, thereby setting limits on LLP mean decay lengths of the order of a few centimetres. At the LHC, searches for Higgs decays into LLP have been carried out by the ATLAS, CMS and LHCb collaborations in different final states, covering a wide range of mean decay lengths. The~LLP mean decay length determines the search strategies and reconstruction techniques that are employed. Below, we discuss the relevant LHC searches, starting with the shortest mean decay lengths considered.

\begin{figure}[!t]
\begin{center}
\includegraphics[width=0.5\textwidth]{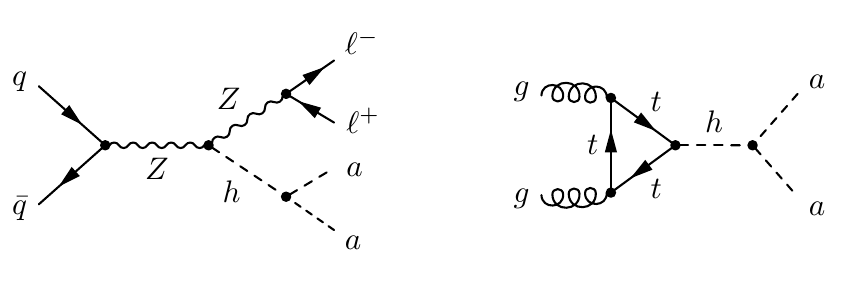} 
\vspace{0mm} 
\caption{\label{fig:llp} Representative diagrams that can give rise to LLP signatures at the LHC. The left and right graph displays associated $Zh$ production followed by $Z \to \ell^+ \ell^-$ and $h \to aa$ and ggF Higgs production with $h \to aa$, respectively. The $a$ dominantly decays to bottom-quark pairs leading in both case to a four-bottom final state. Most existing searches have targeted this final state but analyses that look for multi-jet events have also been performed.}
\end{center}
\end{figure}

ATLAS has performed searches for the decay $h\to aa\to 4 b$ optimised for prompt decays or small proper decay lengths  $c\tau_a\lesssim 6 \cdot 10^{-3} \, {\rm m}$~\cite{ATLAS:2018pvw,Aad:2020rtv}. The~searches select events corresponding to associated  $Vh$  production with decays of the EW bosons into leptons, as displayed on the left-hand side of Figure~\ref{fig:llp}. Since the targeted LLP mean decay lengths were small, standard track/vertex reconstruction and $b$-jet identification techniques  were used.  LHCb  has also performed a search for $gg\to h\to aa$ with $a$ decaying to hadronic jets~\cite{Aaij:2017mic}, which is sensitive to small mean decay lengths in the ballpark of a few millimetres. The corresponding signal process  is illustrated on the right in~Figure~\ref{fig:llp}.  At least one displaced vertex was required in the event due to the limited acceptance of the LHCb vertex detector. The data were recorded by requiring the presence of an energetic charged lepton or hadron in the event in the hardware trigger, and either  one  track with high transverse momentum~($p_T)$ and  a large impact parameter, or a displaced vertex of at least two tracks in the software trigger. Further requirements on the displaced vertex properties were applied in the last stage of the software trigger and in the data analysis.

For mean decay lengths in the range of $10^{-3} \, {\rm m}$ to $1\, {\rm m}$, a substantial fraction of the LLPs is expected to decay inside the inner detector~(ID) of the LHC experiments. This allows for a direct reconstruction of the displaced decay vertex and hence dramatically reduces the SM background rate, which becomes dominated by long-lived hadrons and instrumental backgrounds. At the same time, the macroscopic mean decay lengths significantly decrease the efficiency of standard track reconstruction algorithms and thereby also the identification efficiency of $b$-jets from LLP decays. Searches relying on displaced vertex signatures in the inner tracker systems of ATLAS and CMS are described in the following.

In order to significantly increase the efficiency for reconstructing displaced vertices, dedicated reconstruction techniques, so-called large-radius tracking~(LRT), were developed at ATLAS~\cite{ATL-PHYS-PUB-2017-014}. Properties of displaced vertices reconstructed from LRT tracks like the invariant mass or the number of associated tracks can then be used to discriminate signal from background. Since the LRT reconstruction algorithms are computationally intensive, they were not employed in the ATLAS trigger. Hence, the ATLAS search employing LRT techniques targeted the associated $Zh$ production channel with $h\to aa\to 4 b$ and $Z\to \ell^+\ell^-$. The leptons from the $Z\to \ell^+\ell^-$ decays were used for triggering, and backgrounds were essentially eliminated by requiring two displaced vertices in candidate events~\cite{ATLAS:2021jig}.

CMS has recently searched for $h\to aa\to 4 b$ and $Z\to \ell^+\ell^-$ using the same process~\cite{CMS-PAS-EXO-20-003}. A trigger and selections based on dilepton $Z$-boson decays provide sensitivity to light LLPs with masses of $15 \, {\rm GeV}$ or less. The decays of the LLPs are selected by requiring the presence of displaced jets which are identified using information from the tracking system. CMS in addition searched for the $h\to aa\to 4 b$ process targeting $c\tau_a$ between $10^{-3} \, {\rm m}$ and $3 \, {\rm m}$ by requiring displaced vertices consistent with LLP decays to be reconstructed in the inner tracker~\cite{CMS:2018qxv,CMS:2020iwv}. Unlike the ATLAS analysis~\cite{ATLAS:2021jig}, the CMS search~\cite{CMS:2020iwv} considered the ggF topology and relied on jets for triggering. Two triggers were used, both requiring a large scalar sum of transverse jet energies in the event and at least two energetic jets consistent with a LLP decay, i.e.~at most two associated prompt tracks with $p_T>1 \, {\rm GeV }$ and at least one track consistent with originating from a displaced vertex. Benefiting from the large ggF Higgs production cross section, the CMS~search~\cite{CMS:2020iwv} provides currently the best sensitivity for proper decay lengths between $10^{-3} \, {\rm m}$ and~$10^{-1} \, {\rm m}$.   CMS~has also searched for displaced leptons arising from ggF Higgs production followed by $h\to aa\to 4 \ell$~\cite{CMS-PAS-EXO-18-003}. Since this search assumes that the LLP has equal probability to decay to two muons and two electrons, whereas (\ref{eq:magicgamma}) implies that ${\rm BR} \left (a \to \mu^+ \mu^- \right ) = {\cal O} (10^{-4})$ and ${\rm BR} \left (a \to e^+ e^- \right ) = {\cal O} (10^{-8})$, an interpretation of~\cite{CMS-PAS-EXO-18-003} in the context of neutral naturalness/hidden valleys leads to no meaningful constraint on the $h \to aa$ branching ratio. 

For mean  decay lengths in the range of $1 \, {\rm m}$ to $10^2 \, {\rm m}$, a significant fraction of the LLPs  is expected to decay in the outermost layers of the detector, i.e.~inside the calorimeter~(CM) or the muon spectrometer~(MS). Searches in this regime are described in the following.  

ATLAS searched for $h\to aa \to 4 f$ decays in the CM targeting event topologies compatible with ggF Higgs production~\cite{ATLAS:2019qrr}. The decay of a LLP inside the CM is typically reconstructed as a single jet with striking characteristics, namely a narrow width in pseudorapidity-azimuthal-angle ($\eta\hspace{0.25mm}$--$\hspace{0.25mm} \phi$) space, a high ratio of energy deposited in the hadronic CM to that registered in the electromagnetic CM, and no or only a few low-momentum tracks associated with it. The search uses these characteristics both in dedicated triggers~\cite{ATLAS:2013bsk} and when performing the offline analysis employing an artificial neural network. ATLAS recently performed a search for the same process using the~MS~\cite{ATLAS-CONF-2021-032}. The~tracking capabilities of the MS allow for an explicit reconstruction of a displaced vertex, which dramatically reduces SM and instrumental backgrounds, and allows for a dedicated triggering strategy~\cite{ATLAS:2013bsk} based on the overall activity in the MS. The backgrounds are essentially eliminated by requiring two displaced vertices from candidate $h \to a a \to 4 b$ decays within the MS. The sensitivity for even longer decay times can be extended by requiring only one displaced vertex to be detected in the MS. This strategy has been applied in a similar analysis~\cite{ATLAS:2018tup} and the results were statistically combined with the CM search~\cite{ATLAS:2019qrr}. Finally, ATLAS carried out a search using the combination of the ID and the MS, targeting short and long mean decay lengths~\cite{ATLAS:2019jcm}.

CMS performed a LLP search using the endcap of the MS~\cite{CMS:2021juv}, targeting $c\tau_a$ between $10^{-3} \, {\rm m}$ and $10^2 \, {\rm m}$ and decay chains $h \to aa$ with $a\to b\bar b,c \bar c, \tau^+\tau^-$. While many features are similar to the corresponding ATLAS searches~\cite{ATLAS-CONF-2021-032,ATLAS:2018tup}, the search philosophy differs in one key aspect: the magnetic field return yoke, interleaved with the tracking layers of the MS was used as a sampling CM. Only the endcap region of the CMS MS is considered, as it provides a greater depth of up to 30 nuclear interaction lengths. A dedicated algorithm was applied to cluster the hits in the muon system, and the hit multiplicity in an azimuthal slice close to the $E_T^{\rm miss}$ vector was used as the final discriminant. 

ATLAS and CMS also performed other searches for neutral LLPs decaying to jets targeting SUSY models~\cite{CMS:2017poa,CMS:2018tuo,CMS:2019qjk,ATLAS:2019fwx,CMS:2021tkn} that are not explicitly optimised for $h\to aa\to 4f$ signatures, for example due to very high trigger thresholds, and therefore will not be discussed below.

\begin{figure}[t!]
\centering
\includegraphics[width=1\linewidth]{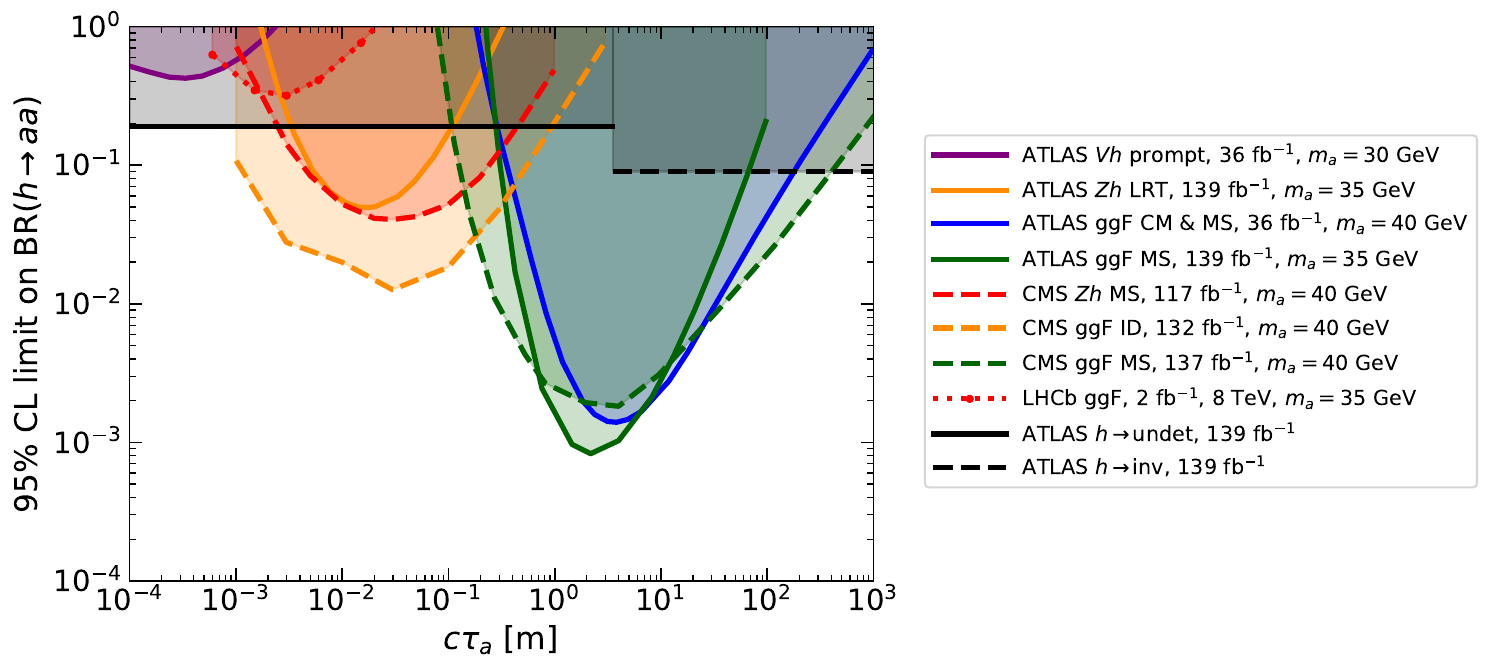}
\vspace{-2mm}
\caption{
\label{fig:HV} 
Observed exclusions at 95\%~CL on the branching ratio ${\rm BR} \left ( h\to aa \right )$ as a function of the proper decay length $c\tau_a$ of the LLP. The shown limits are taken from the ATLAS~\cite{ATLAS:2021jig,ATLAS:2018pvw,ATLAS:2019qrr}, CMS~\cite{CMS:2020iwv,CMS-PAS-EXO-20-003} and LHCb~\cite{Aaij:2017mic} publications, respectively. For comparison the 95\%~CL limits on the branching ratio of undetected and invisible  Higgs decays~(\ref{eq:undetinv}) are also displayed. The~boundary between the constraints from  undetected and  invisible Higgs decays is taken to be $c \tau_a = 3.5 \, {\rm GeV}$ corresponding to $m_a = 35 \, {\rm GeV}$. For~further explanations consult the main text. 
}
\end{figure}

An indirect way to constrain models where the $125 \, {\rm GeV}$ Higgs boson decays into LLPs  is provided by the combination of precision measurements of the  Higgs couplings in visible and invisible final states~\cite{ATLAS:2020kdi,ATLAS-CONF-2020-027,CMS:2020gsy}. In particular,  under the assumption that the coupling modifiers $\kappa_V = g_{hVV}/g_{hVV}^{\rm SM}$ of the $125 \, {\rm GeV}$ Higgs boson to EW gauge bosons satisfy $\kappa_{V}\leq 1$, which generically holds in models with Higgs mixing, the following 95\%~CL constraints 
\begin{equation} \label{eq:undetinv}
{\rm BR} \left ( h \to {\rm undet} \right ) < 0.19 \,, \qquad 
{\rm BR} \left ( h \to {\rm inv} \right )  < 0.09 \,, 
\end{equation}
can be placed~\cite{ATLAS-CONF-2020-027}. Here, the undetected category includes events with undetected BSM particles that do not provide a significant $E_T^{\rm miss}$ contribution such as the ones typically selected by the aforementioned LLP analyses with short mean decay lengths below~${\cal O} (1 \, {\rm m})$. In contrast, LLPs that decay outside the tracker and calorimeters, i.e.~having mean decay lengths larger than~${\cal O} (1 \, {\rm m})$, are not registered by standard reconstruction algorithms, resulting in a $E_T^{\rm miss}$ contribution. Hence, for moderate boosts the above bound on ${\rm BR} \left ( h \to {\rm undet} \right )$ and ${\rm BR} \left ( h \to {\rm inv} \right )$ can be interpreted as an indirect limit on $ {\rm BR} \left ( h \to aa \right )$ for proper decays lengths of $c \tau_a \lesssim 1 \, {\rm m}$ and $c \tau_a \gtrsim 1 \, {\rm m}$, respectively. In~the case of larger boosts, the~boundary between the constraints from undetected and invisible Higgs decays is instead approximately given by $c \tau_a \simeq 0.1 \, {\rm m} \hspace{0.5mm} ( m_a/{\rm GeV}) $~\cite{Curtin:2017izq,ATL-PHYS-PUB-2021-020}.

The 95\%~CL limits of the above searches on the branching ratio ${\rm BR} \left (h\to aa \right )$ as a function of the proper decay length $c\tau_a$ of the LLP are shown in~Figure~\ref{fig:HV}. The displayed masses of the LLP $a$ all fall into the central region of $0^{++}$ mirror glueball masses as predicted by neutral naturalness~(\ref{eq:G0mass}). For proper decay lengths $c \tau_a$ of a few meters the LHC searches are able to set a limit ${\rm BR} \left ( h \to aa \right ) \lesssim 10^{-3}$. In view of~(\ref{eq:BRmagic}) and~(\ref{eq:meanlifeglue}) this is an interesting finding, as it allows to test certain model realisations of neutral naturalness. Notice that the limits from ${\rm BR} \left ( h \to {\rm undet} \right )$ and ${\rm BR} \left ( h \to {\rm inv} \right )$ provide currently the strongest LHC bound on ${\rm BR} \left ( h \to a a \right )$ for $c\tau_a \lesssim 10^{-4} \, {\rm m}$ and $c\tau_a \gtrsim 10^{3} \, {\rm m}$, respectively. Besides these two constraints, proper decay lengths $c\tau_a \gtrsim 10^3 \, {\rm m}$ are currently unexplored by collider measurements. Dedicated detectors like MATHUSLA~\cite{Curtin:2018mvb}, CODEX-b~\cite{Gligorov:2017nwh} and ANUBIS~\cite{Bauer:2019vqk} may address such long mean decay lengths at the HL-LHC.

\subsection{Dark photons} 
\label{sec:darkphoton}

In the case of neutral naturalness the LLP is a composite spin-0 particle. However, hidden sector models with a spin-1 LLP also exist. The model considered in this subsection is based on an extra $U(1)_X$ symmetry in the hidden sector, where the associated vector field $X$ is given a mass via a dark Higgs mechanism involving the singlet scalar field~$S$. As explained below, in certain regions of parameter space this model predicts displaced dilepton vertex signatures that arise from the exotic decays of the $125 \, {\rm GeV}$ Higgs boson to a pair of dark photons $Z_d$ followed by $Z_d \to \ell^+ \ell^-$. This feature makes the discussed hidden sector model experimentally distinct from theories of neutral naturalness where the LLP decay products consist primarily of hadrons. We will not discuss the dark photon searches~\cite{CMS:2019ajt,CMS:2020krr,ATLAS:2021pdg} which are motivated by the theoretical works~\cite{Gabrielli:2013jka,Gabrielli:2014oya,Biswas:2016jsh,Tsai:2016lfg}.  We emphasise that the dark photon models that are discussed in this subsection do not have a DM~candidate. For comprehensive discussions of dark photon models with an additional DM candidate see for example the reviews~\cite{Essig:2013lka,Alexander:2016aln,Fabbrichesi:2020wbt}.

\subsubsection{Theory}

The interaction terms of the hidden sector model that we consider include  both a hypercharge and a Higgs portal. See~\cite{Curtin:2013fra,Curtin:2014cca} for details and further relevant literature.  We write these terms in the following way 
\begin{equation} \label{eq:Ldp}
{\cal L} = \frac{\epsilon}{2\cos \theta_w} \hspace{0.25mm} B_{\mu \nu}  \hspace{0.25mm}  X^{\mu \nu} - \kappa  \hspace{0.25mm}  (H^\dagger H)  \hspace{0.25mm}  S^2 \,, 
\end{equation}
where $B^{\mu \nu} = \partial^\mu B^\nu - \partial^\nu B^\mu$ and  $X^{\mu \nu} = \partial^\mu X^\nu - \partial^\nu X^\mu$ is the $U(1)_Y$ and $U(1)_X$ gauge field strength tensor, respectively,  $\epsilon$ denotes the hypercharge mixing parameter, while $\kappa$ is the Higgs portal coupling. After EW symmetry breaking $\langle H \rangle = (0, v/\sqrt{2})^T$ and spontaneous symmetry breaking of the $U(1)_X$ symmetry by $\langle S \rangle = v_S/\sqrt{2}$, the mass spectrum of the hidden sector model contains two heavy neutral gauge bosons and two neutral Higgs bosons. We will denote these states by $Z$, $Z_d$, $h$ and $h_d$. For small $\epsilon$ ($\kappa$) the $Z$ ($h$) is essentially the SM $Z$ boson (Higgs boson), while the dark photon $Z_d$ (dark~Higgs~$h_d$) is mostly $X$-like ($S$-like).  

\begin{figure}[!t]
\begin{center}
\includegraphics[width=0.6\textwidth]{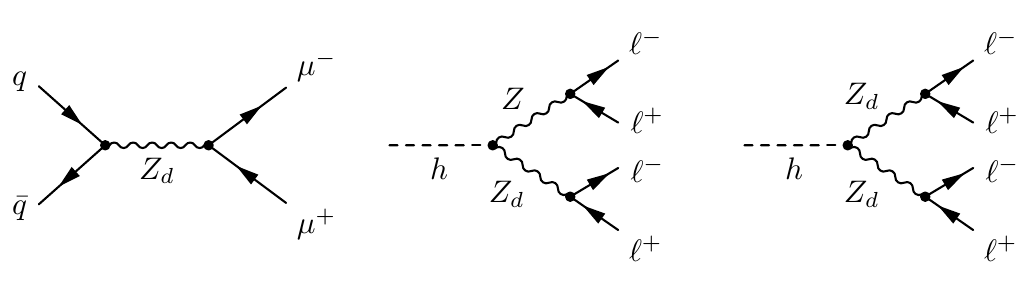} 
\vspace{0mm} 
\caption{\label{fig:darkphoton} Example contributions in the dark photon model~(\ref{eq:Ldp}) to DY dimuon production~(left diagram) and  four-lepton final state production arising from the exotic Higgs decays $h \to ZZ_d$~(middle diagram) and $h \to Z_dZ_d$~(right diagram), respectively. In the latter two cases the production mechanism of the $125 \, {\rm GeV}$ Higgs boson is not shown.}
\end{center}
\end{figure}

The hypercharge portal leads to a modification of the couplings of the neutral gauge bosons to fermions. In the case of the $Z$-boson couplings the corrections start at~${\cal O}(\epsilon^2)$, while the dark photon vector couplings receive corrections already at ${\cal O} (\epsilon)$. Explicitly, one has
\begin{equation}
g_{Z_d f \bar f} = \epsilon \hspace{0.125mm} e \hspace{0.125mm} Q_f + \epsilon \hspace{0.125mm} e   \hspace{0.25mm}  \left ( Q_f - \frac{Y_f}{  \cos^2 \theta_w} \right ) \hspace{0.25mm} \frac{m_{Z_d}^2}{m_{Z}^2}  \,, 
\end{equation}
where $e = \sqrt{4 \pi \alpha}$ is the electromagnetic coupling and $Q_f$ ($Y_f$) is the electric charge (hypercharge) of the relevant fermion. Notice that for $m_{Z_d} \ll m_{Z}$ the $Z_d$-boson coupling to  fermions  is photon-like, while for  $m_{Z_d} \simeq m_{Z}$ the $Z_d$  boson couples to fermions like the SM Z boson. Referring to the $Z_d$ boson as dark photon is hence a bit of a misnomer, because a massive $Z_d$ boson always couples to neutrinos. We will however follow this established naming convention. At~${\cal O} (\epsilon)$ the hypercharge portal also leads to a coupling between the $125 \, {\rm GeV}$ Higgs, a $Z$ and a~$Z_d$ boson. To this order the coupling takes the form
\begin{equation}
g_{hZZ_d} = \frac{2\hspace{0.125mm} \epsilon \hspace{-0.125mm} \tan \theta_w}{v} \frac{m_{Z_d}^2 m_Z^2}{m_Z^2 - m_{Z_d}^2} \,.
\end{equation}
For $\epsilon \gtrsim {\cal O} (10^{-4})$ a  dark photon with mass $m_{Z_d} \gtrsim 1 \, {\rm GeV}$ decays promptly~\cite{Curtin:2013fra,Curtin:2014cca}.  At~the LHC, such dark photons can be searched for in Drell-Yan~(DY) dimuon production $pp \to Z_d \to \mu^+ \mu^-$ and in four-lepton final states that arise from the process $pp \to h \to Z Z_d \to 4 \ell$. The corresponding Feynman diagrams are shown on the left-hand side and in the middle of~Figure~\ref{fig:darkphoton}.

The Higgs portal gives rise to a coupling between the $125 \, {\rm GeV}$ Higgs and two dark photons. To lowest order in $\kappa$ this coupling can be written as 
\begin{equation}
g_{hZ_dZ_d} = \frac{ 2\hspace{0.125mm}   \kappa \hspace{0.125mm}  v \hspace{0.125mm}  m_{Z_d}^2}{m_{h_d}^2 - m_h^2} \,.
\end{equation}
For $m_{Z_d} < m_h/2$ this coupling allows for Higgs decays of the form $h \to Z_d Z_d$~---~a~representative Feynman diagram is shown on the right in~Figure~\ref{fig:darkphoton}. At the same order in $\kappa$ also  interactions between the $125 \, {\rm GeV}$ Higgs and two dark Higgses and a single~dark~Higgs and two dark photons exist. In the following, we will assume that the dark~Higgs is heavy,~i.e.~$m_{h_d} \gg m_h/2$. In such a case the exotic Higgs decay $h \to h_d h_d$ is kinematically forbidden and the production cross section for $pp \to h_d \to Z_d Z_d$ is always smaller than  $pp \to h \to Z_d Z_d$. As a result, for sufficiently light dark photons the  best probe of the Higgs portal parameter $\kappa$ is the exotic Higgs decay $h  \to Z_d Z_d$.    The corresponding partial decay width is given  to leading order in $\kappa$ by the following expression~\cite{Curtin:2013fra,Curtin:2014cca} 
\begin{equation} \label{eq:GahZdZd}
\Gamma \left ( h \to Z_d Z_d \right )=  \frac{\kappa^2 \hspace{0.125mm} v^2 }{32 \hspace{0.125mm} \pi \hspace{0.125mm} m_h} \, \frac{(1-4 x_{Z_d/h})^{1/2}}{ \left ( 1-  x_{h_d/h} \right )^2} \, \left ( 1 - 4  x_{Z_d/h} +12 x_{Z_d/h}^2 \right )  \,.
\end{equation}
Notice that given the small  total width of the SM Higgs boson, the branching ratio following from~(\ref{eq:GahZdZd}) can easily reach the few percent level for values of $\kappa \ll 1$. For example, taking $\kappa = 0.07$, $m_{Z_d} = 30 \, {\rm GeV}$ and $m_{h_d} = 300 \, {\rm GeV}$  leads to  ${\rm BR} ( h \to Z_d Z_d ) = 0.15$,  which is close to the model-independent  limit on ${\rm BR} ( h \to {\rm undet})$  reported in~(\ref{eq:undetinv}). 

The best way to constrain the Higgs portal coupling now depends on the size of kinetic mixing. For $\epsilon \gtrsim {\cal O} (10^{-4})$ the dark photon decays promptly and one can again study four-lepton final states~\cite{ATLAS:2018coo,CMS-PAS-HIG-19-007} to constrain the $h \to Z_d Z_d $ branching ratio and thereby~$\kappa$. For $\epsilon \lesssim {\cal O} (10^{-4})$ the decay length of the dark photon starts to become macroscopic. In fact, for ${\cal O} (10^{-8}) \lesssim \epsilon \lesssim {\cal O} (10^{-4})$ the decays of the $Z_d$ bosons are displaced with a large fraction of events ending up in the LHC detectors~\cite{Curtin:2013fra,Curtin:2014cca}. This~leads to the exciting opportunity to probe very small values of~$\epsilon$ that are inaccessible by other means, provided some Higgs mixing is present in the dark photon model. Below we summarise the LHC searches that have considered the case of LLPs in dark photon models with both a hypercharge and a Higgs portal.

\subsubsection{Experimental constraints}

The ATLAS, CMS and LHCb collaborations have carried out searches for prompt dark photon decays targeting hypercharge mixing parameters $\epsilon \gtrsim 10^{-4}$ in both the DY channel $pp \to Z_d \to \mu^+ \mu^-$~\cite{LHCb:2017trq,LHCb:2019vmc,CMS:2019buh} and in four-lepton production associated with $h \to Z Z_d$ and $h \to Z_d Z_d$~\cite{ATLAS:2018coo,CMS-PAS-HIG-19-007}. ATLAS has also searched for LLPs in the $h \to Z Z_d$ channel~\cite{ATLAS:2018niw}. In addition, smaller $\epsilon$ values have been tested by dedicated searches for long-lived dark photons in final states with displaced dimuon vertices arising from $h \to Z_d Z_d$. Building on earlier LHC~Run~1 analyses~\cite{CMS:2014hka,ATLAS:2015oan}, the latest ATLAS~and CMS results of this kind can be found in~\cite{ATLAS:2018rjc} and~\cite{CMS-PAS-EXO-20-014}, respectively. In the following we review only LHC searches for LLP signatures involving dark photons.

From all possible dark photon decay modes, the $Z_d\to \mu^+\mu^-$ process is experimentally the most accessible one. First, the LHC experiments can trigger on muons with a high efficiency of up to 90\%. Second, the rate of SM background processes is moderate, allowing for relatively low $p_T$ thresholds in the trigger. Third, the sophisticated MSs have a large acceptance for muon tracks with large impact parameters coming from displaced vertices. As a result, a wide range of proper lifetimes can be covered by the combination of the inner tracker employing LRT techniques and the MS.

\begin{figure}[!t]
\begin{center}
\includegraphics[width=.9\linewidth]{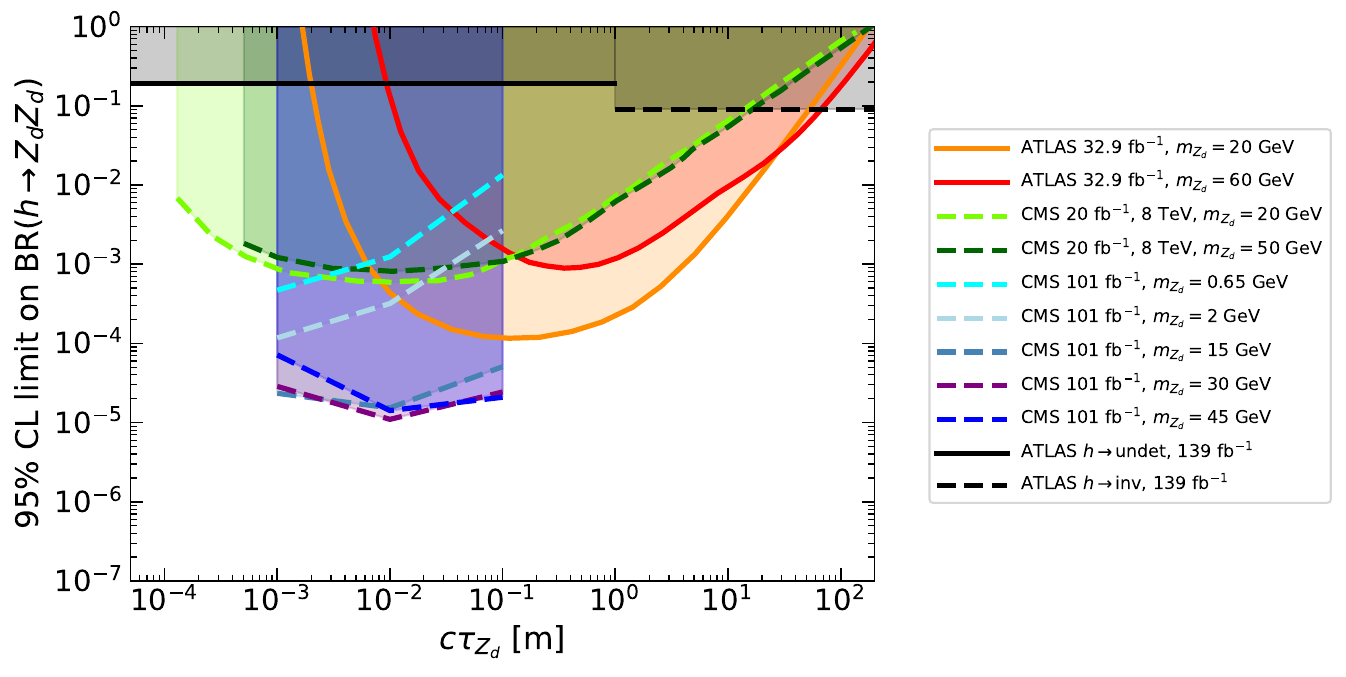} 
\vspace{2mm} 
\caption{\label{fig:darkphoton1}   95\%~CL upper limits on the branching ratio $h \to Z_d Z_d$ as a function of the decay length of the LLP. The displayed exclusions  correspond to the ATLAS~\cite{ATLAS:2018rjc} and CMS~\cite{CMS:2014hka,CMS-PAS-EXO-20-014} results for different mass hypotheses of the dark photon. For comparison the model-independent limits~(\ref{eq:undetinv}) found by ATLAS in~\cite{ATLAS-CONF-2020-027}  are also shown. The~boundary between the constraints from  undetected and  invisible Higgs decays is taken to be $c \tau_{Z_d} = 1 \, {\rm GeV}$  corresponding to $m_{Z_d} = 10 \, {\rm GeV}$.  See text for additional details. }
\end{center}
\end{figure} 

ATLAS searched for dark photons in the $Z_d\to \mu^+\mu^-$ decay mode targeting proper decay lengths of $10^{-3} \, {\rm m} <c\tau_{Z_d}<10^{3} \, {\rm m}$ by requiring displaced vertices consistent with LLP decays to be reconstructed using the muon system~\cite{ATLAS:2018rjc}. A combination of single muon, dimuon and trimuon triggers with progressively lower thresholds down to $p_T>6 \, {\rm GeV} $, as well as a $E_T^{\rm miss}$ trigger have been employed. The reconstruction efficiency for muon tracks from displaced vertices ranges from 70\% for small impact parameters to 10\% at the acceptance limit of $4 \, {\rm m}$. Finally, displaced vertices of oppositely charged muons with an invariant mass above $15 \, {\rm GeV}$ are selected within the fiducial volume.

CMS  also performed a search for dark photons in $\mu^+\mu^-$ final states~\cite{CMS-PAS-EXO-20-014}.  In contrast to the  aforementioned ATLAS search, this search analysed muon tracks reconstructed in the inner tracker, targeting small proper decay lengths of $10^{-3} \, {\rm m} <c\tau_{Z_d}<10^{-1} \, {\rm m}$. The novelty of this analysis is that it searches for pairs of oppositely charged muons with an invariant mass down to the dimuon production threshold. This is achieved using the so-called data scouting technique~\cite{CMS:2016ltu,CMS:2019buh}, where only partial event information, as reconstructed by the high-level trigger system, is recorded. This dramatically reduces the event size, thereby allowing to store orders of magnitude more candidate events for analysis, and lower the muon $p_T$ thresholds down to $4 \, {\rm GeV}$. Finally, candidate muon pairs from the $Z_d\to\mu^+\mu^-$ decay are selected in events with two muons targeting  DY production, and four muons with an invariant mass consistent with the $125 \, {\rm GeV}$ Higgs boson aiming for  $pp\to h\to Z Z_d$ and $pp\to h\to Z_dZ_d$. A sliding window fit is then applied to the  distributions of dimuon invariant masses in several signal regions to look for a potential excess from $Z_d\to\mu^+\mu^-$ decays away from the known SM resonances like the $J\!/\!\psi$  and $\Upsilon$ mesons.  

The region with $c\tau_{Z_d}<10^{-3} \, {\rm m}$ is covered by the CMS search for displaced $Z_d \to e^+ e^-, \mu^+ \mu^-$ decays performed at $8 \, {\rm TeV}$~\cite{CMS:2014hka}, which targets the proper decay length range between $10^{-4} \, {\rm m}$ and $10^2 \, {\rm m}$. This analysis is similar to the $13 \, {\rm TeV}$ CMS search~\cite{CMS:2019buh}, except that it does not apply data scouting techniques and considers both the dielectron and the dimuon channel. Both search channels rely on the dilepton triggers.

In Figure~\ref{fig:darkphoton1} we summarise the existing 95\%~CL bounds on ${\rm BR} \left ( h \to Z_d Z_d \right )$ as a function of the proper decay length $c\tau_{Z_d}$ of the dark photon. The shown ATLAS limits~\cite{ATLAS:2018rjc} correspond to the different dark photon masses $m_{Z_d} = 20 \, {\rm GeV}, 60 \, {\rm GeV}$, while in the case of CMS we display bounds for $m_{Z_d} = 20 \, {\rm GeV}, 50 \, {\rm GeV}$ from~\cite{CMS:2014hka} and for $m_{Z_d} = 0.65 \, {\rm GeV}, 2 \, {\rm GeV}, 15 \, {\rm GeV}, 30 \, {\rm GeV}$, $50 \, {\rm GeV}$ from~\cite{CMS-PAS-EXO-20-014}. Notice that for dark photons with masses of a few tens of GeV, the strongest bound on ${\rm BR} \left ( h \to Z_d Z_d \right )$ typically arises for the lightest choice of the dark photon mass. For $m_{Z_d} \lesssim 5 \, {\rm GeV}$ the limits on ${\rm BR} \left ( h \to Z_d Z_d \right )$ however become significantly weaker with decreasing dark photon mass even if data scouting techniques are employed. Specifically, for dark photons with masses above $15 \, {\rm GeV}$ the ATLAS (CMS) measurements exclude ${\rm BR} \left ( h \to Z_d Z_d \right )\gtrsim10^{-4}$~($10^{-5}$) for proper decay lengths of around $10^{-1} \, {\rm m}$ ($10^{-2} \, {\rm m}$), while for~GeV-scale dark photons the corresponding CMS exclusions are weaker by about two orders of magnitude. One also observes that for proper decay lengths in the range of around $10^{-3} \, {\rm m}$ to $10^{2} \, {\rm m}$ the bound on ${\rm BR} \left ( h \to Z_d Z_d \right )$ that derives from the LLP searches~\cite{ATLAS:2018rjc,CMS-PAS-EXO-20-014} is stronger than the model-independent limits quoted in~(\ref{eq:undetinv}). For shorter or longer proper decay lengths the ATLAS Higgs coupling measurement~\cite{ATLAS-CONF-2020-027} however provides currently the best constraint on the branching ratio of $h \to Z_d Z_d$.

\begin{figure}[!t]
\begin{center}
\includegraphics[width=.95\linewidth]{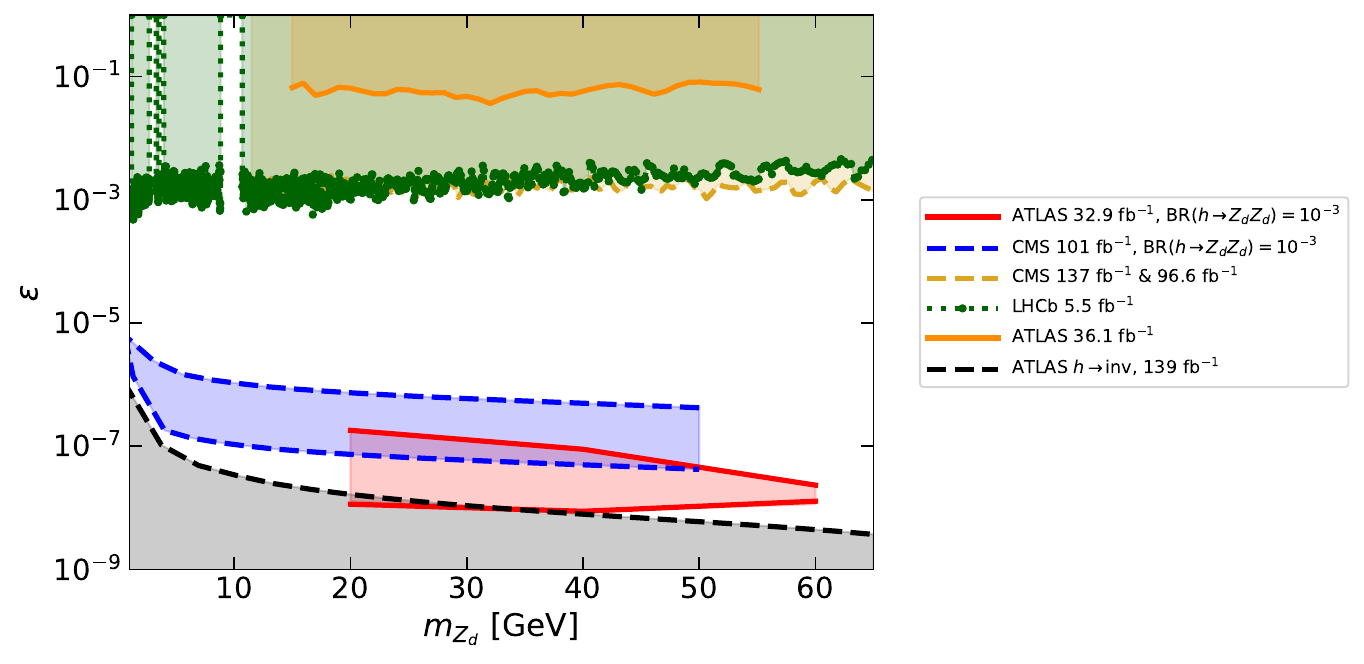}
\vspace{2mm} 
\caption{\label{fig:darkphoton2}   95\%~CL excluded regions in the $m_{Z_d}\hspace{0.25mm}$--$\hspace{0.5mm}\epsilon$  plane following from the ATLAS~\cite{ATLAS:2018rjc} and CMS~\cite{CMS-PAS-EXO-20-014} measurement. In the case of the ATLAS (CMS) results the shown solid red (dashed~blue) bound corresponds to ${\rm BR} \left ( h \to Z_d Z_d \right ) = 10^{-3}$. The 90\%~CL upper limit on the hypercharge mixing parameter that derives from DY dimuon production by LHCb~\cite{LHCb:2019vmc} and CMS~\cite{CMS:2019buh} is also displayed for comparison as a solid green and a dashed yellow line, respectively. The 95\%~CL limit on $\epsilon$ from the ATLAS search~\cite{ATLAS:2018coo} for $pp \to h \to Z Z_d \to  4\ell$ is finally indicated by a solid orange line. The latter three bounds hold for any value of the $h \to Z_d Z_d$ branching ratio. Finally, the indirect bound that follows from the limit~(\ref{eq:undetinv}) on invisible Higgs decays  is shown as a dashed black line. For further details consult the text. }
\end{center}
\end{figure}

The limits presented  in Figure~\ref{fig:darkphoton1} can also be translated into 95\%~CL exclusions in the $m_{Z_d}\hspace{0.25mm}$--$\hspace{0.5mm}\epsilon$  plane. This is done in Figure~\ref{fig:darkphoton2} where we show the constraints that derive from the ATLAS~\cite{ATLAS:2018rjc} and CMS~\cite{CMS-PAS-EXO-20-014} measurement assuming ${\rm BR} \left ( h \to Z_d Z_d \right ) = 10^{-3}$. For the same $h \to Z_d Z_d $ branching ratio, the exclusion that follows from~\cite{CMS:2014hka} almost exactly resembles that of~\cite{CMS-PAS-EXO-20-014} for dark photon masses between $20 \, {\rm GeV}$ and $50 \, {\rm GeV}$. This limit is therefore not displayed in the figure. One observes that $\epsilon$ values of order $10^{-8}$ and $10^{-7}$ are excluded by these searches. Notice that the CMS search leads to weaker limits in terms of $\epsilon$ than the ATLAS search, because CMS does not consider decay lengths beyond $10^{-1} \, {\rm m}$. 
 For comparison we also depict in~Figure~\ref{fig:darkphoton2} the 90\%~CL bounds on the hypercharge mixing parameter  that follow from the searches by LHCb and CMS in DY~dimuon production~\cite{LHCb:2019vmc,CMS:2019buh} as well as  the ATLAS search~\cite{ATLAS:2018coo} that focuses on the $pp \to h \to Z Z_d \to 4 \ell$ channel. These prompt limits are a few orders of magnitude weaker than the bounds that arise from the LLP searches, but they  make no assumption about the amount of Higgs mixing in the dark photon model. In addition, searches for $pp \to h \to Z_d Z_d \to 4 \ell$~\cite{ATLAS:2018coo,CMS-PAS-HIG-19-007} also provide some sensitivity to $\epsilon$, but the resulting upper limits are significantly weaker than those that stem from the DY searches and hence not reported in the figure. 
 
 An indirect constraint on the $m_{Z_d}\hspace{0.25mm}$--$\hspace{0.5mm}\epsilon$ plane can also be derived from the available limits on the invisible Higgs branching ratio~\cite{ATLAS:2020kdi,ATLAS-CONF-2020-027}. Imposing ${\rm BR} \left ( h \to {\rm inv} \right ) = 0.1$, we obtain the dashed black contour shown in Figure~\ref{fig:darkphoton2} under the assumption that the dark photon is not registered by standard reconstruction algorithms for $c \tau_{Z_d} > 0.1 m \hspace{0.5mm} ( m_{Z_d}/{\rm GeV})$, resulting in a~$E_T^{\rm miss}$ contribution. Another assumption is that the kinematic distributions such as the $E_T^{\rm miss}$ spectrum are the same for $pp \to h + X \to Z_d Z_d +X$ and for the SM Higgs production channels that go into the ${\rm BR} \left ( h \to {\rm inv} \right )$ bounds given in~(\ref{eq:BRhinv}) and~(\ref{eq:undetinv}). While~\cite{ATL-PHYS-PUB-2021-020} suggests that these are good assumptions, we cannot quantify the associated systematic uncertainties. The shown indirect bound from invisible Higgs decays has therefore only an indicative character. However, it should be straightforward for ATLAS and CMS to directly reinterpret their Higgs to invisible searches in the context of long-lived dark photons thereby improving on our naive estimate.

\subsection{Models with a vector and a fermion portal} 
\label{sec:FRVZ}

Hidden sector models with both a vector and a fermion portal such as the Falkowski-Ruderman-Volansky-Zupan~(FRVZ) model~\cite{Falkowski:2010cm,Falkowski:2010gv} also include a hidden photon as a possible LLP and can lead to signatures with displaced charged leptons. However, in models of this type, the hidden photons are not directly produced in the exotic decays of the $125 \, {\rm GeV}$ Higgs boson, but through a cascade involving SUSY and hidden sector particles with masses of ${\cal O} ( 10 \, {\rm GeV})$ or below. This implies that the hidden photons have to be even lighter than the other particles in the decay chain, with masses of ${\cal O} (1 \, {\rm GeV})$ in order to be kinematically accessible. Due~to their small mass, the hidden photons are preferentially produced with large boosts at the LHC, resulting in so-called lepton-jets~\cite{Arkani-Hamed:2008kxc,Baumgart:2009tn,Cheung:2009su},~i.e.~collimated groups of leptons in a jet-like structure. These lepton-jet events are accompanied by varying amounts of~$E_{T}^{\rm miss}$ depending on the precise pattern of the Higgs decay.

\begin{figure}[!t]
\begin{center}
\includegraphics[width=0.75\textwidth]{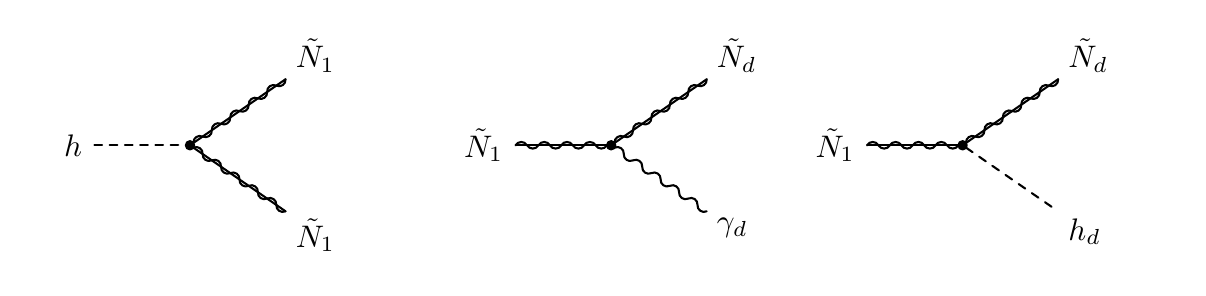} 
\vspace{-6mm} 
\caption{\label{fig:FRVZdiagrams1} Left diagram: Higgs decay to a pair of the lightest MSSM neutralinos. Middle and right diagram: Possible decays of the MSSM bino to the hidden sector, which follow from gaugino kinetic mixing~(\ref{eq:FRVZint}). See text for further details.}
\end{center}
\end{figure}

\subsubsection{Theory}

A minimal model that realises the general idea proposed in~\cite{Falkowski:2010cm,Arkani-Hamed:2008kxc,Baumgart:2009tn,Cheung:2009su,Falkowski:2010gv} contains a massive hidden photon $\gamma_d$ that communicates with the visible sector through mixing with the hypercharge field~---~see the first term in~(\ref{eq:Ldp}). In the FRVZ model, the particle content of the visible sector is that of the minimal supersymmetric SM~(MSSM). Supersymmetrising the hypercharge portal leads to a mixing of the visible bino ($\tilde B$) and the hidden gaugino~($\tilde \gamma_d$). Removing the kinetic mixing between the $\tilde B$ and the $\tilde \gamma_d$ then gives rise to interactions between all hidden fields charged under $U(1)_d$ and the visible neutralinos that are proportional to the hypercharge mixing parameter $\epsilon$. In particular, one obtains an interaction term of the following form
\begin{equation} \label{eq:FRVZint}
{\cal L} = -\frac{\epsilon  \hspace{0.25mm}  g_d}{\cos \theta_w} \, \tilde B  \hspace{0.25mm}  \sum_i  \hspace{0.25mm}  q_i \hspace{0.25mm} h_d^{i{ \dagger}} \hspace{0.25mm}  \tilde h_d^i \,, 
\end{equation}
where $g_d$ is the $U(1)_d$ gauge coupling, $h_d^i$ are the hidden scalar fields, $\tilde h_d^i$ are the hidden gauginos and $q_i$ is the relevant $U(1)_d$ charge.

The interactions~(\ref{eq:FRVZint}) lead to vertices involving a visible neutralino, a hidden neutralino and a hidden photon or a hidden Higgs boson.  An exotic Higgs decay signal can therefore arise in the FRVZ model as follows. Initially the Higgs decays into a pair of the lightest visible neutralinos ($\tilde N_1$), as indicated by the Feynman diagram on the left-hand side of Figure~\ref{fig:FRVZdiagrams1}. In the pure MSSM without the hypercharge portal, the $\tilde N_1$ could be a DM candidate,~i.e.~the lightest SUSY particle or LSP, but in the FRVZ model the presence of the term~(\ref{eq:FRVZint}) allows the $\tilde N_1$ to decay into hidden sector states. Example diagrams are shown in the middle and on the right in~Figure~\ref{fig:FRVZdiagrams1}. In the first case, the~$\tilde N_1$ decays to the lightest hidden sector gaugino ($\tilde N_d$) and a hidden photon, while in the second case the final-state~$\tilde N_d$ is accompanied by a hidden Higgs boson ($h_d$). The lightest hidden sector neutralino~$\tilde N_d$ is stable and therefore a  DM candidate, but the  $\gamma_d$ and the $h_d$ decay further. If the hidden photon is the lightest hidden state then  $\gamma_d \to \ell^+ \ell^-$ and $h_d \to \gamma_d \gamma_d \to 4 \ell$ are possible decay chains that will lead to the aforementioned lepton-jet signatures. Since both  the hypercharge portal in~(\ref{eq:Ldp}) as well as the gaugino kinetic mixing~(\ref{eq:FRVZint}) are proportional to~$\epsilon$, depending on the magnitude of this parameter the lepton-jets can be either prompt or displaced.  

\begin{figure}[!t]
\begin{center}
\includegraphics[width=0.65\textwidth]{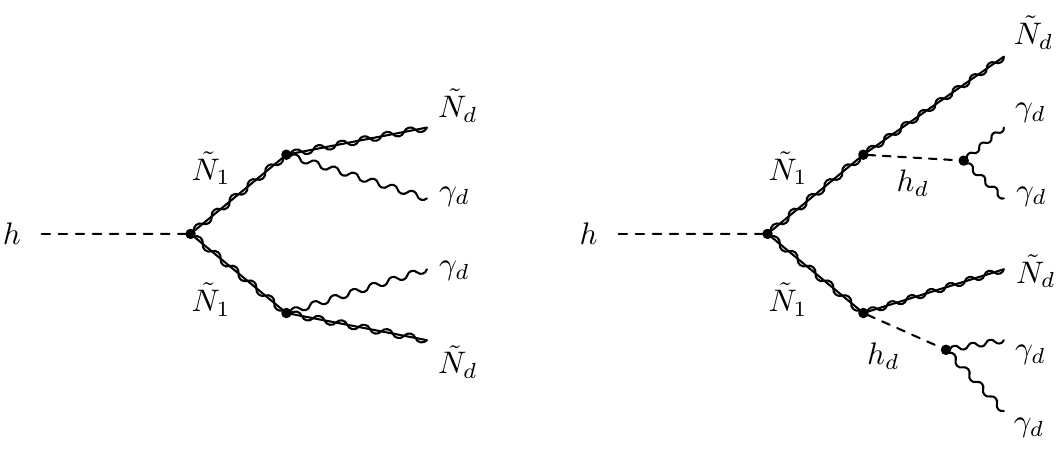} 
\vspace{2mm} 
\caption{\label{fig:FRVZdiagrams2} Benchmark topologies that have been studied in the context of the FRVZ model at the LHC. In the first process (left diagram) both neutralinos produced in $h \to \tilde N_1 \tilde N_1$ decay through $\tilde N_1  \to \tilde N_d \gamma_d$. The hidden photon further decays to all kinematically accessible SM fermions, leading to four-fermion  final states. In the second process (right diagram) the neutralinos instead decay  via $\tilde N_1  \to \tilde N_d h_d$ followed by the decay of the hidden Higgs $h_d \to \gamma_d \gamma_d$. After the decay of the hidden photons the associated final state contains eight SM fermions. The hidden neutralinos~$\tilde N_d$ escape the LHC detectors undetected and therefore appear as a $E_T^{\rm miss}$ signature. }
\end{center}
\end{figure}

Notice that models of inelastic DM (see for instance~\cite{Tucker-Smith:2001myb,Izaguirre:2015zva,Izaguirre:2017bqb,Berlin:2018jbm,Duerr:2019dmv,Duerr:2020muu}) also possess many of the features discussed above. Such models typically contain two fermionic states~$\chi_1$ and $\chi_2$ with a small mass splitting $\Delta m = m_{\chi_2} - m_{\chi_1} > 0$. The role of $\tilde N_1$ ($\tilde N_d$) is played by~$\chi_2$ ($\chi_1$) in inelastic DM models, and the DM candidate $\chi_1$ can be excited to its heavier twin~$\chi_2$ by absorbing a massive dark photon $\gamma_d$. The~simplest realisation of such a scenario consists in assuming a hypercharge portal and postulating a $U(1)_X$ symmetry that is spontaneously broken by a dark Higgs $h_d$. This dark Higgs couples to a pair of dark photons. The particle content and coupling structure of inelastic DM models therefore resemble quite closely those of the simplest FRVZ models. The difference between the two types of models is that, unlike the $\gamma_d \tilde N_1 \tilde N_d$ coupling (\ref{eq:FRVZint}), the $ \gamma_d \chi_1 \chi_2$ coupling is not proportional to the hypercharge mixing parameter~$\epsilon$, but only involves the $U(1)_X$ coupling constant. The~simplest inelastic DM models are hence in some sense a generalisation of the FRVZ idea. In fact, all the existing FRVZ interpretations of LHC searches correspond to such a generalisation, where the $\gamma_d \chi_1 \chi_2$, $\gamma_d f \bar f$, $h_d \chi_1 \chi_2$ and $h \chi_2 \chi_2$ couplings are effectively treated as free parameters. As a concrete example, the ATLAS collaboration has used the following parameters in their interpretations
\begin{equation} \label{eq:iDMbench}
\begin{split}
& \hspace{12mm} g_{ \gamma_d \chi_1 \chi_2} = 0.31  \,, \quad 
g_{ h_d \chi_1 \chi_2} = 0.1  \,, \\[2mm]
& m_{\chi_1} = 2 \, {\rm GeV} \,, \quad 
m_{\chi_2} = 5 \, {\rm GeV} \,, \quad 
m_{h_d} = 2 \, {\rm GeV} \,, 
\end{split}
\end{equation}
where the choice of $g_{ \gamma_d \chi_1 \chi_2}$ corresponds to $g_{ \gamma_d \chi_1 \chi_2} = e = \sqrt{4 \pi \alpha}$ with $\alpha$ the electromagnetic fine structure constant at the EW scale. The hypercharge mixing parameter~$\epsilon$ and the coupling~$g_{ h \chi_2 \chi_2}$ are not directly used as external parameters, but expressed through~$c \tau_{\gamma_d}$ and ${\rm BR} \left ( h \to \chi_2 \chi_2 \right)$, respectively, which then serve as input. Notice that for the parameter choices (\ref{eq:iDMbench}) and sufficiently large ${\rm BR} \left ( h \to \chi_2 \chi_2 \right)$ values, the decays $ h \to \chi_2 \chi_2$ and $\chi_2 \to \chi_1 \gamma_d$ are necessarily prompt. From a theoretical point of view the generalised~FRVZ model is therefore quite similar to the dark photon model discussed in~Section~\ref{sec:darkphoton} if~${\rm BR} \left ( h \to \chi_2 \chi_2 \right)$~is identified with ${\rm BR} \left ( h \to Z_d Z_d \right)$.

\subsubsection{Experimental constraints}

The ATLAS and CMS collaborations  have searched for collimated groups of charged leptons or light hadrons in a jet-like structure to constrain exotic Higgs decays by exploring both prompt~\cite{ATLAS:2015itk,CMS:2015nay} and displaced \cite{ATLAS:2019tkk,CMS:2018jidc} signatures. The results were interpreted in the  generalised FRVZ framework described above, as well as in the context of other portal models. In the case of the generalised FRVZ  model, the published LHC searches have focused on the two benchmark processes illustrated in Figure~\ref{fig:FRVZdiagrams2}, with the Higgs boson produced in the ggF topology. 

The searches for prompt Higgs decays targeted the $h\to \gamma_d\gamma_d+X$ topology (cf.~left diagram in Figure~\ref{fig:FRVZdiagrams2}), where the hidden photon is assumed to decay into charged leptons. The ATLAS search~\cite{ATLAS:2015itk} included $\gamma_d$ decays to $e^+e^-$ and $\mu^+\mu^-$, while the CMS search~\cite{CMS:2015nay} only considered dimuons. Hadronic decays of the hidden photon were not included, since they cannot be easily separated from the QCD multijet background. These searches probe signal hypotheses with hidden photon masses in the ranges of $0.1 \, {\rm GeV}$ to $2 \, {\rm GeV}$ for ATLAS and $0.25 \, {\rm GeV}$ and $2 \, {\rm GeV}$ for CMS. In the ATLAS analysis, events are required to have at least two lepton-jets, which are reconstructed by clustering tracks from the primary vertex within a radius $\Delta R = \sqrt{ (\Delta \eta)^2 + (\Delta \phi)^2 }=0.5$ of the highest $p_T$ track, 
  and subsequently matching them to electron or muon candidates. In the CMS analysis, events are required to have at least two dimuon pairs, which are reconstructed by combining muon-candidate tracks into a common vertex.
  
In the domain of searches targeting LLP signatures, the ATLAS analysis~\cite{ATLAS:2019tkk} has probed the $h\to 4\gamma_d+X$ topology (cf.~right diagram in Figure~\ref{fig:FRVZdiagrams2}) considering both leptonic and hadronic decays of the hidden photons. The mass of the hidden photon was set to $0.4 \, {\rm GeV}$, resulting in a 10\% branching ratio into pions, and the rest of the branching ratio divided equally between electrons and muons~\cite{Falkowski:2010cm}. The dark photon decays into muon pairs were targeted by reconstructing jets of muons that were reconstructed using the MS only, while the decays into electron pairs and hadrons were reconstructed as calorimeter jets with a high fraction of energy deposited in the hadronic calorimeters. In both channels, the jets were required to be isolated from any ID activity, as expected for LLPs, and multivariate analysis techniques using timing and topological information of the jet were employed to discriminate signal from background. The CMS search~\cite{CMS:2018jidc} explored the $h\to \gamma_d\gamma_d+X$ topology (cf.~left diagram in Figure~\ref{fig:FRVZdiagrams2}), considering only $\gamma_d\to\mu^+\mu^-$ decays. Hidden photon masses in the range $0.25 \, {\rm GeV}$ and $8.5 \, {\rm GeV}$ were explored, with the upper limit set by the requirement $m_{\mu \mu} < 9 \, {\rm GeV}$ in order to sufficiently suppress the background from DY and $\Upsilon \to \mu^+\mu^-$ production. Proper decay lengths below $c\tau_{\gamma_d}<0.1 \, {\rm m}$ were targeted, explicitly including both prompt and displaced signatures. Candidate events were required to have exactly two $\gamma_d\to\mu^+\mu^-$ candidates that are isolated from significant activity in the tracking system and have an invariant mass consistent with each other.

\begin{figure}[!t]
\begin{center}
\includegraphics[width=\linewidth]{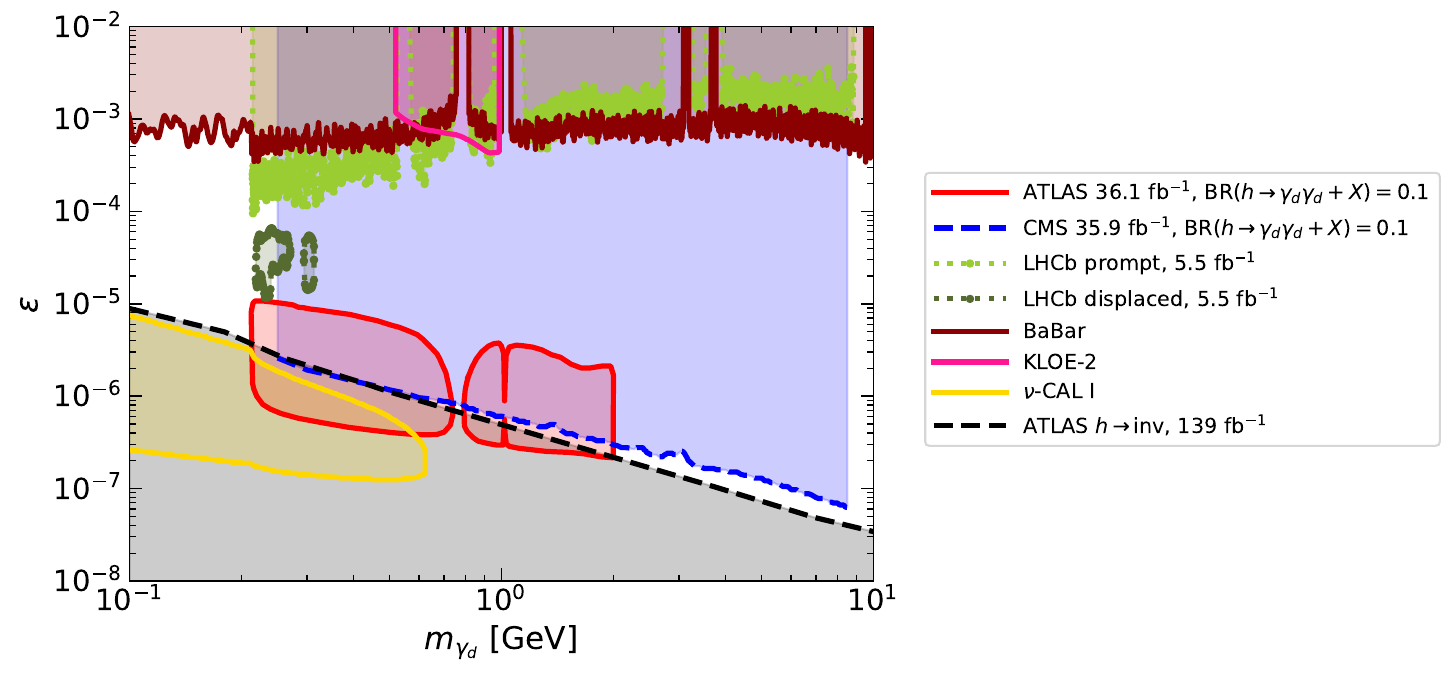} 
\vspace{-2mm} 
\caption{\label{fig:FRVZsummary}  
90\%~CL upper limits that follow from the ATLAS search~\cite{ATLAS:2019tkk}  and the CMS search~\cite{CMS:2018jidc} for the exotic Higgs boson decay $h \to \gamma_d \gamma_d +X$ in the generalised FRVZ model. Both exclusions assume ${\rm BR} \left ( h \to \gamma_d \gamma_d + X\right ) = 0.1$. For comparison also  constraints in the  $m_{\gamma_d}\hspace{0.25mm}$--$\hspace{0.5mm}\epsilon$  plane  from LHCb~\cite{LHCb:2019vmc},  BaBar~\cite{BaBar:2014zli}, KLOE-2~\cite{KLOE-2:2018kqf} and  $\nu$-CAL~I~\cite{Blumlein:1991xh} are shown. The latter bounds hold for any value of the $h \to \gamma_d \gamma_d + X$ branching ratio. Finally, the indirect  upper limit that follows from (\ref{eq:undetinv}) is  depicted.    Consult the text for further explanations.}
\end{center}
\end{figure}

Figure~\ref{fig:FRVZsummary} contains a summary of 90\%~CL exclusions in the $m_{\gamma_d}\hspace{0.25mm}$--$\hspace{0.5mm}\epsilon$~plane in the generalised FRVZ model for hidden photon masses in the range of $0.1 \, {\rm GeV}$ to $10 \, {\rm GeV}$. The constraints following from the ATLAS~\cite{ATLAS:2019tkk} and CMS~\cite{CMS:2018jidc} search for the exotic Higgs decay $h \to \gamma_d \gamma_d +X$ apply in the generalised FRVZ model and assume ${\rm BR} \left ( h \to \gamma_d \gamma_d + X\right ) = 0.1$. The ATLAS analysis employs~(\ref{eq:iDMbench}) and focuses on the mass range $0.2 \, {\rm GeV} \lesssim m_{\gamma_d} \lesssim 3.6 \, {\rm GeV}$ and small kinetic mixings that lead to LLP signatures. Depending on $m_{\gamma_d}$ it excludes hypercharge mixing parameters within $1 \cdot 10^{-5} \lesssim \epsilon \lesssim 3 \cdot 10^{-7}$. The CMS search probes hidden photon masses in the range $0.25 \, {\rm GeV} \leq m_{\gamma_d} \leq 8.5 \, {\rm GeV}$ and targets proper decay lengths of $c \tau_{\gamma_d} \leq 0.1 \, {\rm m}$. This analysis is able to exclude hypercharge mixing parameters $\epsilon \lesssim 3 \cdot 10^{-6}$ ($\epsilon \lesssim 7 \cdot 10^{-8}$) at low~(high) $m_{\gamma_d}$. The~displayed LHCb~\cite{LHCb:2019vmc}, BaBar~\cite{BaBar:2014zli}, KLOE-2~\cite{KLOE-2:2018kqf} and $\nu$-CAL~I~\cite{Blumlein:1991xh} bounds have been taken from the {\tt Darkcast} package developed in~\cite{Ilten:2018crw}. They hold irrespectively of the value of the $h \to \gamma_d \gamma_d + X$ branching ratio. The exclusion that follows from the limit~(\ref{eq:undetinv}) is obtained from our recast as described in Section~\ref{sec:darkphoton} and assumes ${\rm BR} \left ( h \to {\rm inv} \right) =0.1$. The limit arising from the recent ATLAS analysis~\cite{ATL-PHYS-PUB-2021-020} of mono-jet signatures is not shown in the figure because it is not sensitive to ${\rm BR} \left ( h \to \gamma_d \gamma_d + X \right ) = 0.1$. Figure~\ref{fig:FRVZsummary} clearly shows that in the context of the generalised FRVZ model, LHC searches for exotic Higgs decays provide an opportunity to probe values of $\epsilon$ that are at present inaccessible by other means. Future LLP experiments like MATHUSLA, CODEX-b, ANUBIS, FASER~\cite{FASER:2018eoc} and FASER2~\cite{Anchordoqui:2021ghd} located at the LHC, in combination with Belle~II~\cite{Duerr:2019dmv,Duerr:2020muu}, are expected to set additional stringent constraints on the $m_{\gamma_d}\hspace{0.25mm}$--$\hspace{0.5mm}\epsilon$~plane in the generalised FRVZ or inelastic DM frameworks.

\section{Outlook}
\label{sec:conclusions}

In this article we have reviewed the status of the LHC constraints on DM scenarios where the $125 \, {\rm GeV}$ Higgs boson plays a prominent role. Specifically, we have covered the case of SM Higgs portals~(Section~\ref{sec:EFTHiggsPortal}), models with extended Higgs and gauge sectors with one mediator (Section~\ref{sec:extendedhiggsportals}) or two mediators (Section~\ref{sec:extendedhiggsgaugeportal}) as well as theories that can lead to LLP signatures induced by the decay of the $125 \, {\rm GeV}$ Higgs~(Section~\ref{sec:LLP}). In~all cases, we have collected the existing constraints from the different LHC experiments to obtain state-of-the-art summary plots that show the current LHC sensitivities to the relevant model parameters in various benchmark scenarios. In some cases, we have also indicated in these summary plots the restrictions that other measurements impose, including those of the properties of the $125 \, {\rm GeV}$ Higgs boson, flavour physics,  EW precision measurements, DM~DD and ID experiments, and the relic density to just name a few.

While our review mostly focuses on ``known knowns'', i.e.~existing experimental constraints and their interpretations, we have also tried to discuss aspects of the considered models that are in principle known but that might be unknown to at least some of the readers, i.e.~``unknown knowns''. For~instance, in the case of the Higgs portal models we have emphasised in Section~\ref{sec:2.4} that $E_T^{\rm miss}$ searches for off-shell Higgs production in the VBF, $t \bar t$ and $tW$, and potentially other channels provide sensitivity to the parameter space where $m_{\rm DM} > m_h/2$. In fact, in the case of the kinetic Higgs portal such searches are the only known way to test DM masses above the Higgs threshold, making dedicated experimental searches and/or interpretations of the relevant mono-$X$ signatures in our opinion an important goal for future LHC runs. In the context of the Higgs portal models, we have furthermore argued in Section~\ref{subsec:further} that interpreting LHC limits on the invisible Higgs branching ratio in terms of a SI~DM-nucleon cross section leads to somewhat unintuitive results, in the sense that the indirectly obtained $\sigma_{\rm SI}$ bounds do not become constant in the limit $m_{\rm DM} \ll m_h/2$, as one would naively expect. We have explained the reason for this unexpected feature. To avoid potential confusion, we suggest to present the LHC limits on ${\rm BR } \left ( h \to {\rm inv} \right)$ as well as the limits on $\sigma_{\rm SI}$ obtained by DD experiments in terms of the effective interaction strength of the different Higgs portal models, for example in terms of $|c_m|$ in the case of the marginal Higgs portal. This is the standard presentation in the theoretical literature on EFT Higgs portals, and adding it to the interpretation of results from the LHC and/or DM DD experiments would further promote a fruitful exchange between these communities as well as the theory community.

We have stressed at the end of Section~\ref{sec:2HDMa} that all existing collider studies of the 2HDM+$a$ model have assumed a Yukawa sector of~type-II. For~this choice, the bounds from FCNC processes and LHC searches for heavy Higgs bosons are strong, pushing the masses of the additional 2HDM Higgs bosons above the $500 \, {\rm GeV}$ range. In~fermiophobic 2HDM models of type-I the constraints on the additional Higgs bosons can be significantly relaxed, thereby allowing for new scalars and pseudoscalars with EW-scale masses. The fermiophobic nature of the Higgs bosons can lead to unconventional production mechanisms and decay patterns of the non-SM spin-0 particles. In~our opinion, the mono-$X$ phenomenology in a fermiophobic 2HDM+$a$ deserves dedicated studies. The Yukawa sector of the model also plays an important role for the 2HDM+$s$ model considered in Section~\ref{sec:2HDMs}. In particular, we have shown that a suitable choice of the Yukawa sector allows to tame excessive 2HDM+$s$ corrections to the SI DM-nucleon cross section, opening up the possibility to test model realisations at the~LHC which lead to a viable DM phenomenology that is consistent with the observed relic density.

The existing LHC interpretations of the 2HDM-$Z^\prime$ model and the 2MDM model discussed in Section~\ref{sec:2HDMZp} and Section~\ref{sec:darkHiggs}, respectively, have all focused on the mono-Higgs channel with $h \to b \bar b, \gamma \gamma$ in the former case and on the mono-$s$ channel with $s \to b \bar b, VV$ in the latter case. In order to present a more global picture of the LHC sensitivity to these models, we have reinterpreted existing LHC searches for heavy spin-1 resonances decaying to visible particles. Our studies show that mono-$X$ searches are not the only way to probe these two-mediator models at the~LHC. In fact, both discussed DM models are in general more tightly constrained by resonance searches for the $Z^\prime$ boson in dijet or $t\bar{t}$ final states than by mono-$X$ searches. We~believe this to be a rather generic feature of DM models with a heavy spin-1 mediator. Despite the dominant sensitivity of resonance searches in the context of the 2HDM-$Z^\prime$ and 2MDM models, searches for $X + E_T^{\rm miss}$ signals are important given their distinct experimental signature. We have furthermore pointed out that the 2HDM-$Z^\prime$ model can also be probed by a search for $Z^\prime \to Zh$~resonances. Similarly, we advocate the exploration of the $hh + E_T^{\rm miss}$ and $tt + E_T^{\rm miss}$ topologies in the context of the 2MDM model. We are convinced that only through a combined exploration of the whole suite of searches in various channels can the full LHC potential be exploited.

Searches for BSM LLPs have gained a lot of momentum in LHC~Run~2. In Section~\ref{sec:NN}, Section~\ref{sec:darkphoton} and Section~\ref{sec:FRVZ}, we have presented state-of-the-art summary plots for models of neutral naturalness/hidden valleys, dark photon models and BSM theories with both a hypercharge and a fermion portal, respectively, that can lead to LLP signatures in the decay of the $125 \, {\rm GeV}$ Higgs boson. We have stressed that an indirect way to constrain models of this type  is provided by the  precision measurements of the $125 \, {\rm GeV}$ Higgs properties in visible and invisible final states.  For instance, in the case of neutral naturalness we find  that the limits from  ${\rm BR} \left ( h \to {\rm undet} \right )$ and ${\rm BR} \left ( h \to {\rm inv} \right )$ provide currently the strongest LHC bound on ${\rm BR} \left ( h \to a a \right )$ for $c\tau_a \lesssim 10^{-4} \, {\rm m}$ and $c\tau_a \gtrsim 10^{3} \, {\rm m}$, respectively. To further emphasise the possibility to test LLP scenarios via $E_T^{\rm miss}$ searches, we have in the case of the two models with a hypercharge portal performed recasts of the limits imposed by the latest $h \to {\rm inv}$ results to derive upper limits on the hypercharge mixing parameter $\epsilon$ as a function of the dark photon mass. 

The ``known knowns'' and the ``unknown knowns'' discussed in this review are clearly only a snapshot of the broad landscape of collider searches for DM through the Higgs lens. We tried our best to represent a complete picture of experimental searches and the relevant theoretical works, and apologise for any potential omission. We hope that our review can trigger the experimental exploration of new search strategies and possibly even groundbreaking new ideas or discoveries in the years leading to the~HL-LHC era.

\acknowledgments{We would like to thank Andreas~Albert, Martino~Borsato, Karri Folan Di Petrillo and Felix~Kahlhoefer for their useful comments on the manuscript. This work is supported by the German Research Foundation (DFG) under grant No.~AR~1321/1-1 and the Isaac Newton Trust under grant No.~G101121}


\end{paracol}
\reftitle{References}




%

\end{document}